\documentclass[a4paper,onecolumn,accepted=2022-03-22]{quantumarticle}
\pdfoutput=1

\usepackage[utf8]{inputenc}
\usepackage[english]{babel}
\usepackage{amsmath,amsbsy,bm,amssymb,mathtools,dsfont,etoolbox,braket}
\usepackage{xspace}
\usepackage{graphicx,float}
\usepackage[font=small,labelfont=bf]{caption}
\usepackage[subrefformat=parens,labelformat=parens]{subcaption}
\usepackage{tikz,pgfplots,pgfplotstable}
\usepackage[mode=build]{standalone}
\usepackage{multirow,makecell,array,tabularx,booktabs}
\usepackage{algorithm,algpseudocode}
\usepackage{siunitx}
\usepackage{csquotes}
\usepackage[bibstyle=numeric,citestyle=numeric-comp,sorting=none,backend=biber]{biblatex}
\usepackage{hyperref}

\usepackage{amsmath,amsbsy,bm,amssymb,mathtools,dsfont,etoolbox,braket}
\usepackage{siunitx}
\usepackage{xspace}
\usepackage{tikz}
\usepackage{pgfplots}
\usepackage{pgfplotstable}
\pgfplotsset{compat=newest}
\usetikzlibrary{backgrounds}
\pgfdeclarelayer{background}
\pgfdeclarelayer{foreground}
\usepgfplotslibrary{groupplots}
\usepgfplotslibrary{colormaps}
\pgfsetlayers{background,main,foreground}
\usetikzlibrary{calc,external}
\usetikzlibrary{shapes.misc}
\usetikzlibrary{patterns}
\usetikzlibrary{decorations.pathreplacing}
\usetikzlibrary{arrows,arrows.meta}
\usetikzlibrary{quantikz}
\usepgfplotslibrary{fillbetween}
\usepackage{cleveref}

\newcommand{\neworrenewcommand}[1]{\providecommand{#1}{}\renewcommand{#1}}
\newcommand{\cercle}[6]{%
	\node[circle,inner sep=0,minimum size={2*#2},xshift=#5,yshift=#6](a) at (#1) {};
	\draw[gray,thick] (a.#3) arc (#3:{#3+#4}:#2);
}
\newcommand{\measurement}[1]{%
	\node[draw,thick,fill=white,minimum width=.8cm,minimum height=.45cm,anchor=west] (m) at (#1) {};
	\cercle{#1}{.3cm}{0}{180}{.4cm}{-.15cm}
	\draw[thick,->,>=latex] ($(m.south)+(0,.05)$) -- ($(m.south)+(0,.05)+(-.2,.3)$);
}
\newcommand{\treepval}[2]{\hat{p}_{#1}=\num{#2}}
\newcommand{\Treepval}[2]{\hat{p}_{#1}=\num{#2}^*}
\newcommand{\tttboard}[9]{%https://tex.stackexchange.com/questions/2132/how-to-define-a-command-that-takes-more-than-9-arguments
	\neworrenewcommand{\tttboardinner}[3]{%
		\coordinate (c00) at ##1;
		\coordinate (c10) at ($(c00)+.5*(1,0)$);
		\coordinate (c20) at ($(c10)+.5*(1,0)$);
		\coordinate (c01) at ($(c00)-.5*(0,1)$);
		\coordinate (c11) at ($(c01)+.5*(1,0)$);
		\coordinate (c21) at ($(c11)+.5*(1,0)$);
		\coordinate (c02) at ($(c01)-.5*(0,1)$);
		\coordinate (c12) at ($(c02)+.5*(1,0)$);
		\coordinate (c22) at ($(c12)+.5*(1,0)$);
		\node[##2] (f00) at (c00) {#1};
		\node[##2] (f10) at (c10) {#2};
		\node[##2] (f20) at (c20) {#3};
		\node[##2] (f01) at (c01) {#4};
		\node[##2] (f11) at (c11) {#5};
		\node[##2] (f21) at (c21) {#6};
		\node[##2] (f02) at (c02) {#7};
		\node[##2] (f12) at (c12) {#8};
		\node[##2] (f22) at (c22) {#9};
		\draw[##3] ($(f00)+.5*(-.5,.5)$) -- ($(f20)+.5*(.5,.5)$);
		\draw[##3] ($(f01)+.5*(-.5,.5)$) -- ($(f21)+.5*(.5,.5)$);
		\draw[##3] ($(f02)+.5*(-.5,.5)$) -- ($(f22)+.5*(.5,.5)$);
		\draw[##3] ($(f02)+.5*(-.5,-.5)$) -- ($(f22)+.5*(.5,-.5)$);
		\draw[##3] ($(f00)+.5*(-.5,.5)$) -- ($(f02)+.5*(-.5,-.5)$);
		\draw[##3] ($(f10)+.5*(-.5,.5)$) -- ($(f12)+.5*(-.5,-.5)$);
		\draw[##3] ($(f20)+.5*(-.5,.5)$) -- ($(f22)+.5*(-.5,-.5)$);
		\draw[##3] ($(f20)+.5*(.5,.5)$) -- ($(f22)+.5*(.5,-.5)$);
		\coordinate (boardcenter) at (c11);
	}
	\tttboardinner%
}
\newcommand{\tttX}{$\bigtimes$\xspace}
\newcommand{\tttO}{$\boldsymbol{\bigcirc}$\xspace} %$\bigcirc$
\newcommand{\tttXc}[1]{{\color{#1}\tttX}}
\newcommand{\tttOc}[1]{{\color{#1}\tttO}}

\tikzset{
treenode/.style 		= {draw,circle,fill=black,inner sep=0pt,minimum width=.1cm},
leafnode/.style 		= {draw,circle,fill=white,thick,inner sep=0pt,minimum width=.1cm},
edgenode/.style 		= {shape=rectangle, rounded corners, thick, draw, fill=white, font=\footnotesize, inner sep=1.65pt},
leafnodelabel/.style 	= {shape=rectangle, align=left, font=\footnotesize, inner sep=0pt, outer sep=0pt, anchor=west, xshift=.15cm}}

\definecolor{parimary}{RGB}{0,0,255}
\definecolor{secondary}{RGB}{255,0,0}
\definecolor{ternary}{RGB}{0,255,0}
\definecolor{color0}{RGB}{171,217,233}
\definecolor{color1}{RGB}{215,25,28}
\definecolor{color2}{RGB}{44,123,182}
\definecolor{color3}{RGB}{253,174,97}
\definecolor{color4}{RGB}{255,255,191}
\definecolor{colora}{RGB}{55,126,184}
\definecolor{colorb}{RGB}{228,26,28}
\definecolor{colorc}{RGB}{77,175,74}
\definecolor{colord}{RGB}{152,78,163}
\definecolor{colore}{RGB}{255,127,0}
\definecolor{color41}{RGB}{202,0,32}
\definecolor{color42}{RGB}{244,165,130}
\definecolor{color43}{RGB}{146,197,222}
\definecolor{color44}{RGB}{5,113,176}
\definecolor{color51}{RGB}{228,26,28}
\definecolor{color52}{RGB}{55,126,184}
\definecolor{color53}{RGB}{77,175,74}
\definecolor{color54}{RGB}{152,78,163}
\definecolor{color55}{RGB}{255,127,0}
\colorlet{boardcol1}{red!50!black}
\colorlet{boardcol2}{blue!50!black}
\definecolor{Qv}{HTML}{53257F} %Quantum violet
\definecolor{Qg}{HTML}{555555} %Quantum gray

\crefname{figure}{Fig.}{Figs.}
\Crefname{figure}{Fig.}{Figs.}
\crefname{table}{Tab.}{Tabs.}
\Crefname{table}{Tab.}{Tabs.}
\crefname{equation}{Eq.}{Eqs.}
\Crefname{equation}{Eq.}{Eqs.}
\crefname{section}{Sec.}{Secs.}
\Crefname{section}{Sec.}{Secs.}
\crefname{subsection}{Sec.}{Secs.}
\Crefname{subsection}{Sec.}{Secs.}
\crefname{algorithm}{Alg.}{Algs.}
\Crefname{algorithm}{Alg.}{Algs.}
\crefname{appendix}{App.}{App.}
\Crefname{appendix}{App.}{App.}

\newcommand{\ie}{i.\,e.\xspace}
\newcommand{\eg}{e.\,g.\xspace}
\newcommand{\cf}{cf.\xspace}

\newcommand{\bnum}[1]{\textbf{\num[separate-uncertainty=false]{#1}}}
\newcommand{\tnum}[1]{{\small\num[separate-uncertainty=false]{#1}}}
\newcommand{\Tnum}[1]{{\small\bnum{#1}}}

\DeclareMathOperator*{\argmax}{arg\,max}

\DeclareMathOperator*{\CP}{C}
\newcommand{\composition}[2]{\CP_{#1}^{#2}}

\newcommand{\orderprod}[2]{\prod_{#1}^{#2}}
\DeclareMathOperator*{\NOT}{NOT}
\DeclareMathOperator*{\CNOT}{CNOT}
\newcommand{\MCNOT}[1]{\operatorname{C}_{#1}\!\operatorname{NOT}}
\DeclareMathOperator*{\SWAP}{SWAP}
\DeclareMathOperator*{\CSWAP}{CSWAP}
\newcommand{\MCSWAP}[1]{\operatorname{C}_{#1}\!\operatorname{SWAP}}
\DeclareRobustCommand{\unitop}{\text{\usefont{U}{bbold}{m}{n}1}}
\newcommand{\RZ}{\operatorname{R}_z}
\newcommand{\SX}{\sqrt{\NOT}}
\DeclareMathOperator*{\MOD}{mod}
\newcommand{\BigO}[1]{\mathcal{O}(#1)}
\DeclareMathOperator*{\TR}{Tr}
\newcommand{\trace}[1]{\TR\!\left\{#1\right\}}
\DeclareMathOperator*{\EX}{\mathbb{E}}
\newcommand{\expectation}[1]{\EX\!\left[#1\right]}
\DeclareMathOperator*{\STD}{\sigma}
\newcommand{\stdev}[1]{\STD\!{#1}}
\newcommand{\stdeva}[1]{\hat{\STD}\!{#1}}

\DeclareRobustCommand\treenode{\tikz[baseline=-.6ex]{\node[anchor=base,draw,circle,fill=black,inner sep=0pt,minimum width=.1cm]at(0,0){};}\xspace}
\DeclareRobustCommand\treeleaf{\tikz[baseline=-.6ex]{\node[anchor=base,draw,circle,fill=white,thick,inner sep=0pt,minimum width=.1cm]at(0,0){};}\xspace}
\DeclareRobustCommand\treerulezero{\tikz[baseline=-.75ex]{\node[draw,fill=white,circle,midway,font=\tiny,inner sep=1pt]at(0,0){$0$};}\xspace}
\DeclareRobustCommand\treeruleone{\tikz[baseline=-.75ex]{\node[draw,fill=white,circle,midway,font=\tiny,inner sep=1pt]at(0,0){$1$};}\xspace}
\DeclareRobustCommand\circlednumber[1]{\tikz[baseline=-.75ex]{\node[draw,circle,fill=white,inner sep=1pt,font=\small,xshift=-.1cm] at (0,0) {#1};}\xspace}

\title{Representation of binary classification trees with binary features by quantum circuits}
\author{Raoul Heese}
\affiliation{Fraunhofer ITWM, 67663 Kaiserslautern, Germany}
\orcid{0000-0001-7479-3339}
\email{raoul.heese@itwm.fraunhofer.de}
\author{Patricia Bickert}
\affiliation{Fraunhofer ITWM, 67663 Kaiserslautern, Germany}
\orcid{0000-0003-0737-7604}
\author{Astrid Elisa Niederle}
\affiliation{BASF SE, 67063 Ludwigshafen, Germany}
\orcid{0000-0003-2171-8639}

\begin{document}

\maketitle

\begin{abstract}
We propose a quantum representation of binary classification trees with binary features based on a probabilistic approach. By using the quantum computer as a processor for probability distributions, a probabilistic traversal of the decision tree can be realized via measurements of a quantum circuit. We describe how tree inductions and the prediction of class labels of query data can be integrated into this framework. An on-demand sampling method enables predictions with a constant number of classical memory slots, independent of the tree depth. We experimentally study our approach using both a quantum computing simulator and actual IBM quantum hardware. To our knowledge, this is the first realization of a decision tree classifier on a quantum device.
\end{abstract}

\section{Introduction} \label{sec:introduction}
Decision trees are well-known predictive models commonly used in data mining and machine learning for a wide area of applications \cite{kotsiantis2011,maimon2014,breiman2017}. In general, a decision tree can be viewed as a flowchart-like structure that can be used to query data. Starting from the root, each internal node represents a test on the query data and each outgoing branch represents a possible outcome of this test. For a binary tree, the test result is a Boolean value and can therefore be either true or false (\ie, there are two branches from each internal node). Each leaf of the tree can be associated with a decision. Therefore, the path from the root to the leaf implies a set of decision rules for the query data in the sense of a sequential decision process. In particular, we consider \emph{binary classification trees} where the decision of a leaf determines the membership of a data point to a predefined discrete set of classes.\par
The inference of a decision tree from a given data set is a supervised machine learning task also known as decision tree \emph{induction} (or decision tree learning). However, finding a globally optimal solution is NP-hard \cite{hyafil1976,bertsimas2017} and therefore heuristic recursion algorithms are favored in practice \cite{zharmagambetov2020}. Such algorithms typically work in a greedy top-down fashion \cite{quinlan1986}: Starting from the root, the best test is estimated for each internal node by minimizing a data impurity function. The data set is splitted accordingly into two subsets along each of the two outgoing branches. This process is repeated recursively for each internal node until a stopping criterion terminates the tree traversal and results in a leaf with a classification decision based on the majority class present in the data subset within the node. The algorithm ends when all paths lead to a leaf. A heuristically created decision tree is not guaranteed to be globally optimal but might still be suitable for practical purposes.\par
In the context of quantum computing, decision trees can be assigned to the field of \emph{quantum machine learning} \cite{biamonte2017}. Several previous papers consider the interplay between decision trees and quantum computing. In \cite{fahri1998}, the traversal speed of decision trees is studied and classical and quantum approaches are compared. The authors find no benefit of one over the other. An heuristic algorithm to induce quantum classification trees is presented in \cite{lu2014}, where data points are encoded as quantum states and measurements are used to find the best splits. However, some parts of the hybrid quantum-classical algorithm seem to be incomplete and can be considered ``enigmatic'' \cite{schuld2015}. In \cite{khadiev2019}, a conceptional quantum extension is presented for the classical decision tree induction algorithm \emph{C5.0} \cite{maimon2014} such that an almost quadratic speed-up can be gained by implementing a suitable quantum subroutine. However, there is no publication that we know of that considers both induction and evaluation of a quantum decision tree in a consistent manner. In this sense, ``the potential of a quantum decision tree is still to be established'' \cite{schuld2015}.\par
Decision trees have also been used to realize efficient data management strategies in a quantum computing context. In \cite{giovannetti2008}, a ``bucket-brigade'' quantum random access memory (qRAM) is proposed, which achieves an exponential reduction in access overhead compared to a conventional classical random access memory design. In this approach, three-level memory elements are structured in a binary tree to organize the flow of information to the memory cells in the leaves (and back). Moreover, a whole ensemble of decision trees can also be used to store structured data. In \cite{kerenidis2016}, binary decision trees are used as building blocks for a data structure to store a preference matrix for a quantum recommendation system. To this end, the preference matrix is decomposed into row vectors, which are then stored in a binary tree such that each leaf holds the amplitude of one vector component and each internal node stores the respective amplitude sums of its branches, which enables efficient computations.\par
The term ``quantum decision tree'' is also used in the context of query complexity analysis of quantum algorithms in analogy to the decision tree view of classical query complexity. Here, the quantum algorithm is treated as a sequence of unitary operators followed by a measurement \cite{barnum2001,buhrmann2002,shi2002,beigi2020}. However, such a configuration has a different purpose than a decision tree in the sense of a predictive model, which is why the term ``quantum query algorithm'' is also used instead. On the other hand, a classical decision tree query can in principle be expressed by a quantum query algorithm.\par
In this manuscript, we follow a different strategy and propose a quantum representation of decision trees, which we call \emph{Q-tree}, by using the quantum computer as a processor for probability distributions. For this purpose, we first take a probabilistic perspective to describe the decisions in a (classical) decision tree as conditions of the probability distribution of the training data. This approach allows us to associate a marginalized conditional probability distribution to each leaf that describes the probability distribution to reach the leaf in a probabilistic traversal of the tree and the corresponding class label probability distribution, respectively. In this sense, we \emph{load} a classical decision tree into a quantum circuit.
\par
The set of conditions in the tree can also be parameterized to obtain a tuple of vectors that uniquely define the structure of the tree. Based on these parameters, an appropriate quantum circuit can be built that fully represents the probabilistic model of the tree. By performing repeated measurements, this Q-tree can then be used for a probabilistic traversal of the tree in order to sample from the probability distribution of a randomly reached leaf. This approach enables a truly random representation of the tree \cite{calude2010,kofler2010}. The results from such a random sampling can be used for a prediction of the class labels of query data. A similar, but slightly modified circuit is also able to provide predictions for uncertain query data. Moreover, a hybrid quantum-classical algorithm can be used for a tree induction by optimizing the corresponding parameters. We propose the usage of a \emph{genetic algorithm} \cite{katoch2021} for this purpose.\par
In short, the conceptional foundation for our approach is a straightforward processing of probability distributions in which we utilize the quantum computer to perform conditional marginalizations based on the parameterized tree structure. Related concepts have already be pursued, \eg, in \cite{low2014,borujeni2020,deoliveira2021} in the context of Bayesian networks.\par
The remaining paper is structured as follows. We start in \cref{sec:problem description} by describing the considered machine learning problem. In \cref{sec:classical representation}, we discuss preliminary considerations regarding purely classical decision trees, which are used in \cref{sec:quantum representation} to present our proposed Q-tree quantum representation. Subsequently, we demonstrate the usage of this quantum representation in \cref{sec:experiments} on a quantum computing simulator and actual quantum hardware. We close with conclusions and an outlook in \cref{eqn:conclusions}.

\section{Problem description} \label{sec:problem description}
We consider data points of the form $\mathbf{d} \equiv (\mathbf{x}, \mathbf{y})$ with $k$-dimensional binary input features $\mathbf{x} \in \mathbb{B}^k$ and $m$-dimensional binary (class) labels $\mathbf{y} \in \mathbb{B}^m$ with $\mathbb{B} \equiv \{0,1\}$. Based on a given training data set 
\begin{align} \label{eqn:D:training}
	\mathbf{D}_{\mathrm{train}} \equiv \{ (\mathbf{x}^1, \mathbf{y}^1), \dots, (\mathbf{x}^t, \mathbf{y}^t) \}
\end{align}
consisting of $t$ data points, a decision tree $T$ can be inducted using an algorithm $A$ such that
\begin{align} \label{eqn:induction}
	\mathbf{D}_{\mathrm{train}}, A \mapsto T.
\end{align}
The algorithm $A$ can for example represent a global optimization method or a heuristic approach as discussed in \cref{sec:introduction}.\par
We limit ourselves to \emph{binary classification trees with binary features} $T$, which are also known as \emph{binary decision diagrams} \cite{akers1978}. Starting from the root, such a tree consists of a set of nodes connected with branches, where each node has exactly one branch going in. Internal nodes have two branches going out, whereas leaves have no outgoing branches. In each internal node, the test
\begin{align} \label{eqn:test}
	x_i \overset{!}{=} 1
\end{align}
is performed for a specific feature $i \in \{1,\dots,k\}$ and the two outgoing branches represent the rules $x_i=1$ (\ie, test passed) and $x_i=0$ (\ie, test failed), respectively. Each node can consequently be associated with a data set representing a subset of the training data that fulfills the decision rules of the path from the root to the respective node.\par
\begin{figure}
	\begin{center}
		\begin{tikzpicture}
\coordinate (n00) at (0,0);
\coordinate (n10) at ($(n00)+(-2,-.5)$);
\coordinate (n11) at ($(n00)+(2,-.5)$);
\coordinate (n20) at ($(n10)+(-1,-.5)$);
\coordinate (n21) at ($(n10)+(1,-.5)$);
\coordinate (n22) at ($(n11)+(-1,-.5)$);
\coordinate (n23) at ($(n11)+(1,-.5)$);
\coordinate (n30) at ($(n20)+(-.5,-.5)$);
\coordinate (n31) at ($(n20)+(.5,-.5)$);
\coordinate (n32) at ($(n21)+(-.5,-.5)$);
\coordinate (n33) at ($(n21)+(.5,-.5)$);
\coordinate (n34) at ($(n22)+(-.5,-.5)$);
\coordinate (n35) at ($(n22)+(.5,-.5)$);
\coordinate (n36) at ($(n23)+(-.5,-.5)$);
\coordinate (n37) at ($(n23)+(.5,-.5)$);
\draw[thick,Qv,dotted] (-4,-.25) -- (5,-.25);
\draw[thick,Qv,dotted] (-4,-.75) -- (5,-.75);
\draw[thick,Qv,dotted] (-4,-1.25) -- (5,-1.25);
\draw[thick] (n00) -- (n10) node [draw,fill=white,circle,midway,font=\tiny,inner sep=1pt] {$0$};
\draw[thick] (n00) -- (n11) node [draw,fill=white,circle,midway,font=\tiny,inner sep=1pt] {$1$};
\draw[thick] (n10) -- (n20) node [draw,fill=white,circle,midway,font=\tiny,inner sep=1pt] {$0$};
\draw[thick] (n10) -- (n21) node [draw,fill=white,circle,midway,font=\tiny,inner sep=1pt] {$1$};
\draw[thick] (n11) -- (n22) node [draw,fill=white,circle,midway,font=\tiny,inner sep=1pt] {$0$};
\draw[thick] (n11) -- (n23) node [draw,fill=white,circle,midway,font=\tiny,inner sep=1pt] {$1$};
\draw[thick] (n20) -- (n30) node [draw,fill=white,circle,midway,font=\tiny,inner sep=1pt] {$0$};
\draw[thick] (n20) -- (n31) node [draw,fill=white,circle,midway,font=\tiny,inner sep=1pt] {$1$};
\draw[thick] (n21) -- (n32) node [draw,fill=white,circle,midway,font=\tiny,inner sep=1pt] {$0$};
\draw[thick] (n21) -- (n33) node [draw,fill=white,circle,midway,font=\tiny,inner sep=1pt] {$1$};
\draw[thick] (n22) -- (n34) node [draw,fill=white,circle,midway,font=\tiny,inner sep=1pt] {$0$};
\draw[thick] (n22) -- (n35) node [draw,fill=white,circle,midway,font=\tiny,inner sep=1pt] {$1$};
\draw[thick] (n23) -- (n36) node [draw,fill=white,circle,midway,font=\tiny,inner sep=1pt] {$0$};
\draw[thick] (n23) -- (n37) node [draw,fill=white,circle,midway,font=\tiny,inner sep=1pt] {$1$};
\node[draw,circle,fill=black,inner sep=0pt,minimum width=.1cm] at (n00) {};
\node[draw,circle,fill=black,inner sep=0pt,minimum width=.1cm] at (n10) {};
\node[draw,circle,fill=black,inner sep=0pt,minimum width=.1cm] at (n11) {};
\node[draw,circle,fill=black,inner sep=0pt,minimum width=.1cm] at (n20) {};
\node[draw,circle,fill=black,inner sep=0pt,minimum width=.1cm] at (n21) {};
\node[draw,circle,fill=black,inner sep=0pt,minimum width=.1cm] at (n22) {};
\node[draw,circle,fill=black,inner sep=0pt,minimum width=.1cm] at (n23) {};
\node[draw,circle,fill=white,thick,inner sep=0pt,minimum width=.1cm] at (n30) {};
\node[draw,circle,fill=white,thick,inner sep=0pt,minimum width=.1cm] at (n31) {};
\node[draw,circle,fill=white,thick,inner sep=0pt,minimum width=.1cm] at (n32) {};
\node[draw,circle,fill=white,thick,inner sep=0pt,minimum width=.1cm] at (n33) {};
\node[draw,circle,fill=white,thick,inner sep=0pt,minimum width=.1cm] at (n34) {};
\node[draw,circle,fill=white,thick,inner sep=0pt,minimum width=.1cm] at (n35) {};
\node[draw,circle,fill=white,thick,inner sep=0pt,minimum width=.1cm] at (n36) {};
\node[draw,circle,fill=white,thick,inner sep=0pt,minimum width=.1cm] at (n37) {};
\node[] (d0) at (4.5,0) {$d=0$};
\node[] (d1) at (4.5,-.5) {$d=1$};
\node[] (d2) at (4.5,-1) {$d=2$};
\node[] (d4) at (4.5,-1.5) {$d=3$};
\draw[decorate,decoration={brace,amplitude=5pt},xshift=0pt] ($(d0)+(.6,.2)$) -- ($(d2)+(.6,-.2)$) node[midway,anchor=west,xshift=.3cm] {internal nodes};
\draw[decorate,decoration={brace,amplitude=5pt},xshift=0pt] ($(d4)+(.6,.2)$) -- ($(d4)+(.6,-.2)$) node[midway,anchor=west,xshift=.3cm] {leaves};
\draw[decorate,decoration={brace,amplitude=5pt},xshift=0pt] let \p1=(n00), \p2=(n30) in ($(\x2,\y2)+(-.6,-.2)$) -- ($(\x2,\y1)+(-.6,.2)$) node[midway,anchor=east,xshift=-.3cm] {$T$};
\node[anchor=south,fill=white,inner sep=0pt,outer sep=0pt,yshift=.15cm] at (n00) {root};
%\node[anchor=south,fill=white,inner sep=0pt,outer sep=0pt,yshift=.15cm] at (n10) {0};
%\node[anchor=south,fill=white,inner sep=0pt,outer sep=0pt,yshift=.15cm] at (n11) {1};
%\node[anchor=south,fill=white,inner sep=0pt,outer sep=0pt,yshift=.15cm] at (n20) {00};
%\node[anchor=south,fill=white,inner sep=0pt,outer sep=0pt,yshift=.15cm] at (n21) {01};
%\node[anchor=south,fill=white,inner sep=0pt,outer sep=0pt,yshift=.15cm] at (n22) {10};
%\node[anchor=south,fill=white,inner sep=0pt,outer sep=0pt,yshift=.15cm] at (n23) {11};
%\node[anchor=north,fill=white,inner sep=0pt,outer sep=0pt,yshift=-.15cm] at (n30) {000};
%\node[anchor=north,fill=white,inner sep=0pt,outer sep=0pt,yshift=-.15cm] at (n31) {001};
%\node[anchor=north,fill=white,inner sep=0pt,outer sep=0pt,yshift=-.15cm] at (n32) {010};
%\node[anchor=north,fill=white,inner sep=0pt,outer sep=0pt,yshift=-.15cm] at (n33) {011};
%\node[anchor=north,fill=white,inner sep=0pt,outer sep=0pt,yshift=-.15cm] at (n34) {100};
%\node[anchor=north,fill=white,inner sep=0pt,outer sep=0pt,yshift=-.15cm] at (n35) {101};
%\node[anchor=north,fill=white,inner sep=0pt,outer sep=0pt,yshift=-.15cm] at (n36) {110};
%\node[anchor=north,fill=white,inner sep=0pt,outer sep=0pt,yshift=-.15cm] at (n37) {111};
\end{tikzpicture}
		\caption{Schematic sketch of a binary classification tree with binary features. Starting from the root, each path along the internal nodes (\treenode) to a leaf (\treeleaf) represents a set of decision rules for the data features in the form of \cref{eqn:test} to predict the corresponding labels (\ie, to classify the query data). The outgoing branches represent the two cases $x_i=0$ (denoted by \treerulezero) and $x_i=1$ (denoted by \treeruleone), respectively. The depth $d$ indicates the number of decision tests required to reach a node. We only consider trees with no repeating decision rules along a path and no missing internal nodes. The whole tree (of maximum depth $3$ in this example) is denoted by $T$.} \label{fig:sample-tree}
	\end{center}
\end{figure}
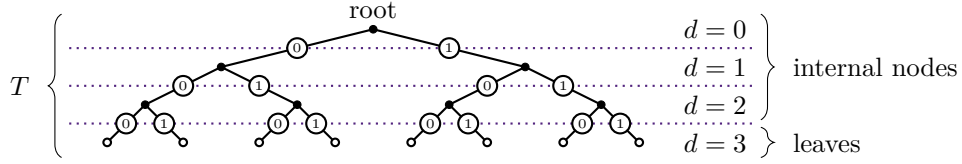

The \emph{depth} $d$ of a node indicates the number of decision tests required to reach it, where the root corresponds to $d=0$. For reasons of simplicity, we limit ourselves to trees with the following two properties: 
\begin{enumerate}
	\item A feature cannot be split twice along a path, \ie, decision rules must not repeat.
	\item All leaves have the same depth, \ie, the tree is not missing any internal nodes.
\end{enumerate}
The first presumption implies that $d<k$ such that in a tree of depth $k-1$, all features have to be evaluated once to reach a leaf. At depth $d$, there are $2^d$ nodes, each with $k-d$ possible splitting decisions, which in total sum up to $\sum_{l=0}^d 2^l(k-l)$ possible decisions along all paths up to this depth. A schematic tree is sketched in \cref{fig:sample-tree}.\par
After its induction, a decision tree $T$ of this form can subsequently be used to predict the labels $\mathbf{y}$ for a query vector of input features $\mathbf{x}$ (\ie, it can be used to classify the query data), which is in general not part of the training data set. However, instead of a crisp prediction value, each leaf of the tree $T$ is labeled with the probabilities of predicting either the label $y_i=0$ or the label $y_i=1$ for all $i \in \{1,\dots,m\}$ as given by the corresponding ratios of labels in the associated subset of the training data. In this sense, we consider \emph{probability estimation trees} that can also be used to make fuzzy predictions \cite{provost2003}. Specifically, the tree $T$ can predict the label probability distribution $p(\mathbf{y}) \in [0,1]$ from features $\mathbf{x}$ according to
\begin{align} \label{eqn:prediction}
	T, \mathbf{x} \mapsto p(\mathbf{y}),
\end{align}
which also allows to predict the most probable labels $\argmax_{\mathbf{y}} p(\mathbf{y}) \in \mathbb{B}^m$.\par
It can be assumed that all data points from the training data set and all query data points are sampled from an underlying ``true'' (but usually unknown) probability distribution in the sense that
\begin{align} \label{eqn:p:true}
	\mathbf{d} \sim \tilde{p}(x_1,\dots,x_k,y_1,\dots,y_m) \equiv \tilde{p}(\mathbf{x},\mathbf{y})
\end{align}
for each data point. The induction of a decision tree therefore corresponds to finding a representative estimation for this probability distribution based on the information contained in the training data set.\par
\section{Classical representation} \label{sec:classical representation}
In this section, preliminary considerations regarding purely classical decision trees are presented. First, we describe how a decision tree can be parameterized by a tuple of vectors. This parameterization can the be used for the tree induction. Subsequently, we discuss the probabilistic view of decision trees, in which a marginalized conditional probability distribution can be assigned to each node based on the training data. This formulation can be directly realized with a quantum circuit and is therefore of central importance for our proposed Q-tree approach. Finally, we briefly explain how the prediction of query data fits into the probabilistic framework.

\subsection{Tree parameterization}
A binary decision tree, as shown in \cref{fig:sample-tree}, has a highly symmetric structure, uniquely defined by the feature indices for which a splitting decision is performed in each internal node. We therefore present two possible tree parameterizations in the following, which are particularly useful for the subsequent considerations.

\subsubsection{Decision configuration}
The structure of a tree $T$ can be parameterized as a tuple of vectors defining the tests in each node. For this purpose, we introduce the \emph{decision configuration}
\begin{align} \label{eqn:E}
	E \equiv (\mathbf{e}^0, \dots, \mathbf{e}^d)
\end{align}
for a tree of maximum depth $d$ as a $(d)$-tuple of layer (\ie, depth-layer) decisions 
\begin{align} \label{eqn:e}
	\mathbf{e}^i \equiv (e^i_1, \dots, e^i_{2^i}) \in \{1,\dots,k\}^{2^i} \,\forall\, i \in \{0,\dots,d\}.
\end{align}
The vectors $\mathbf{e^i}$ represents all decisions in depth $i$, which are defined by the feature indices $e^i_j$ for which a split is made, one for each each node $j \in \{1,\dots,2^i\}$ in this depth, sorted in a specific order.\par
As a convenient sorting order, we propose that given a bit string $\bar{x}_{0}\cdots\bar{x}_{l-1}$ of split decisions from the root (\ie, the single node with depth $0$) to a node of depth $l\geq1$ corresponding to the decision rules $x_{e^{0}}=\bar{x}_{0} \land \dots \land x_{e^{l-1}}=\bar{x}_{l-1}$, the decision in this node is determined by $e^l_{\nu}$ with
\begin{align} \label{eqn:nu}
	\nu \equiv \nu_l(\bar{x}_{0}\cdots\bar{x}_{l-1}) \equiv \sum_{j=0}^{l-1} 2^{j} \bar{x}_{l-1-j} + 1
\end{align}
so that $\nu \in \{1,\dots,2^l\}$. Here we assume that a valid sequence of decisions
\begin{align} \label{eqn:s}
	e^{i} \in \{e^i_1,\dots,e^i_{2^i}\} \,\forall\, i \in \{0,\dots,l-1\}
\end{align}
has been chosen (in the sense that this path is actually present in the tree $T$). For the root, we define $\nu_0 \equiv 1$. Thus, any node in the tree $T$ can be identified by its depth $l \in \{0,\dots,d\}$ and an index $\nu$.\par
In a valid tree, all feature indices $e^i_j$ must be chosen so that no repeated splits occur along any path. For a tree of depth $d$, this means that the condition
\begin{align} \label{eqn:E:condition}
	e^0_1 \neq e^d_j \land e^l_{\nu_l(\bar{x}_{0}\cdots\bar{x}_{l-1})} \neq e^d_j \,\forall\, j \in \{1,\dots,2^d\} \,\forall\, l \in \{1,\dots,d-1\} \,\forall\, \bar{x}_{0},\dots,\bar{x}_{d-1} \in \mathbb{B}
\end{align}
has to be fulfilled by the corresponding decision configuration $E$.\par
\begin{figure}
	\begin{center}
		\begin{tikzpicture}
\coordinate (n00) at (0,0);
\coordinate (n10) at ($(n00)+(-2,-.5)$);
\coordinate (n11) at ($(n00)+(2,-.5)$);
\coordinate (n20) at ($(n10)+(-1,-.5)$);
\coordinate (n21) at ($(n10)+(1,-.5)$);
\coordinate (n22) at ($(n11)+(-1,-.5)$);
\coordinate (n23) at ($(n11)+(1,-.5)$);
\coordinate (n30) at ($(n20)+(-.5,-.5)$);
\coordinate (n31) at ($(n20)+(.5,-.5)$);
\coordinate (n32) at ($(n21)+(-.5,-.5)$);
\coordinate (n33) at ($(n21)+(.5,-.5)$);
\coordinate (n34) at ($(n22)+(-.5,-.5)$);
\coordinate (n35) at ($(n22)+(.5,-.5)$);
\coordinate (n36) at ($(n23)+(-.5,-.5)$);
\coordinate (n37) at ($(n23)+(.5,-.5)$);
%
%\draw[thick,blue,dotted] (-4,-.25) -- (5,-.25);
%\draw[thick,blue,dotted] (-4,-.75) -- (5,-.75);
%\draw[thick,blue,dotted] (-4,-1.25) -- (5,-1.25);
%
\draw[thick] (n00) -- (n10) node [draw,fill=white,circle,midway,font=\tiny,inner sep=1pt] {$0$};
\draw[thick] (n00) -- (n11) node [draw,fill=white,circle,midway,font=\tiny,inner sep=1pt] {$1$};
\draw[thick] (n10) -- (n20) node [draw,fill=white,circle,midway,font=\tiny,inner sep=1pt] {$0$};
\draw[thick] (n10) -- (n21) node [draw,fill=white,circle,midway,font=\tiny,inner sep=1pt] {$1$};
\draw[thick] (n11) -- (n22) node [draw,fill=white,circle,midway,font=\tiny,inner sep=1pt] {$0$};
\draw[thick] (n11) -- (n23) node [draw,fill=white,circle,midway,font=\tiny,inner sep=1pt] {$1$};
\draw[thick] (n20) -- (n30) node [draw,fill=white,circle,midway,font=\tiny,inner sep=1pt] {$0$};
\draw[thick] (n20) -- (n31) node [draw,fill=white,circle,midway,font=\tiny,inner sep=1pt] {$1$};
\draw[thick] (n21) -- (n32) node [draw,fill=white,circle,midway,font=\tiny,inner sep=1pt] {$0$};
\draw[thick] (n21) -- (n33) node [draw,fill=white,circle,midway,font=\tiny,inner sep=1pt] {$1$};
\draw[thick] (n22) -- (n34) node [draw,fill=white,circle,midway,font=\tiny,inner sep=1pt] {$0$};
\draw[thick] (n22) -- (n35) node [draw,fill=white,circle,midway,font=\tiny,inner sep=1pt] {$1$};
\draw[thick] (n23) -- (n36) node [draw,fill=white,circle,midway,font=\tiny,inner sep=1pt] {$0$};
\draw[thick] (n23) -- (n37) node [draw,fill=white,circle,midway,font=\tiny,inner sep=1pt] {$1$};
\node[draw,circle,fill=black,inner sep=0pt,minimum width=.1cm] at (n00) {};
\node[draw,circle,fill=black,inner sep=0pt,minimum width=.1cm] at (n10) {};
\node[draw,circle,fill=black,inner sep=0pt,minimum width=.1cm] at (n11) {};
\node[draw,circle,fill=black,inner sep=0pt,minimum width=.1cm] at (n20) {};
\node[draw,circle,fill=black,inner sep=0pt,minimum width=.1cm] at (n21) {};
\node[draw,circle,fill=black,inner sep=0pt,minimum width=.1cm] at (n22) {};
\node[draw,circle,fill=black,inner sep=0pt,minimum width=.1cm] at (n23) {};
\node[draw,circle,fill=white,thick,inner sep=0pt,minimum width=.1cm] at (n30) {};
\node[draw,circle,fill=white,thick,inner sep=0pt,minimum width=.1cm] at (n31) {};
\node[draw,circle,fill=white,thick,inner sep=0pt,minimum width=.1cm] at (n32) {};
\node[draw,circle,fill=white,thick,inner sep=0pt,minimum width=.1cm] at (n33) {};
\node[draw,circle,fill=white,thick,inner sep=0pt,minimum width=.1cm] at (n34) {};
\node[draw,circle,fill=white,thick,inner sep=0pt,minimum width=.1cm] at (n35) {};
\node[draw,circle,fill=white,thick,inner sep=0pt,minimum width=.1cm] at (n36) {};
\node[draw,circle,fill=white,thick,inner sep=0pt,minimum width=.1cm] at (n37) {};
%
%\node[] (d0) at (4.5,0) {$d=0$};
%\node[] (d1) at (4.5,-.5) {$d=1$};
%\node[] (d2) at (4.5,-1) {$d=2$};
%\node[] (d4) at (4.5,-1.5) {$d=3$};
%
\node[anchor=south,fill=white,inner sep=0pt,outer sep=0pt,yshift=.15cm] at (n00) {$x_{e^0_1}$};
\node[anchor=south,fill=white,inner sep=0pt,outer sep=0pt,yshift=.15cm] at (n10) {$x_{e^1_1}$};
\node[anchor=south,fill=white,inner sep=0pt,outer sep=0pt,yshift=.15cm] at (n11) {$x_{e^1_2}$};
\node[anchor=south,fill=white,inner sep=0pt,outer sep=0pt,yshift=.15cm] at (n20) {$x_{e^2_1}$};
\node[anchor=south,fill=white,inner sep=0pt,outer sep=0pt,yshift=.15cm] at (n21) {$x_{e^2_2}$};
\node[anchor=south,fill=white,inner sep=0pt,outer sep=0pt,yshift=.15cm] at (n22) {$x_{e^2_3}$};
\node[anchor=south,fill=white,inner sep=0pt,outer sep=0pt,yshift=.15cm] at (n23) {$x_{e^2_4}$};
\node[anchor=north,fill=white,inner sep=0pt,outer sep=0pt,yshift=-.15cm,align=center] at (n30) {000 \\ 1};
\node[anchor=north,fill=white,inner sep=0pt,outer sep=0pt,yshift=-.15cm,align=center] at (n31) {001 \\ 2};
\node[anchor=north,fill=white,inner sep=0pt,outer sep=0pt,yshift=-.15cm,align=center] at (n32) {010 \\ 3};
\node[anchor=north,fill=white,inner sep=0pt,outer sep=0pt,yshift=-.15cm,align=center] at (n33) {011 \\ 4};
\node[anchor=north,fill=white,inner sep=0pt,outer sep=0pt,yshift=-.15cm,align=center] at (n34) {100 \\ 5};
\node[anchor=north,fill=white,inner sep=0pt,outer sep=0pt,yshift=-.15cm,align=center] at (n35) {101 \\ 6};
\node[anchor=north,fill=white,inner sep=0pt,outer sep=0pt,yshift=-.15cm,align=center] at (n36) {110 \\ 7};
\node[anchor=north,fill=white,inner sep=0pt,outer sep=0pt,yshift=-.15cm,align=center] at (n37) {111 \\ 8};
\node[anchor=north,fill=white,inner sep=0pt,outer sep=0pt,yshift=-.1725cm,xshift=-1.16cm,align=right] at (n30) {$\bar{x}_0\bar{x}_1\bar{x}_2 = $ \\ $\nu = $};
\end{tikzpicture}
		\caption{Tree parameterization $E$, \cref{eqn:E}, for the tree from \cref{fig:sample-tree}. For each internal node, we show the features for which the respective splitting test, \cref{eqn:test}, is performed. In addition, we also show the bit strings of split decisions $\bar{x}_0\bar{x}_1\bar{x}_2$ for the leaves and the corresponding node index $\nu$, \cref{eqn:nu}. Any node in the tree can be uniquely identified by such a bit string corresponding to the decision rules along the path from the root to this node.} \label{fig:sample-tree-param}
	\end{center}
\end{figure}
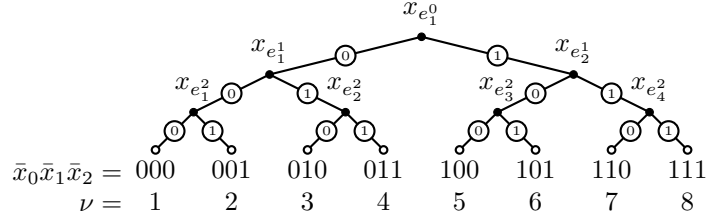

Summarized, the decision configuration $E$ uniquely defines the structure of the tree $T$ as an ordered nested sequence of $\sum_{l=0}^d 2^l$ feature indices in total. In addition to its structural information, $T$ also contains information about the training data $\mathbf{D}_{\mathrm{train}}$, \cref{eqn:D:training}, since each node can be associated with the subset of training data that fulfills the corresponding decision rules, which is particularly used in the leaves for the inference of the labels of query data. Combining $E$ and $\mathbf{D}_{\mathrm{train}}$, a one-on-one correspondence
\begin{align} \label{eqn:TE}
	T \leftrightarrow (E, \mathbf{D}_{\mathrm{train}})
\end{align}
can be established. An example is sketched in \cref{fig:sample-tree-param}.

\subsubsection{Compressed decision configuration}
In each layer of the tree, the sequence of decisions, \cref{eqn:nu}, grows by one feature, which is equivalent to reducing the available number of split choices by one. However, this is not directly reflected by the encoding $E$, \cref{eqn:E}, in which the feature indices $e^i_j$ can attain $k$ possible values independent of the depth. We fix this defect by imposing an additional condition given by \cref{eqn:E:condition}. It states that only such values of $e^i_j$ are allowed that produce a tree without repeated splits. Consequently, additional dependencies between the elements of $E$ emerge.\par
In the context of this work, it turns out to be more practical to switch to a dependency-free parameterization, which naturally reflects the reduction of choices by having less degrees of freedom in each layer. For this purpose, we introduce the \emph{compressed decision configuration}
\begin{align} \label{eqn:C}
	C \equiv (\mathbf{c}^0, \dots, \mathbf{c}^d)
\end{align}
for a tree of maximum depth $d$ as a $(d)$-duple of independent vectors
\begin{align} \label{eqn:c}
	\mathbf{c}^i \equiv (c^i_1, \dots, c^i_{2^i}) \in \{i+1,\dots,k\}^{2^i} \,\forall\, i \in \{0,\dots,d\}
\end{align}
in analogy to \cref{eqn:E,eqn:e}. The elements $c^i_j$ (with $j \in \{1,\dots,2^i\}$) are mutually independent and correspond to a valid tree without the need for additional constraints. There exists a bijective transformation
\begin{align} \label{eqn:EC}
	E \leftrightarrow C
\end{align}
between \cref{eqn:E,eqn:C} as we show in \cref{sec:app:compressed decision configuration} such that decision tree $T$ can be parameterized by either $E$, \cref{eqn:E,eqn:E:condition}, or $C$, \cref{eqn:C}. We consequently use $E$ and $C$ interchangeably.

\subsection{Tree induction} \label{sec:classical representation:tree induction}
The relations in \cref{eqn:TE,eqn:EC} allow us to rephrase \cref{eqn:induction} as
\begin{align} \label{eqn:induction:C}
	\mathbf{D}_{\mathrm{train}}, A \mapsto C
\end{align}
in the sense that the induction of a tree $T$ based on the training data $\mathbf{D}_{\mathrm{train}}$, \cref{eqn:D:training}, and an algorithm $A$ can equivalently be considered as an induction of the corresponding compressed decision configuration $C$, \cref{eqn:C}. As an example for $A$, which is particularly useful for our quantum decision tree approach, we propose a genetic algorithm \cite{mitchell1999,katoch2021}, which we briefly discuss in the following.\par
In general, a genetic algorithm is a gradient-free optimization strategy inspired by natural selection in which a population of candidate solutions is iteratively modified to find the best solution (among the set of evaluated candidates) that maximizes a given fitness function. Typically, a genetic algorithm is defined by three components:
\begin{enumerate}
	\item A chromosome representation of the solution domain in the sense that each candidate can be uniquely identified by their chromosome.
	\item A set of genetic operators to alter the population of candidate solutions (typically in the form of selection, crossover, and mutation).
	\item A fitness function, which can be evaluated for each candidate solution.
\end{enumerate}
The resulting solution is only the best solution from the set of evaluated candidates and it cannot be generally guaranteed that it is also globally optimal.\par
In our case, the goal of the genetic algorithm is to find a decision tree $T$ of a predefined maximal depth $d$ as its solution. Therefore, a possible chromosome representation of a candidate solution is given by the vector
\begin{align} \label{eqn:ga:sol}
	\mathbf{C} \equiv (c^0_1, c^1_1, c^1_2, c^2_1, \dots, c^d_{2^d}) \in \mathcal{S}
\end{align}
consisting of the swap indices $c^i_j$ of the corresponding compressed decision configuration $C$, \cref{eqn:C}. The elements of $\mathbf{C}$, which are also called chromosome attributes, are mutually independent and constrained by lower and upper bounds as defined in \cref{eqn:c}. These limits define the solution domain
\begin{align} \label{eqn:ga:S}
	\mathcal{S} \equiv \{ 1, \dots, k \}^1 \times \cdots \times \{ d+1, \dots, k \}^{2^d} \subseteq \{ 1, \dots, k \}^{\sum_{l=0}^d 2^l}
\end{align}
in such a way that $\mathbf{C}$ can represent any possible tree of maximal depth $d$. Furthermore, the fitness function can be any map of the form
\begin{align} \label{eqn:ga:F}
	F(\mathbf{C}, \mathbf{D}_{\mathrm{train}}) \mapsto \mathbb{R}
\end{align}
and should be chosen in such a way that ordering solutions by their ``quality'' corresponds to ordering them by their fitness.\par
A genetic algorithm with these ingredients can be used to find the approximation of the solution
\begin{align} \label{eqn:ga:opt}
	\mathbf{\hat{C}}_{\theta} \approx \argmax_{\mathbf{C}} F(\mathbf{C}, \mathbf{D}_{\mathrm{train}})
\end{align}
depending on its hyperparameters $\theta$. In \cref{sec:app:genetic induction algorithm}, we propose a concrete realization based on this conceptional framework.

\subsection{Probabilistic view} \label{sec:classical representation:probabilistic view}
Since the training data $\mathbf{D}_{\mathrm{train}}$, \cref{eqn:D:training}, is sampled from a probability distribution $\tilde{p}(\mathbf{x},\mathbf{y})$, \cref{eqn:p:true}, we can conversely use the probability distribution
\begin{align} \label{eqn:p}
	p(\mathbf{x},\mathbf{y}) \equiv \frac{\vert\{ \mathbf{d} | (\mathbf{x'},\mathbf{y'}) = \mathbf{d} \in \mathbf{D}_{\mathrm{train}} \land \mathbf{x'} = \mathbf{x} \land \mathbf{y'} = \mathbf{y} \}\vert}{\vert\mathbf{D}_{\mathrm{train}}\vert}
\end{align}
based on a histogram of the samples to estimate
\begin{align} \label{eqn:p:est}
	p(\mathbf{x},\mathbf{y}) \approx \tilde{p}(\mathbf{x},\mathbf{y}),
\end{align}
where $\vert\cdot\vert$ denotes the set cardinality. This estimated probability distribution can be used to establish a probabilistic view of a decision tree by switching from assigning subsets of the data to each node to assigning the corresponding histogram-based probability distributions. However, instead of recounting the histogram for each node, we follow a mathematically equivalent path by successively applying Bayes' rule to construct conditional probability distributions according to the decision rules, as described in the following. A related discussion regarding decision trees and probability distributions can also be found in \cite{correia2020}.\par
Given a tree $T$, we can obtain the corresponding training data $\mathbf{D}_{\mathrm{train}}$ and decision configuration $E$ according to \cref{eqn:TE}. The root (where no previously applied decision rules are present) can be directly associated with $p(\mathbf{x},\mathbf{y})$. The corresponding label probability distribution
\begin{align} \label{eqn:py}
	p(\mathbf{y}) = \sum_{\mathbf{x}} p(\mathbf{x},\mathbf{y})
\end{align}
can be found by marginalizing over all features $\mathbf{x}$. Here and in the following, a sum over binary variables iterates over $\mathbb{B}$ for each variable.\par
To obtain the probability distributions associated with deeper nodes, the splitting decisions have to be taken into account. For example, the first split of the tree with respect to the feature $e^0_1 \in \{1,\dots,k\}$ corresponds to a conditioning on $x_{e^0_1}$ so that the probability distributions assigned to the two branched nodes read
\begin{align}
	p(x_1,\dots,x_{e^0_1-1},x_{e^0_1+1},\dots,x_k,\mathbf{y} \,|\, x_{e^0_1}=0) = \frac{p(x_1,\dots,x_{e^0_1-1},x_{e^0_1}=0,x_{e^0_1+1},\dots,x_k,\mathbf{y})}{p(x_{e^0_1}=0)}
\end{align}
and
\begin{align}
	p(x_1,\dots,x_{e^0_1-1},x_{e^0_1+1},\dots,x_k,\mathbf{y} \,|\, x_{e^0_1}=1) = \frac{p(x_1,\dots,x_{e^0_1-1},x_{e^0_1}=1,x_{e^0_1+1},\dots,x_k,\mathbf{y})}{p(x_{e^0_1}=1)},
\end{align}
respectively. In analogy to \cref{eqn:py}, the corresponding label probability distributions are given by
\begin{align}
	p(\mathbf{y} \,|\, x_{e^0_1}=0) = \sum_{\mathbf{x}_{\setminus\{e^0_1\}}} p(\mathbf{x}_{\setminus\{e^0_1\}},\mathbf{y} \,|\, x_{e^0_1}=0)
\end{align}
and
\begin{align}
	p(\mathbf{y} \,|\, x_{e^0_1}=1) = \sum_{\mathbf{x}_{\setminus\{e^0_1\}}} p(\mathbf{x}_{\setminus\{e^0_1\}},\mathbf{y} \,|\, x_{e^0_1}=1),
\end{align}
respectively, where we have made use of the abbreviation
\begin{align}
	\mathbf{x}_{\setminus S} \equiv \{ x_i \,|\, i \in \{1,\dots,k\} \setminus S \} \subseteq \{ x_1, \dots, x_k \}
\end{align}
for a set of indices $S \subseteq \{1,\dots,k\}$. Here and in the following, sets as arguments of a probability distribution are to be understood in such a way that the corresponding elements are used as arguments to this probability distribution in an appropriate order.\par
Any node of depth $l$ in the tree $T$ can be identified by its index $\nu$, \cref{eqn:nu}. Instead of using the index, the node can also be identified by a set of conditions
\begin{align} \label{eqn:Cnu}
	\mathcal{C}_{\nu} \equiv \mathcal{C}_{\nu_l(\bar{x}_{0},\dots,\bar{x}_{l-1})} \equiv \{ x_{e^0}=\bar{x}_{e^0},\dots,x_{e^{l-1}}=\bar{x}_{e^{l-1}} \}
\end{align}
with feature values $\bar{x}_{i} \in \mathbb{B}$ for all $i \in \{0,\dots,l-1\}$ by following the corresponding sequence of decisions $\{e^0,\dots,e^{l-1}\}$, \cref{eqn:s}.\par
Consequently, any node can be associated with a marginalized conditional probability distribution 
\begin{align} \label{eqn:pn:xy}
	p(\mathbf{x}_{\setminus\{e^0,\dots,e^{l-1}\}},\mathbf{y} \,|\, \mathcal{C}_{\nu}) = \frac{p(\mathbf{x}_{\setminus\{e^0,\dots,e^{l-1}\}},\mathcal{C}_{\nu},\mathbf{y})}{p(\mathcal{C}_{\nu})}
\end{align}
with the probability distribution of the decision set
\begin{align} \label{eqn:pn}
	p(\mathcal{C}_{\nu}) = \sum_{\mathbf{x}_{\setminus\{e^0,\dots,e^{l-1}\}}, \mathbf{y}} p(\mathbf{x}_{\setminus\{e^0,\dots,e^{l-1}\}},\mathcal{C}_{\nu},\mathbf{y})
\end{align}
and the corresponding label probability distribution
\begin{align} \label{eqn:pn:y}
	p(\mathbf{y} \,|\, \mathcal{C}_{\nu}) = \sum_{\mathbf{x}_{\setminus\{e^0,\dots,e^{l-1}\}}} p(\mathbf{x}_{\setminus\{e^0,\dots,e^{l-1}\}},\mathbf{y} \,|\, \mathcal{C}_{\nu}),
\end{align}
which can also be marginalized to get the probability distribution
\begin{align} \label{eqn:pn:yi}
	p(y_i \,|\, \mathcal{C}_{\nu}) = \sum_{y_1,\dots,y_{i-1},y_{i+1},\dots,y_m} p(\mathbf{y} \,|\, \mathcal{C}_{\nu})
\end{align}
of the $i$th label with $i \in \{1,\dots,m\}$. Thus, we have established a probabilistic relation between the training data $\mathbf{D}_{\mathrm{train}}$ and the tree structure $E$, \cref{eqn:E}.\par
A probabilistic tree traversal corresponds to drawing a sample $\bar{\mathbf{x}} \in \mathbb{B}^k$ from 
\begin{align} \label{eqn:px}
	p(\mathbf{x}) = \sum_{\mathbf{y}} p(\mathbf{x},\mathbf{y}),
\end{align}
which is based on \cref{eqn:p:est}, and subsequently using the decisions $\mathcal{C}_{\nu}$ in the tree $T$ to reach a certain node. The probability distribution of reaching this node is then given by $p(\mathcal{C}_{\nu})$ and the corresponding label probability by $p(\mathbf{y} \,|\, \mathcal{C}_{\nu})$. Equivalently, the probabilistic tree traversal can also be understood as drawing a sample $\bar{\mathbf{x}} \in \mathbb{B}^k$ and $\bar{\mathbf{y}} \in \mathbb{B}^m$ directly from
\begin{align} \label{eqn:p:Cnuy}
	p(\mathcal{C}_{\nu}, \mathbf{y}=\bar{\mathbf{y}}) = \sum_{\mathbf{x}_{\setminus\{e^0,\dots,e^{l-1}\}}} p(\mathbf{x}_{\setminus\{e^0,\dots,e^{l-1}\}},\mathcal{C}_{\nu},\mathbf{y})|_{\mathbf{y}=\bar{\mathbf{y}}},
\end{align}
which allows to infer $p(\mathcal{C}_{\nu})$ and $p(\mathbf{y}=\bar{\mathbf{y}} \,|\, \mathcal{C}_{\nu})$, respectively.

\subsection{Tree predictions}
Given a tree $T$ of maximal depth $d$, the prediction of query data $\mathbf{x}^q \in \mathbb{B}^k$ in the sense of \cref{eqn:prediction} can be performed by an iterative traversal of the tree. Starting from the root, the index of the traversed node of depth $1$ (after the first split with feature index $e^0_{1}$) is given by
\begin{align}
	q_1 \equiv q_1(\mathbf{x}^q) = \nu_1(x^q_{e^0_{1}}),
\end{align}
the index of the subsequently traversed node of depth $2$ (after the second split with feature index $e^1_{q_1}$) by
\begin{align}
	q_2 \equiv q_2(\mathbf{x}^q) = \nu_2(x^q_{e^0_{1}} x^q_{e^1_{q_1}})
\end{align}
and so on, where we recall \cref{eqn:nu}. Generally, the the index of the traversed node of depth $l > 1$ reads
\begin{align}
	q_l \equiv q_l(\mathbf{x}^q) = \nu_l(x^q_{e^0_{1}} x^q_{e^1_{q_1}} \cdots x^q_{e^{l-1}_{q_{l-1}}})
\end{align}
until for $l=d$ the leaf with index $q_d$ is reached.\par
The traversed path imposes the set of conditions
\begin{align} \label{eqn:Cnu:q}
	\mathcal{C}_{q_d(\mathbf{x}^q)} = \{ x_{e^0_{1}}=x^q_{e^0_{1}}, \dots, x_{e^{d-1}_{q_{d-1}}}=x^q_{e^{d-1}_{q_{d-1}}} \}
\end{align}
according to \cref{eqn:Cnu}. Consequently, the predicted probability distribution for $i$th label with $i \in \{1,\dots,m\}$ reads
\begin{align} \label{eqn:pyi:pred}
	\hat{p}(y_i \,|\, \mathbf{x}^q) \equiv p(y_i \,|\, \mathcal{C}_{q_d(\mathbf{x}^q)}),
\end{align}
where we recall \cref{eqn:pn:yi}. This expression also allows to infer the most probable label as
\begin{align} \label{eqn:pyi:pred:argmax}
	\hat{y}_i \equiv \argmax_{y_i} p(y_i \,|\, \mathcal{C}_{q_d(\mathbf{x}^q)}),
\end{align}
where we recall \cref{eqn:pn:yi}.
\section{Quantum representation} \label{sec:quantum representation}
So far, our considerations have been completely classical. In this section, we propose Q-trees as a quantum representation of binary classification trees with binary features using quantum circuits \cite{nielsen2010}. Our considerations are mainly based on the probabilistic perspective on decision trees from \cref{sec:classical representation:probabilistic view}.\par
Specifically, we presume that there is a tree $T$ of maximum depth $d$, which is uniquely defined by the training data $\mathbf{D}_{\mathrm{train}}$, \cref{eqn:D:training}, and the compressed decision configuration $C$, \cref{eqn:C}. Any leaf of $T$ can be associated with a probability distribution, \cref{eqn:pn:xy}, that describes the probability distribution of $\mathbf{D}_{\mathrm{train}}$, \cref{eqn:p:est}, conditioned by the decisions along the path from the root to the leaf. Based on this perspective, we can construct a quantum circuit that performs a random traversal of the tree in the sense that it samples the label probability $p(\mathbf{y} \,|\, \mathcal{C}_{\nu})$, \cref{eqn:pn:y}, of a random leaf that is selected with probability $p(\mathcal{C}_{\nu})$, where $\mathcal{C}_{\nu}$, \cref{eqn:Cnu}, denotes the set of decisions that identifies a path in the tree. We achieve this by preparing the probability distribution of $p(\mathbf{x},\mathbf{y})$, \cref{eqn:p:est}, in terms of qubit amplitudes and subsequently apply a set of (conditional) $\SWAP$ gates to alter the amplitudes according to the tree structure. The final projective measurement over a subset of the qubits then yields a sample from $p(\mathcal{C}_{\nu}, \mathbf{y}=\bar{\mathbf{y}})$, \cref{eqn:p:Cnuy}.\par 
In the following, we explain this approach in more detail. We start by describing the corresponding quantum circuit and how it can be constructed from a given tree. In the subsequent part, we briefly discuss how tree inductions and label predictions of query data can be performed within this framework.

\subsection{Tree circuit} \label{sec:tree circuit}
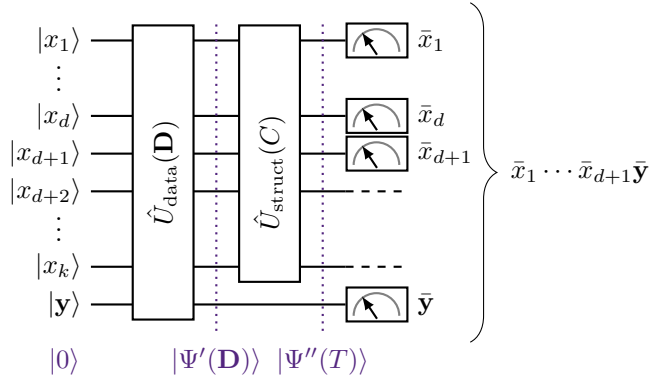
\begin{figure}
	\begin{center}
		\begin{tikzpicture}
\coordinate (q1) at (0,0);
\coordinate (d1) at (0,-.5);
\coordinate (qd) at (0,-1);
\coordinate (qd1) at (0,-1.5);
\coordinate (qd2) at (0,-2);
\coordinate (d2) at (0,-2.5);
\coordinate (qk) at (0,-3);
\coordinate (qy) at (0,-3.5);
\node[anchor=east] at ($(q1)+(-.25,0)$) {$\ket{x_1}$};
\node[anchor=east,xshift=-.25cm,yshift=.1cm] at ($(d1)+(-.25,0)$) {$\vdots$};
\node[anchor=east] at ($(qd)+(-.25,0)$) {$\ket{x_{d}}$};
\node[anchor=east] at ($(qd1)+(-.25,0)$) {$\ket{x_{d+1}}$};
\node[anchor=east] at ($(qd2)+(-.25,0)$) {$\ket{x_{d+2}}$};
\node[anchor=east] at ($(qk)+(-.25,0)$) {$\ket{x_k}$};
\node[anchor=east,xshift=-.25cm,yshift=.1cm] at ($(d2)+(-.25,0)$) {$\vdots$};
\node[anchor=east] (psiy) at ($(qy)+(-.25,0)$) {$\ket{\mathbf{y}}$};
\draw[thick] ($(q1)+(-.25,0)$) -- ($(q1)+(2.5,0)+(.6,0)$);
\draw[thick] ($(qd)+(-.25,0)$) -- ($(qd)+(2.5,0)+(.6,0)$);
\draw[thick] ($(qd1)+(-.25,0)$) -- ($(qd1)+(2.5,0)+(.6,0)$);
\draw[thick] ($(qd2)+(-.25,0)$) -- ($(qd2)+(2.5,0)+(.6,0)$);
\draw[thick] ($(qk)+(-.25,0)$) -- ($(qk)+(2.5,0)+(.6,0)$);
\draw[thick] ($(qy)+(-.25,0)$) -- ($(qy)+(2.5,0)+(.6,0)$);
\draw[thick,draw=black,fill=white] ($(q1)+(.5,0)+(-.4,.2)+(.2,0)$) rectangle ($(qy)+(.5,0)+(+.4,-.2)+(.2,0)$);
\node[rotate=90] (ui) at ($.5*(q1)+.5*(qy)+(.5,0)+(.2,0)$) {$\hat{U}_{\mathrm{data}}(\mathbf{D})$};
\draw[thick,draw=black,fill=white] ($(q1)+(1.7,0)+(-.4,.2)+(.4,0)$) rectangle ($(qk)+(1.7,0)+(+.4,-.2)+(.4,0)$);
\node[rotate=90] (ut) at ($.5*(q1)+.5*(qy)+(1.7,0)+(.4,0)$) {$\hat{U}_{\mathrm{struct}}(C)$};
\coordinate (m1) at ($(q1)+(2.5,0)+(.6,0)$);
\coordinate (md) at ($(qd)+(2.5,0)+(.6,0)$);
\coordinate (md1) at ($(qd1)+(2.5,0)+(.6,0)$);
\coordinate (my) at ($(qy)+(2.5,0)+(.6,0)$);
\measurement{m1}
\node[anchor=west] at (m.east) {$\bar{x}_1$};
\measurement{md}
\node[anchor=west] at (m.east) {$\bar{x}_{d}$};
\measurement{md1}
\node[anchor=west] at (m.east) {$\bar{x}_{d+1}$};
\draw[thick,dashed] ($(qd2)+(2.5,0)+(.6,0)$)-- ($(qd2)+(2.5,0)+(.75,0)+(.6,0)$);
\draw[thick,dashed] ($(qk)+(2.5,0)+(.6,0)$) -- ($(qk)+(2.5,0)+(.75,0)+(.6,0)$);
\measurement{my}
\node[anchor=west] at (m.east) {$\bar{\mathbf{y}}$};
\draw[decorate,decoration={brace,amplitude=10pt,raise=4pt},xshift=0pt,yshift=0pt] ($(q1)+(4,.5)+(.6,0)$) -- ($(qy)+(4,-.5)+(.6,0)$) node [anchor=west,midway,xshift=0.55cm] {$\bar{x}_1 \cdots \bar{x}_{d+1}\bar{\mathbf{y}}$};
\node[Qv] (psi0) at ($(psiy)+(0,-.75)$) {$\ket{0}$};
\draw let \p1=(psi0), \p2=(ui) in node[Qv] (psi1) at ($(\x2,\y1)+(.7,0)$) {$\ket{\Psi'(\mathbf{D})}$};
\draw let \p1=(psi1), \p2=(ut) in node[Qv] (psi2) at ($(\x2,\y1)+(.7,0)$) {$\ket{\Psi''(T)}$};
\draw[thick,Qv,dotted] ($(psi1)+(0,.4)$) -- ($(psi1)+(0,4.5)$);
\draw[thick,Qv,dotted] ($(psi2)+(0,.4)$) -- ($(psi2)+(0,4.5)$);
\end{tikzpicture}
		\caption{Q-tree (circuit). Layout for our proposed quantum representation of the classical tree $T$ of maximum depth $d$, which is uniquely defined by the training data $\mathbf{D}_{\mathrm{train}}$ and the compressed decision configuration $C$, \cref{eqn:C}. The circuit contains $k+m$ qubits in total, where the first $k$ qubits $\ket{x_1},\dots,\ket{x_k}$ represent the features and the next $m$ qubits $\ket{y_1},\dots,\ket{y_m}$ represent the labels. Initially, all qubits are prepared in the ground state $\ket{0}$, \cref{eqn:psi:0}. The first unitary operator $\hat{U}_{\mathrm{data}}(\mathbf{D}_{\mathrm{train}})$ performs an encoding of the training data that leads to the state $\ket{\Psi'(\mathbf{D}_{\mathrm{train}})}$, \cref{eqn:psi:d}. The second unitary operator $\hat{U}_{\mathrm{struct}}(C)$ performs an encoding of the decision configuration that leads to the state $\ket{\Psi''(T)}$, \cref{eqn:psi:t}. Third, a projective measurement on the first $d+1$ qubits and the last $m$ qubits is performed, \cref{eqn:pT:p}. Since the measurement results are drawn from a probability distribution that is associated with a randomly selected leaf of $T$, \cref{eqn:pT:p}, the circuit can be considered a quantum representation of this tree.} \label{fig:tree-circuit}
	\end{center}
\end{figure}

A Q-tree representing the classical tree $T$ can be realized by using the circuit layout sketched in \cref{fig:tree-circuit}. This \emph{Q-tree circuit}, which we also refer to as Q-tree for short, contains $k+m$ qubits in total. The first $k$ qubits (denoted by $\ket{x_1},\dots,\ket{x_k}$ or their respective indices $1,\dots,k$) represent the features, whereas the next $m$ qubits (denoted by $\ket{y_1},\dots,\ket{y_m}$ or their respective indices $k+1,\dots,k+m$) represents the labels.\par
Initially, all qubits are assumed to be prepared in the ground state
\begin{align} \label{eqn:psi:0}
	\ket{\Psi} \equiv \ket{0}.
\end{align}
In the next step, two unitary operators are applied. The first is responsible for the encoding of the training data and the second for the encoding of the structure (\ie, the decision configuration) of the tree. Finally, a projective measurement on the first $d+1$ qubits and the last $m$ qubits is performed. We discuss the components of the circuit in more detail below.

\subsubsection{Data encoding} \label{sec:quantum representation:data encoding}
The first unitary operator $\hat{U}_{\mathrm{data}}(\mathbf{D}_{\mathrm{train}})$ in the Q-tree, \cref{fig:tree-circuit}, is responsible for an encoding of the training data $\mathbf{D}_{\mathrm{train}}$, \cref{eqn:D:training}, in form of quantum amplitudes \cite{grover2002}. By definition, it acts on all $k+m$ qubits and leads to the state
\begin{align} \label{eqn:psi:d}
	\ket{\Psi'(\mathbf{D}_{\mathrm{train}})} \equiv \hat{U}_{\mathrm{data}}(\mathbf{D}_{\mathrm{train}}) \ket{0} = \sum_{\mathbf{x},\mathbf{y}} \sqrt{p(\mathbf{x},\mathbf{y})} \ket{\mathbf{x},\mathbf{y}},
\end{align}
where we recall $p(\mathbf{x},\mathbf{y})$, \cref{eqn:p}. The encoded state $\ket{\Psi'(\mathbf{D}_{\mathrm{train}})}$ contains all $k+m$ (possibly entangled) qubits. This data encoding, where features are encoded in qubits and their respective probabilities in the state amplitudes such that the amplitude vector represents a classical discrete probability distribution, is also referred to as \emph{qsample encoding} \cite{ozols2012,low2014,schuld2018a}.\par
Classically, storing the total probability distribution requires $\BigO{2^{k+m}}$ classical memory slots. Since only $k+m$ qubits are required to store the complete information from $p(\mathbf{x},\mathbf{y})$, the qsample encoding achieves an exponential compression. To obtain $\hat{U}_{\mathrm{data}}(\mathbf{D}_{\mathrm{train}})$ for a given $\mathbf{D}_{\mathrm{train}}$, classical algorithms can be used \cite{mottonen2005,shende2006,plesch2011}. Recent methods include \emph{black-box state preparation without arithmetics} \cite{sanders2019,bausch2021} and \emph{quantum generative adversarial networks} \cite{dallairedemers2018,benedetti2019,hu2019,zoufal2019}.
We do not further discuss these approaches here and instead assume that an appropriate data encoding operator is available to realize \cref{eqn:psi:d}. In general, $\hat{U}_{\mathrm{data}}(\mathbf{D}_{\mathrm{train}})$ consists of $\BigO{ 2^{k+m} }$ gate operations \cite{barenco1995} but can be simplified if the probability distribution is not fully correlated \cite{low2014}. The complexity can be further reduced if ancilla qubits are used \cite{araujo2021,sun2021,zhang2021}, an approach which we do not further pursue in this manuscript.

\subsubsection{Structure encoding} \label{sec:quantum representation:structure encoding}
It is well-known that quantum circuits with a qsample encoding can be used to manipulate probability distributions, for example to perform marginalizations \cite{low2014,schuld2018a}. This circumstance can be exploited to construct a unitary operator with a conditional mapping of the initial probability distribution of the data to marginalized conditional probability distributions in individual leaves of the tree. This is the conceptional idea behind the structure encoding (or splitting decision encoding), which we explain in more detail in the following.\par
The second unitary operator $\hat{U}_{\mathrm{struct}}(C)$ in the Q-tree, \cref{fig:tree-circuit}, is responsible for this structure encoding according to
\begin{align} \label{eqn:psi:t}
	\ket{\Psi''(T)} \equiv \hat{U}_{\mathrm{struct}}(C) \ket{\Psi'(\mathbf{D}_{\mathrm{train}})}.
\end{align}
It only acts on the first $k$ qubits and is determined by the compressed decision configuration $C$, \cref{eqn:C}. Specifically, each layer swap $\mathbf{c}^i$ with $i \in \{0,\dots,d\}$, \cref{eqn:c}, can be associated with $2^i$ (multi-controlled) $\SWAP$ gates, which are partially ``decorated'' by $\NOT$ gates. The $\NOT$ gates are added in such a way that each combination of decorated and non-decorated controlled qubits is placed once, each assigned to one element $c^i_j$ of $\mathbf{c}^i$ with $j \in \{1,\dots,2^i\}$.\par
This sequence of gates can be expressed by
\begin{align} \label{eqn:UC}
	\hat{U}_{\mathrm{struct}}(C) = \orderprod{i=0}{d} \hat{U}_{\mathrm{struct}}^i(\mathbf{c}^i)
\end{align}
with
\begin{align} \label{eqn:UC:i}
	\hat{U}_{\mathrm{struct}}^i(\mathbf{c}^i) \equiv \orderprod{j=1}{2^i} \hat{\Gamma}^i_j(c^i_j)
\end{align}
for each layer swap. The products in \cref{eqn:UC,eqn:UC:i} are ordered products in the sense of
\begin{align} \label{eqn:orderprod}
	\orderprod{i=a}{b} \hat{O}_i \equiv \hat{O}_b \hat{O}_{b+1} \cdots \hat{O}_{a-1} \hat{O}_a
\end{align}
for a set of unitary operators $\hat{O}_i$ for $i \in \{a,\dots,b\}$ and $a \leq b$. Furthermore, we introduce
\begin{align} \label{eqn:Gamma}
	\hat{\Gamma}^i_j(c) \equiv \hat{\mu}^i_j \, \hat{\gamma}^i(c) \, \hat{\mu}^i_j,
\end{align}
with $c \in \{i+1,\dots,k\}$. It contains the abbreviations
\begin{align} \label{eqn:mu}
	\hat{\mu}^i_j \equiv \begin{cases} \hat{\eta}^i_j & \text{for}\,\, i>0 \\ \unitop & \text{for}\,\, i=0 \end{cases}
\end{align}
and
\begin{align} \label{eqn:gamma}
	\hat{\gamma}^i(c) \equiv \begin{cases} \MCSWAP{i}( i+1, c, \{ 1, \dots, i \} ) & \text{for}\,\, i \geq 1 \\ \SWAP( i+1, c ) & \text{for}\,\, i = 0 \end{cases}
\end{align}
based on
\begin{align}
	\hat{\eta}^v_j \equiv \prod_{u=1}^{v} \begin{cases} \NOT(u) & \text{for}\,\, \kappa(j, v, u) = 0 \\ \unitop & \text{otherwise} \end{cases}
\end{align}
with $v \in \{1,\dots,d\}$. The latter also makes use of
\begin{align} \label{eqn:kappa}
	\kappa(j, v, u) \equiv \left\lfloor \frac{j-1}{2^{v-u}} \right\rfloor \MOD 2 \in \mathbb{B}
\end{align}
for $u \in \{1,\dots,v\}$.\par
The unitary operators $\hat{\mu}^i_j$ and $\hat{\gamma}^i(c)$ represent one of four standard gates \cite{nielsen2010}:
\begin{itemize}
	\item An identity gate, denoted by $\unitop$, which has no effect on the state.
	\item A $\NOT$ or Pauli X gate acting on the $q$th qubit with $q \in \{1,\dots,k\}$, which is denoted by $\NOT(q)$.
	\item A $\SWAP$ gate with respect to two target qubits $q_1$ and $q_2$ (with $q_1, q_2 \in \{1,\dots,k\}$), which is denoted by $\SWAP( q_1, q_2 )$. For identical swapping qubits $q_1=q_2$, one has $\SWAP(q_1,q_1) = \unitop$.
	\item A (multi-)controlled $\SWAP$ gate with respect to two target qubits $q_1$ and $q_2$ to swap (with $q_1, q_2 \in \{1,\dots,k\}$) and $v \geq 1$ control qubits defined by the non-empty set $Q \in \{1,\dots,k\}^v$, which is denoted by $\MCSWAP{v}( q_1, q_2 , Q )$ and acts on $v+2$ qubits. $\MCSWAP{1}$ represents a $\CSWAP$ or Fredkin gate. Furthermore, for identical swapping qubits $q_1=q_2$, one has $\MCSWAP{v}(q_1, q_1, Q) = \unitop$. To simplify our notation, we also write $\MCSWAP{0}( q_1, q_2 , \{\} ) \equiv \SWAP( q_1, q_2 )$.
\end{itemize}
The operations in $\hat{U}_{\mathrm{struct}}(C)$ can consequently be considered as a sequence of these standard gates. The $\NOT$, $\SWAP$, and (multi-)controlled $\SWAP$ gates are understood to act as identity gates (\ie, $\unitop$) on the remaining qubits.\par
An example circuit is outlined in \cref{fig:tree-compose}. It shows the connection between the structure of a decision tree $T$ (which is parameterized by the decision configuration $E$, \cref{eqn:E}, and the compressed decision configuration $C$, \cref{eqn:C}, with the correspondence from \cref{eqn:EC}) and its Q-tree encoding via \cref{eqn:UC}. The quantum representation of the tree structure of all nodes of depth $d$ requires up to $d 2^d$ $\NOT$ gates and up to $2^d$ $\MCSWAP{d}$ gates. However, at least $\sum_{l=1}^{d-1} 2^d (1-2^{-l})$ of the $\NOT$ gates are consecutively applied to the same qubit and therefore eliminate each other.\par
\begin{figure}
	\begin{center}
		\includestandalone{fig_tree-compose}{\phantomsubcaption\label{fig:tree-compose:a}}{\phantomsubcaption\label{fig:tree-compose:b}}
		\caption{Correspondence between a decision tree $T$ and the Q-tree encoding of its structure given by $\hat{U}_{\mathrm{struct}}(C)$, \cref{eqn:UC}, which is the second unitary operator in the Q-tree, \cref{fig:tree-circuit}. (a)~Decision tree from \cref{fig:sample-tree-param} with splitting feature indices $e^i_j$ as given by the decision configuration $E$, \cref{eqn:E}, and corresponding swap indices $c^i_j$ as given by the compressed decision configuration $C$, \cref{eqn:C}, where \cref{eqn:EC} holds true. (b)~Structural encoding transformation consisting of a sequence of layer transformations $\hat{U}_{\mathrm{struct}}^i(\mathbf{c}^i)$, \cref{eqn:UC:i}, which in turn are sequences of $\NOT$ and $\hat{\gamma}^i(c)$ gates. The latter represent a $\SWAP$ gate or a $\MCSWAP{i}$ gate as defined in \cref{eqn:gamma}. We denote $\NOT$ gates with $X$ and show the individual qubits affected by the $\SWAP$ and $\MCSWAP{i}$ gates: control qubits are marked by $\alpha$ and $\alpha'$, respectively, whereas qubits to swap and their potential swapping partners are marked by $\beta$ and $\beta'$, respectively. Thus, one has $\hat{\gamma}^0(c^0_j) = \SWAP( \beta, \beta' )$ for $d=0$, $\hat{\gamma}^1(c^1_j) = \MCSWAP{1}( \beta, \beta' , \{ \alpha \} )$ for $d=1$, and $\hat{\gamma}^2(c^2_j) = \MCSWAP{2}( \beta, \beta' , \{ \alpha, \alpha' \} )$ for $d=2$. The parameters $\alpha$, $\alpha'$, and $\beta$ are determined by $d$, whereas $\beta'$ is determined by the corresponding swap index $c^i_j$. Directly consecutive $\NOT$ gates eliminate each other and can therefore be omitted, but we nevertheless show them for the sake of completeness.} \label{fig:tree-compose}
	\end{center}
\end{figure}
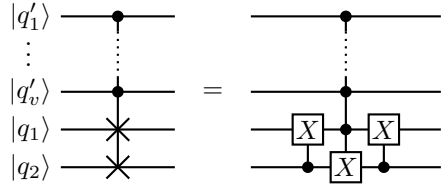
\begin{figure}
	\begin{center}
		\begin{tikzpicture}
% mcswap
\coordinate (q0) at (0,0);
\coordinate (qd) at (0,-.5);
\coordinate (qc) at (0,-1);
\coordinate (q1) at (0,-1.5);
\coordinate (q2) at (0,-2);
\node[anchor=east] at ($(q0)+(-.25,0)$) {$\ket{q'_1}$};
\node[anchor=east,xshift=-.25cm,yshift=.1cm] at ($(qd)+(-.25,0)$) {$\vdots$};
\node[anchor=east] at ($(qc)+(-.25,0)$) {$\ket{q'_v}$};
\node[anchor=east] at ($(q1)+(-.25,0)$) {$\ket{q_1}$};
\node[anchor=east] at ($(q2)+(-.25,0)$) {$\ket{q_2}$};
\draw[thick] ($(q0)+(-.25,0)$) -- ($(q0)+(1.25,0)$);
\draw[thick] ($(qc)+(-.25,0)$) -- ($(qc)+(1.25,0)$);
\draw[thick] ($(q1)+(-.25,0)$) -- ($(q1)+(1.25,0)$);
\draw[thick] ($(q2)+(-.25,0)$) -- ($(q2)+(1.25,0)$);
% cswap
\draw[thick]  ($(q0)+(.5,0)$) -- ($(q0)+(.5,-.2)$);
\draw[thick]  ($(qc)+(.5,0)$) -- ($(qc)+(.5,.2)$);
\draw[thick,dotted]  ($(q0)+(.5,-.2)$) -- ($(qc)+(.5,.2)$);
\draw[thick]  ($(qc)+(.5,0)$) -- ($(q2)+(.5,0)$);
\node[draw,thick,fill=black,circle,minimum width=.125cm,minimum height=.125cm,inner sep=0pt] at ($(q0)+(.5,0)$) {};
\node[draw,thick,fill=black,circle,minimum width=.125cm,minimum height=.125cm,inner sep=0pt] at ($(qc)+(.5,0)$) {};
\coordinate (swap) at ($(q1)+(.5,0)$);
\draw[thick] ($(swap)-(.15,.15)$) -- ($(swap)+(.15,.15)$);
\draw[thick] ($(swap)-(.15,-.15)$) -- ($(swap)+(.15,-.15)$);
\coordinate (swap) at ($(q2)+(.5,0)$);
\draw[thick] ($(swap)-(.15,.15)$) -- ($(swap)+(.15,.15)$);
\draw[thick] ($(swap)-(.15,-.15)$) -- ($(swap)+(.15,-.15)$);
% =
\node[] at ($.5*(q0)+.5*(q2)+(1.75,0)$){$=$};
% decomposition
\coordinate (q0) at ($(q0)+(2.5,0)$);
\coordinate (qd) at ($(qd)+(2.5,0)$);
\coordinate (qc) at ($(qc)+(2.5,0)$);
\coordinate (q1) at ($(q1)+(2.5,0)$);
\coordinate (q2) at ($(q2)+(2.5,0)$);;
\draw[thick] ($(q0)+(-.25,0)$) -- ($(q0)+(2.25,0)$);
\draw[thick] ($(qc)+(-.25,0)$) -- ($(qc)+(2.25,0)$);
\draw[thick] ($(q1)+(-.25,0)$) -- ($(q1)+(2.25,0)$);
\draw[thick] ($(q2)+(-.25,0)$) -- ($(q2)+(2.25,0)$);
% cnot
\draw[thick]  ($(q1)+(.5,0)$) --  ($(q2)+(.5,0)$);
\node[draw,thick,fill=black,circle,minimum width=.125cm,minimum height=.125cm,inner sep=0pt] at ($(q2)+(.5,0)$) {};
\node[draw,thick,fill=white,minimum width=.4cm,minimum height=.4cm,inner sep=0pt] at ($(q1)+(.5,0)$) {$X$};
% mcnot
\draw[thick]  ($(q0)+(1,0)$) -- ($(q0)+(1,-.2)$);
\draw[thick]  ($(qc)+(1,0)$) -- ($(qc)+(1,.2)$);
\draw[thick,dotted]  ($(q0)+(1,-.2)$) -- ($(qc)+(1,.2)$);
\draw[thick]  ($(qc)+(1,0)$) -- ($(q2)+(1,0)$);
\node[draw,thick,fill=black,circle,minimum width=.125cm,minimum height=.125cm,inner sep=0pt] at ($(q0)+(1,0)$) {};
\node[draw,thick,fill=black,circle,minimum width=.125cm,minimum height=.125cm,inner sep=0pt] at ($(qc)+(1,0)$) {};
\node[draw,thick,fill=black,circle,minimum width=.125cm,minimum height=.125cm,inner sep=0pt] at ($(q1)+(1,0)$) {};
\node[draw,thick,fill=white,minimum width=.4cm,minimum height=.4cm,inner sep=0pt] at ($(q2)+(1,0)$) {$X$};
% cnot
\draw[thick]  ($(q1)+(1.5,0)$) --  ($(q2)+(1.5,0)$);
\node[draw,thick,fill=black,circle,minimum width=.125cm,minimum height=.125cm,inner sep=0pt] at ($(q2)+(1.5,0)$) {};
\node[draw,thick,fill=white,minimum width=.4cm,minimum height=.4cm,inner sep=0pt] at ($(q1)+(1.5,0)$) {$X$};
\end{tikzpicture}
		\caption{Decomposition of a (multi-controlled) $\SWAP$ gate into two $\CNOT$ gates and a \mbox{(multi-)controlled} $\NOT$ gate as given by \cref{eqn:swapdec}.} \label{fig:decomposition-mcswap}
	\end{center}		
\end{figure}

To estimate the complexity of the structure encoding circuit, we decompose it into one- and two-qubit gate operations \cite{barenco1995}. For this purpose, we make use of the relation \cite{liu2020}
\begin{align} \label{eqn:swapdec}
	\MCSWAP{v}( q_1, q_2 , Q ) = \CNOT(q_2, q_1) \, \MCNOT{v+1}( q_1, Q \cup \{q_1\} ) \, \CNOT(q_2, q_1)
\end{align}
for $q_1 \neq q_2$ with $v \geq 0$ control qubits, which is also sketched in \cref{fig:decomposition-mcswap}. Here, $\MCNOT{v}( q,  Q )$ denotes a (multi-)controlled $\NOT$ gate acting on qubit $q$ with $v \geq 1$ control qubits defined by the non-empty set $Q \equiv \{q'_1,\dots,q'_v\}^v$ with $q,q'_1,\dots,q'_v \in \{1,\dots,k\}$. For $v=1$, $\MCNOT{1}( q, \{ q'_1 \} )$ represents a two-qubit $\CNOT$ gate, which is also denoted by $\CNOT( q,  q'_1 )$. For $v>2$, $\MCNOT{v}( q,  Q )$ is also called (multi-controlled) Toffoli gate. The $\MCNOT{v}$ gate can be further decomposed in various ways \cite{biswal2019}, depending on whether additional ancilla qubits are to be used or not. We particularly consider an ancilla-free decomposition into $8 v - 20$ two-qubit controlled-rotation gates \cite{saeedi2013}. Thus, the structure encoding of a tree $T$ with depth $d$ requires a total of $\BigO{d 2^d}$ $\NOT$ gates and $\BigO{d}$ two-qubit controlled-rotation gates, resulting in an effective complexity of $\BigO{d 2^d}$ gates.\par
As visualized in \cref{fig:tree-compose}, each unitary operator $\hat{\Gamma}^i_j(c)$, \cref{eqn:Gamma}, can be associated with a node of depth $i$ and index $j$ in $T$. According to the definition of the operator, $\NOT$ gates appear only in pairs and therefore effectively only $\MCSWAP{i}$ gates are applied. In particular, the $\NOT$ decorations are exclusively attached to control qubits before and after a $\MCSWAP{i}$ and are therefore responsible for flipping the controls temporarily. The placement of $\NOT$ gates is determined by $\kappa(j, v, u)$, \cref{eqn:kappa}, which decides whether the $u$th qubit acts as an open control gate (for $\kappa(j, v, u) = 0$) or a closed control gate \cite{liu2020}. The application of $\hat{\Gamma}^i_j(c)$ to an arbitrary entangled $k$-qubit state $\sum_{\mathbf{x}} a(\mathbf{x}) \ket{\mathbf{x}}$ with amplitudes $a(\mathbf{x})$ consequently yields
\begin{align}
	\hat{\Gamma}^i_j(c) \sum_{\mathbf{x}} a(\mathbf{x}) \ket{\mathbf{x}} = \sum_{\mathbf{x}} a_{\Gamma}(\mathbf{x}) \ket{\mathbf{x}}
\end{align}
with
\begin{align}
	a_{\Gamma}(\mathbf{x}) \equiv \begin{cases} a(x_1,\dots,x_{i},x_{c},x_{i+2},\dots,x_{c-1},x_{i+1},x_{c+1},\dots,x_k) & \text{for}\,\, x_q=\kappa(j, i, u) \,\forall\, u \in \{1,\dots,i\} \\ a(\mathbf{x}) & \text{otherwise} \end{cases}
\end{align}
for $i \geq 1$ and, without loss of generality, $i+1 < c$. Clearly, the only transformation of the amplitudes is a conditional swap of indices. In case of $i=0$, the index swap is performed unconditioned and for $i+1 = c$, one has $a_{\Gamma}(\mathbf{x}) = a(\mathbf{x})$.

\subsubsection{Measurement} \label{sec:quantum representation:measurement}
The projective measurement at the end of the Q-tree yields a bit string
\begin{align} \label{eqn:pT:bitstring}
	\bar{x}_1\cdots\bar{x}_{d+1}\bar{\mathbf{y}} \sim p_{T}(\bar{x}_{1},\dots,\bar{x}_{d+1},\bar{\mathbf{y}})
\end{align}
with
\begin{align} \label{eqn:pT:p}
	p_{T}(\bar{x}_{1},\dots,\bar{x}_{d+1},\bar{\mathbf{y}}) & \equiv p_{T}(x_1=\bar{x}_{1},\dots,x_{d+1}=\bar{x}_{d+1},\mathbf{y}=\bar{\mathbf{y}}) \nonumber\\
	& = \operatorname{Tr} \{ \braket{\bar{x}_1,\dots,\bar{x}_{d+1}, \bar{\mathbf{y}} | \Psi''(T)} \braket{\Psi''(T) | \bar{x}_1,\dots,\bar{x}_{d+1},\bar{\mathbf{y}}} \}_{\bar{x}_{d+2},\dots,\bar{x}_{k}} \nonumber\\
	& = \sum_{x_{d+2},\dots,x_k} p_{T}(\mathbf{x},\bar{\mathbf{y}})|_{x_{1}=\bar{x}_1,\dots,x_{d+1}=\bar{x}_{d+1}} \nonumber\\
	& = \sum_{\mathbf{x}_{\setminus\{e^0,\dots,e^{d}\}}} p(\mathbf{x},\bar{\mathbf{y}})|_{x_{e^0}=\bar{x}_1,\dots,x_{e^{d}}=\bar{x}_{d+1}} \nonumber\\
	& = p(x_{e^0}=\bar{x}_1,\dots,x_{e^{d}}=\bar{x}_{d+1},\mathbf{y}=\bar{\mathbf{y}}) = p(\mathcal{C}_{\nu}, \mathbf{y}=\bar{\mathbf{y}}),
\end{align}
as shown in \cref{sec:app:q-tree measurements}, where we recall $\mathcal{C}_{\nu}=\mathcal{C}_{\nu(\bar{x}_{1},\dots,\bar{x}_{d+1})}$, \cref{eqn:Cnu}, and $p(\mathcal{C}_{\nu}, \mathbf{y}=\bar{\mathbf{y}})$, \cref{eqn:p:Cnuy}. Thus, we can infer the probability distribution of reaching a certain leaf, \cref{eqn:pn}, from
\begin{align} \label{eqn:pT:pn}
	p_{T}(\bar{x}_{1},\dots,\bar{x}_{d+1}) = \sum_{\mathbf{y}} p_{T}(\bar{x}_{1},\dots,\bar{x}_{d+1},\mathbf{y}) = p(\mathcal{C}_{\nu}),
\end{align}
and the label probability distribution, \cref{eqn:pn:y}, from
\begin{align} \label{eqn:pT:pn:y}
	p_{T}(\bar{\mathbf{y}} | \bar{x}_{1},\dots,\bar{x}_{d+1}) = \frac{p_{T}(\bar{x}_{1},\dots,\bar{x}_{d+1},\mathbf{y})}{p_{T}(\bar{x}_{1},\dots,\bar{x}_{d+1})} = p(\mathbf{y} | \mathcal{C}_{\nu})|_{\mathbf{y}=\bar{\mathbf{y}}},
\end{align}
respectively. In addition, we can also obtain the probability distribution
\begin{align} \label{eqn:pT:pn:yi}
	p_{T}(\bar{y_i} | \bar{x}_{1},\dots,\bar{x}_{d+1}) = \sum_{y_1,\dots,y_{i-1},y_{i+1},\dots,y_m} 	p_{T}(\mathbf{y} | \bar{x}_{1},\dots,\bar{x}_{d+1}) |_{\mathbf{y}=\bar{\mathbf{y}}} = p(y_i | \mathcal{C}_{\nu})|_{y_i=\bar{y}_i},
\end{align}
of the $i$th label with $i \in \{1,\dots,m\}$, \cref{eqn:pn:yi}.\par
Summarized, we have shown that measuring the Q-tree (\ie, the circuit shown in \cref{fig:tree-circuit} with $k+m$ qubits and a complexity of $\BigO{2^{k+m} + d 2^d}$ gates) corresponds to drawing a sample from the conditional probability distribution $p(\mathcal{C}_{\nu}, \mathbf{y}=\bar{\mathbf{y}})$, \cref{eqn:p:Cnuy}, that is associated with the randomly selected leaf of depth $d$ and index $\nu$ of the respective decision tree $T$. The probability that a certain leaf is selected is given by the corresponding probability distribution $p(\mathcal{C}_{\nu})$, \cref{eqn:pn}. Consequently, the Q-tree can be considered a quantum representation of $T$.\par
Conversely, from multiple measurements we can estimate $p(\mathcal{C}_{\nu})$ and $p(y_i | \mathcal{C}_{\nu})$, respectively. For this purpose, an ensemble of bit strings $\{b_1,\dots,b_N\}$ is collected from a sequence of $N$ measurements. For each bit string, the corresponding set of conditions $\mathcal{C}_{\nu}$, \cref{eqn:Cnu}, is extracted from the fist $d+1$ bits and the corresponding labels $\bar{\mathbf{y}}$ from the next $m$ bits. The number of measurements $n(\mathcal{C}_{\nu})$ that fulfills the set of conditions $\mathcal{C}_{\nu}$ is collected and divided into $2m$ bins. First, $m$ bins $n_j^{0}(\mathcal{C}_{\nu})$ with $j \in \{1,\dots,m\}$, for which the $(d+1+j)$th bit is $0$ (\ie, $\bar{y}_j=0$). And second, $m$ bins $n_j^{1}(\mathcal{C}_{\nu})$, for which the $(d+1+j)$th bit is $1$. By definition, one has $\sum_{\mathcal{C}_{\nu}} n(\mathcal{C}_{\nu}) = N$ and $n(\mathcal{C}_{\nu})=n_j^{0}(\mathcal{C}_{\nu})+n_j^{1}(\mathcal{C}_{\nu})$ for all $j \in \{1,\dots,m\}$, respectively.\par
Based on these measurement results, we can obtain the approximations
\begin{align} \label{eqn:pn:est}
	p(\mathcal{C}_{\nu}) \approx \hat{p}(\mathcal{C}_{\nu}) \equiv \frac{n(\mathcal{C}_{\nu})}{N}
\end{align}
and
\begin{align} \label{eqn:pyi:est}
	p(y_i \,|\, \mathcal{C}_{\nu}) \approx \hat{p}(y_i \,|\,\mathcal{C}_{\nu}) \equiv \frac{n_i^{y_i}(\mathcal{C}_{\nu})}{n(\mathcal{C}_{\nu})} \,\,\text{if}\,\, n(\mathcal{C}_{\nu}) > 0,
\end{align}
respectively. For $n(\mathcal{C}_{\nu}) = 0$ (\ie, none of the measured bit strings fulfills the set of conditions $\mathcal{C}_{\nu}$), \cref{eqn:pyi:est} is undefined. The (statistical) uncertainties of \cref{eqn:pn:est,eqn:pyi:est} are discussed in \cref{sec:app:uncertainty}.\par
The sampling from measurements corresponds to a truly random tree traversal the sense that it relies on intrinsic quantum randomness \cite{bera2017}. If we assume that the training data $\mathbf{D}_{\mathrm{train}}$ is sufficiently large such that \cref{eqn:p:est} holds true, such samples are in fact approximately drawn from the underlying distribution $\tilde{p}(\mathcal{C}_{\nu}, \mathbf{y}=\bar{\mathbf{y}})$, \cref{eqn:p:true}. In the following, we briefly explain how we can use the proposed setup to induce trees or to predict the labels of (uncertain) query data.

\subsection{Tree induction} \label{sec:quantum representation:tree induction} 
To induce a Q-tree in the sense of \cref{eqn:induction:C}, we consider it as a parameterized (or variational) circuit \cite{benedetti2019}. Specifically, the Q-tree is according to \cref{eqn:psi:d,eqn:psi:t} parameterized by the training data $\mathbf{D}_{\mathrm{train}}$, \cref{eqn:D:training}, and the compressed decision configuration $C$, \cref{eqn:C}. Since the training data is supposed to be immutable, the variable parameters are represented only by $C$. Consequently, a variational quantum algorithm can be used to optimized the parameterized circuit according to some optimization goal \cite{cerezo2020}.
Following this strategy, we employ a genetic algorithm in the same way as for the classical tree. For this purpose, we replace the fitness function, \cref{eqn:ga:F}, by
\begin{align} \label{eqn:ga:F:q}
	\hat{F}(b_1,\dots,b_N,\mathbf{D}_{\mathrm{train}}) \mapsto \mathbb{R},
\end{align}
where $\{b_1,\dots,b_N\}$ represents an ensemble of $N$ bit strings, \cref{eqn:pT:bitstring}, collected from a sequence of $N$ measurements of the Q-tree. Due to the noisy nature of the measurement results from quantum circuits, $\hat{F}(b_1,\dots,b_N,\mathbf{D}_{\mathrm{train}})$ can in fact be considered a noisy fitness function \cite{zhai1996}. A concrete example for \cref{eqn:ga:F:q} is proposed in \cref{sec:app:genetic induction algorithm}.\par
The use of genetic algorithms in the context of quantum computing is a well-known field \cite{sofge2006,zhang2011,lahozbeltra2016}. Recent results particularly highlight the potential of evolutionary methods for parameterized circuit optimization in the presence of noisy quantum hardware \cite{grimsley2019,rattew2020,franken2021}. However, there are still unresolved practical challenges such as barren plateaus \cite{arrasmith2020}.

\subsection{Tree predictions} \label{sec:quantum representation:tree predictions}
To predict the labels of query data $\mathbf{x}^Q \in \mathbb{B}^k$, we assume that the estimated label probability distribution $\hat{p}(y_i \,|\, \mathcal{C}_{\nu})$, \cref{eqn:pyi:est}, is known for all $\nu \in \{1,\dots,2^d\}$ from a previous sampling. In this case, there are two alternative prediction approaches, which we present in the following.\par
The first prediction approach corresponds to a classical evaluation of the tree in the sense that the knowledge about the decision configuration $E$, \cref{eqn:E}, allows a conditional traversal of the tree according to the elements of $\mathbf{x}^q$ until a leaf is reached. Thus, the predicted probability distribution for the $i$th label with $i \in \{1,\dots,m\}$ reads
\begin{align} \label{eqn:pyi:pred:cl}
	\hat{p}_{\mathrm{cl}}(y_i \,|\, \mathbf{x}^q) \equiv \hat{p}(y_i \,|\, \mathcal{C}_{q_d(\mathbf{x}^q)})
\end{align}
in analogy to \cref{eqn:pyi:pred}, where we recall $\mathcal{C}_{q_d(\mathbf{x}^q)}$, \cref{eqn:Cnu:q}. Furthermore, we can use 
\begin{align} \label{eqn:pyi:pred:cl:argmax}
	\hat{y}_i \equiv \argmax_{y_i} \hat{p}_{\mathrm{cl}}(y_i \,|\, \mathbf{x}^q)
\end{align}
in analogy to \cref{eqn:pyi:pred:argmax} to predict labels.\par
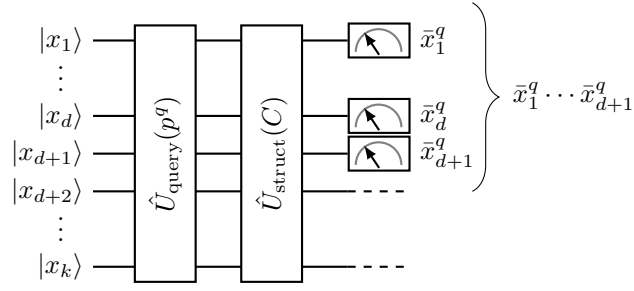
\begin{figure}
	\begin{center}
		\begin{tikzpicture}
\coordinate (q1) at (0,0);
\coordinate (d1) at (0,-.5);
\coordinate (qd) at (0,-1);
\coordinate (qd1) at (0,-1.5);
\coordinate (qd2) at (0,-2);
\coordinate (d2) at (0,-2.5);
\coordinate (qk) at (0,-3);
\node[anchor=east] at ($(q1)+(-.25,0)$) {$\ket{x_1}$};
\node[anchor=east,xshift=-.25cm,yshift=.1cm] at ($(d1)+(-.25,0)$) {$\vdots$};
\node[anchor=east] at ($(qd)+(-.25,0)$) {$\ket{x_{d}}$};
\node[anchor=east] at ($(qd1)+(-.25,0)$) {$\ket{x_{d+1}}$};
\node[anchor=east] at ($(qd2)+(-.25,0)$) {$\ket{x_{d+2}}$};
\node[anchor=east] at ($(qk)+(-.25,0)$) {$\ket{x_k}$};
\node[anchor=east,xshift=-.25cm,yshift=.1cm] at ($(d2)+(-.25,0)$) {$\vdots$};
\draw[thick] ($(q1)+(-.25,0)$) -- ($(q1)+(2.5,0)+(.6,0)$);
\draw[thick] ($(qd)+(-.25,0)$) -- ($(qd)+(2.5,0)+(.6,0)$);
\draw[thick] ($(qd1)+(-.25,0)$) -- ($(qd1)+(2.5,0)+(.6,0)$);
\draw[thick] ($(qd2)+(-.25,0)$) -- ($(qd2)+(2.5,0)+(.6,0)$);
\draw[thick] ($(qk)+(-.25,0)$) -- ($(qk)+(2.5,0)+(.6,0)$);
\draw[thick,draw=black,fill=white] ($(q1)+(.5,0)+(-.4,.2)+(.2,0)$) rectangle ($(qk)+(.5,0)+(+.4,-.2)+(.2,0)$);
\node[rotate=90] (ui) at ($.5*(q1)+.5*(qk)+(.5,0)+(.2,0)$) {$\hat{U}_{\mathrm{query}}(p^{q})$};
\draw[thick,draw=black,fill=white] ($(q1)+(1.7,0)+(-.4,.2)+(.4,0)$) rectangle ($(qk)+(1.7,0)+(+.4,-.2)+(.4,0)$);
\node[rotate=90] (ut) at ($.5*(q1)+.5*(qk)+(1.7,0)+(.4,0)$) {$\hat{U}_{\mathrm{struct}}(C)$};
\coordinate (m1) at ($(q1)+(2.5,0)+(.6,0)$);
\coordinate (md) at ($(qd)+(2.5,0)+(.6,0)$);
\coordinate (md1) at ($(qd1)+(2.5,0)+(.6,0)$);
%\coordinate (my) at ($(qy)+(2.5,0)+(.6,0)$);
\measurement{m1}
\node[anchor=west] at (m.east) {$\bar{x}_1^q$};
\measurement{md}
\node[anchor=west] at (m.east) {$\bar{x}_{d}^q$};
\measurement{md1}
\node[anchor=west] at (m.east) {$\bar{x}_{d+1}^q$};
\draw[thick,dashed] ($(qd2)+(2.5,0)+(.6,0)$)-- ($(qd2)+(2.5,0)+(.75,0)+(.6,0)$);
\draw[thick,dashed] ($(qk)+(2.5,0)+(.6,0)$) -- ($(qk)+(2.5,0)+(.75,0)+(.6,0)$);
\draw[decorate,decoration={brace,amplitude=10pt,raise=4pt},xshift=0pt,yshift=0pt] ($(q1)+(4,.5)+(.6,0)$) -- ($(qd1)+(4,-.5)+(.6,0)$) node [anchor=west,midway,xshift=0.55cm] {$\bar{x}_1^q \cdots \bar{x}_{d+1}^q$};
\end{tikzpicture}
		\caption{Query circuit. Layout for the prediction of labels for a probability distribution of query data $p^{q}(\mathbf{x}^q)$, \cref{eqn:pq}, using a Q-tree, \cref{fig:tree-circuit}. The circuit contains $k$ qubits and consists of the unitary operator $\hat{U}_{\mathrm{query}}(p^{q})$ for the qsample encoding of $p^{q}(\mathbf{x}^q)$, the unitary operator $\hat{U}_{\mathrm{struct}}(C)$, \cref{eqn:UC}, for the structural encoding of the tree $T$ and a subsequent projective measurement of the first $d+1$ qubits. The measurement result, \cref{eqn:pq:measured}, can be used for the prediction of labels of uncertain query data, \cref{eqn:pq:pyi:pred}.} \label{fig:pred-circuit}
	\end{center}
\end{figure}

The second prediction approach corresponds to a semi-quantum evaluation of the tree. This approach allows to process a probability distribution of query data
\begin{align} \label{eqn:pq}
	p^{q} \equiv p^{q}(\mathbf{x}^q)
\end{align}
with support $\mathbb{B}^k$ instead of a single data point, which enables the consideration of uncertainty in the query data. For this purpose, the evaluation of the quantum circuit outlined in \cref{fig:pred-circuit} is required, which we refer to as \emph{query circuit}. It contains $k$ qubits, which are assumed to be initially prepared in the ground state $\ket{0}$, \cref{eqn:psi:0}, and consists of two consecutive unitary operators followed by a projective measurement of the first $d+1$ qubits. The first unitary operator $\hat{U}_{\mathrm{query}}(p^{q})$ is responsible for a qsample encoding of $p^{q}(\mathbf{x}^q)$ in analogy to $\hat{U}_{\mathrm{data}}(\mathbf{D}_{\mathrm{train}})$, \cref{eqn:psi:d}, whereas the second unitary operator corresponds to $\hat{U}_{\mathrm{struct}}(C)$, \cref{eqn:UC}, and is therefore responsible for the structural encoding of the tree $T$. In particular, $p^{q}$ can also represent a single query data point $\mathbf{x}^q$ by setting $p^{q}(\mathbf{x}^q) = 1$. The qsample data encoding can in this case be achieved straightforwardly by applying a suitably chosen array of $\NOT$ gates.\par
A measurement yields the a string
\begin{align}
	\bar{x}_1^q\cdots\bar{x}_{d}^q \sim p_{(p^{q},C)}(\bar{x}_{1},\dots,\bar{x}_{d+1})
\end{align}
with
\begin{align} \label{eqn:pq:measured}
	p_{(p^{q},C)}(\bar{x}_{1},\dots,\bar{x}_{d+1}) = \operatorname{Tr} \{ \braket{\bar{x}^q_1,\dots,\bar{x}^q_{d+1} | \Psi^{q}(p^{q},C)} \braket{\Psi^{q}(p^{q},C) | \bar{x}^q_1,\dots,\bar{x}^q_{d+1}} \}_{\bar{x}_{d+2},\dots,\bar{x}_{k}}
\end{align}
and
\begin{align}
	\ket{\Psi^{q}(p^{q},C)} \equiv \hat{U}_{\mathrm{struct}}(C) \hat{U}_{\mathrm{query}}(p^{q}) \ket{0}.
\end{align}
We obtain an ensemble $\{b_1,\dots,b_N\}$ of such bit strings from $N$ measurements and count the number of measurements $n(\mathcal{C}_{\nu})$ that fulfill a set of conditions $\mathcal{C}_{\nu}$, \cref{eqn:Cnu}, such that $\sum_{\mathcal{C}_{\nu}} n(\mathcal{C}_{\nu}) = N$. Thus, we can estimate the probability distribution of reaching a certain leaf
\begin{align} \label{eqn:pq:pn:est}
	\hat{p}^{q}(\mathcal{C}_{\nu} \,|\, p^q) \equiv \frac{n(\mathcal{C}_{\nu})}{N},
\end{align}
in the sense of \cref{eqn:pn}.\par
The predicted probability distribution for $i$th label with $i \in \{1,\dots,m\}$ consequently reads
\begin{align} \label{eqn:pq:pyi:pred}
	\hat{p}_{\mathrm{qm}}(y_i \,|\, p^q) \equiv \sum_{\nu = 1}^{2^d} \hat{p}^{q}(\mathcal{C}_{\nu} \,|\, p^q) \hat{p}(y_i \,|\, \mathcal{C}_{\nu}),
\end{align}
which represents an average of $\hat{p}(y_i \,|\, \mathcal{C}_{\nu})$ over all leaves in the sense of an approximation
\begin{align} \label{eqn:pq:pyi:pred:approx}
	\hat{p}_{\mathrm{qm}}(y_i \,|\, p^q) \approx p(y_i \,|\, p^q) \equiv \sum_{\nu = 1}^{2^d} p^q(\mathcal{C}_{\nu}) p(y_i \,|\, \mathcal{C}_{\nu}),
\end{align}
where we recall \cref{eqn:pyi:pred}. The labels can be obtained in analogy to \cref{eqn:pyi:pred:cl:argmax}. The (statistical) uncertainties of \cref{eqn:pyi:pred:cl,eqn:pq:pn:est,eqn:pq:pyi:pred} are discussed in \cref{sec:app:uncertainty}.\par
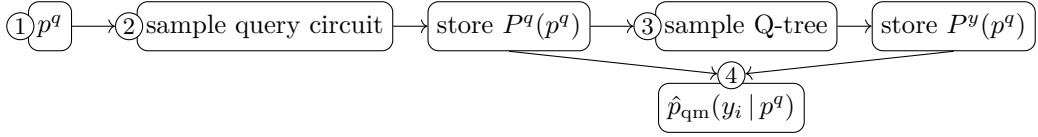
\begin{figure}
	\begin{center}
		\begin{tikzpicture}
\node[draw,rounded corners,minimum height=.65cm,align=center] (pq) at (0,0) {$p^q$};
\node[draw,rounded corners,minimum height=.65cm,align=center,anchor=west] (s1) at ($(pq.east)+(.85,0)$) {sample query circuit};
\node[draw,rounded corners,minimum height=.65cm,align=center,anchor=west] (Pq) at ($(s1.east)+(.45,0)$) {store $P^q(p^q)$};
\node[draw,rounded corners,minimum height=.65cm,align=center,anchor=west] (s2) at ($(Pq.east)+(.85,0)$) {sample Q-tree};
\node[draw,rounded corners,minimum height=.65cm,align=center,anchor=west] (Py) at ($(s2.east)+(.45,0)$) {store $P^y(p^q)$};
\node[draw,rounded corners,minimum height=.65cm,anchor=north,align=center] (pr) at ($.5*(Pq)+.5*(Py)+(0,-.75)$) {$\hat{p}_{\mathrm{qm}}(y_i \,|\, p^q)$};
\node[draw,circle,fill=white,inner sep=1pt,font=\small,xshift=-.1cm] (Pqn) at (pq.west) {1};
\node[draw,circle,fill=white,inner sep=1pt,font=\small,xshift=-.1cm] (s1n) at (s1.west) {2};
\node[draw,circle,fill=white,inner sep=1pt,font=\small,xshift=-.1cm] (s2n) at (s2.west) {3};
\node[draw,circle,fill=white,inner sep=1pt,font=\small,yshift=.1cm] (prn) at (pr.north) {4};
\draw[->] (pq) -- (s1n);
\draw[->] (s1) -- (Pq);
\draw[->] (Pq) -- (s2n);
\draw[->] (s2) -- (Py);
\draw[->] (Pq.south) -- (prn);
\draw[->] (Py.south) -- (prn);
\end{tikzpicture}
		\caption{Sketch of the proposed on-demand sampling method to perform Q-tree predictions with a constant number of classical memory slots, which consists of four steps. First (\circlednumber{1}), the probability distribution of query data $p^{q}(\mathbf{x}^q)$, \cref{eqn:pq}, is chosen. Second (\circlednumber{2}), the query circuit from \cref{fig:pred-circuit} is measured $N$ times to sample $\hat{p}^{q}(\mathcal{C}_{\nu})$, \cref{eqn:pq:pn:est}. The results are stored in $P^q(p^q)$, \cref{eqn:Pq}. Third (\circlednumber{3}), the Q-tree, \cref{fig:tree-circuit}, is measured $N$ times to sample $\hat{p}(y_i \,|\, \mathcal{C}_{\nu})$, \cref{eqn:pyi:pred:cl}, where only leaves present in $\Lambda(p^q)$, \cref{eqn:Lambdaq}, are considered. The results are stored in $P^y(p^q)$, \cref{eqn:Py}. Finally (\circlednumber{4}), the stored results can be used for a prediction of the query data via \cref{eqn:pq:pyi:pred}. For a single query data point, this approach requires $\BigO{1}$ classical memory slots instead of $\BigO{2^d}$ when also storing the vanishing elements.}
		\label{fig:on-demand-sampling}
	\end{center}
\end{figure}

The disadvantage of these two prediction methods is that they presume that the whole estimated label probability distribution $\hat{p}(y_i \,|\, \mathcal{C}_{\nu})$, \cref{eqn:pyi:est}, is available beforehand, which requires $\BigO{2^d}$ classical memory slots. As an alternative, we propose an \emph{on-demand sampling} prediction method to reduce this requirement to $\BigO{1}$ classical memory slots. This method consists of four steps, which are sketched in \cref{fig:on-demand-sampling}.\par
First, a probability distribution of query data $p^{q}(\mathbf{x}^q)$, \cref{eqn:pq}, is prescribed, which can in particular also represent a single query data point. Second, the corresponding query circuit, \cref{fig:pred-circuit}, is measured $N$ times to sample the probability distribution of reaching a certain leaf $\hat{p}^{q}(\mathcal{C}_{\nu})$, \cref{eqn:pq:pn:est}. All node indices of non-zero elements of this probability distribution are collected in the set
\begin{align} \label{eqn:Lambdaq}
	\Lambda(p^q) \equiv \left\{ \nu \,|\, \nu \in \{1,\dots,2^d\} \land \hat{p}^{q}(\mathcal{C}_{\nu} \,|\, p^q) > 0 \right\},
\end{align}
where $\nu$ iterates over all leaves. For a single query data point, one has $\vert \Lambda(p^q) \vert=1$ such that the set of corresponding non-zero elements
\begin{align} \label{eqn:Pq}
	P^q(p^q) \equiv \left\{ \hat{p}^{q}(\mathcal{C}_{\nu}) \,|\, \hat{p}^{q}(\mathcal{C}_{\nu}) \,\forall\, \nu \in \Lambda(p^q) \right\}
\end{align}
requires $\BigO{1}$ classical memory slots (instead of $\BigO{2^d}$ when also storing the vanishing elements). Third, the Q-tree of interest, \cref{fig:tree-circuit}, is measured $N$ times to sample the predicted probability distribution for the $i$th label $\hat{p}(y_i \,|\, \mathcal{C}_{\nu})$, \cref{eqn:pyi:pred:cl}, in such a way that only leaves present in $\Lambda(p^q)$ are considered. That is, for a single query data point, only $\BigO{1}$ classical memory slots are required to store the resulting set of entries
\begin{align} \label{eqn:Py}
	P^y(p^q) \equiv \left\{ \hat{p}(y_i \,|\, \mathcal{C}_{\nu}) \,|\, \hat{p}(y_i \,|\, \mathcal{C}_{\nu}) \,\forall\, i \in \{1,\dots,m\} \,\forall\, \nu \in \Lambda(p^q) \right\}
\end{align}
(instead of $\BigO{2^d}$ when considering all leaves). Finally, \cref{eqn:pq:pyi:pred} can be used (omitting the vanishing summands) with the results from \cref{eqn:Pq,eqn:Py} to perform the prediction of the prescribed distribution of query data. In total, $\BigO{1}$ classical memory slots are required for a single query data point. Therefore, an exponential data compression can be achieved for the prediction similar to the exponential data compression for the data encoding from \cref{sec:quantum representation:data encoding}. The disadvantage of this approach is, of course, that the sampling must be performed anew for each query data based on the evaluation of the query circuit. Thus, the reduction of classical memory is traded off for additional quantum computations.\par
For a sufficient number of measurements, the on-demand sampling leads to the same predictions as the presampled approach. On noisy quantum hardware, however, the predictions might deviate. For example, one might collect measurements from the query circuit such that $\vert \Lambda(p^q) \vert>1$ despite considering only a single query data point. In this case, choosing the largest probability 
\begin{align}
	\tilde{\Lambda}(p^q) \equiv \left\{ \argmax_{\nu \in \{1,\dots,2^d\}} \hat{p}^{q}(\mathcal{C}_{\nu} \,|\, p^q) \right\} \approx \Lambda(p^q)
\end{align}
could serve as a suitable method for noise mitigation. A more detailed discussion of this topic goes beyond the scope of this paper, but can be considered as a possible research direction.
\section{Experiments} \label{sec:experiments}
In the present section, we test the proposed Q-tree approach for quantum representations of decision trees in practice. As our study example, we consider two data sets. First, the \emph{tic-tac-toe endgame} data set \cite{aha2021} from the UCI machine learning repository \cite{dua2017}, which we use for experiments on a quantum computing simulator (on classical hardware), and second, a simple toy data set, which we use to run experiments on actual quantum hardware. In the following, we start by describing the data. Subsequently, we present the experimental results on the simulator and the hardware, respectively.

\subsection{Data} \label{sec:experiments:data}
We consider two data sets for our experiments. The first data set is the tic-tac-toe endgame data set \cite{aha2021}, which contains the complete set of possible board configurations at the end of tic-tac-toe games. Each of the $\num{958}$ instances corresponds to one legal tic-tac-toe endgame board. A board is encoded by a nine-dimensional vector of ternary attributes $\mathbf{x'} \in \{0,1,2\}^9$, which determine whether each of the nine board fields is taken by one of the two players or left blank ($0$ represents starting player, $1$ represents the other player and $2$ represents a blank field) with the allocation shown in \cref{fig:experiments:board:x}. We convert $\mathbf{x'}$ into a $\num{15}$-dimensional feature vector $\mathbf{x} \in \mathbb{B}^{15}$ (\ie, $k=15$) according to
\begin{align} \label{eqn:experiments:sim:xconversion}
	\sum_{i=0}^{9} x'_{9-i} 3^i = \sum_{i=0}^{15} x_{15-i} 2^i.
\end{align}
In addition, a binary label $\mathbf{y} \in \mathbb{B}^{1}$ (\ie, $m=1$) is assigned to each board instance to decide the player who won the game ($y_1=1$ represents a victory of the starting player and $y_1=0$ a victory of the other player), where a victory is achieved by a ``three-in-a-row.''\par
\begin{figure}
	\begin{center}
		\begin{tikzpicture} 
\tttboard%
{$x'_1$}{$x'_2$}{$x'_3$}%
{$x'_4$}{$x'_5$}{$x'_6$}%
{$x'_7$}{$x'_8$}{$x'_9$}%
{(0,0)}{}{}
\end{tikzpicture}
		\caption{Allocation of the nine ternary attributes $x'_1,\dots,x'_9 \in \{0,1,2\}^9$ to the nine tic-tac-toe board game fields. The attributes determine whether a field $i \in \{1,\dots,9\}$ is taken by one of the two players ($x'_i=0$ for the starting player or $x'_i=1$ for the other player) or left blank ($x'_i=2$).}
		\label{fig:experiments:board:x}
	\end{center}
\end{figure}
The second data set we consider is a toy data set, which consists of five seven-dimensional binary feature vectors (\ie, $k=7$) and five corresponding binary labels (\ie, $m=1$). In total, the data set reads
\begin{align} \label{eqn:experiments:XY}
	\mathbf{X} \equiv
	\begin{pmatrix}
		1&1&0&0&0&0&0\\
		0&1&1&0&0&0&0\\
		0&0&1&0&0&0&0\\
		0&0&0&0&0&0&0\\
		1&1&1&0&0&0&0\\
	\end{pmatrix}
	\,\,\text{and}\,\,
	\mathbf{Y} \equiv
	\begin{pmatrix}
		0\\0\\1\\0\\1
	\end{pmatrix},
\end{align}
where each row represents a data point and each column a feature vector and label, respectively. By definition, $y_1=1$ if $\mathbf{x}$ contains an odd number of ones and $y_1=0$ otherwise (\ie, the label represents an \emph{exclusive or} operation over all features). In particular, the label depends only on the first three features, whereas the last four features carry no information.

\subsection{Simulator} \label{sec:experiments:simulator}
We use the quantum computing simulator \emph{Qiskit} (Quantum Information Software Kit) \cite{qiskit2019} to evaluate quantum circuits on classical hardware. In particular, we empirically test Q-trees using the data-based decision problem to predict the winner of a tic-tac-toe match $\mathbf{y} \in \mathbb{B}^1$ based on the board configuration $\mathbf{x} \in \mathbb{B}^{15}$ as described in \cref{sec:experiments:data}. For this purpose, the tic-tac-toe endgame data set is randomly split into a training data set $\mathbf{D}_{\mathrm{train}}$, \cref{eqn:D:training}, and a test data set $\mathbf{D}_{\mathrm{test}}$ of equal size (\ie, $\num{479}$ instances each) such that both splitted data sets contain the same proportion of labels. We use the training data set to induce decision trees and the test data set to verify the respective tree performances as explained in the following in more detail.\par
To begin with, we induce a decision tree of maximum depth $d=\num{3}$ as described in \cref{sec:quantum representation:tree induction} using the genetic algorithm from \cref{sec:app:genetic induction algorithm} with the hyperparameters $\theta=(\num{16},\num{20},\num{3},\num{.3},\num{.5},\num{.15},\num{1},\num{1})$, \cref{eqn:app:ga:theta}, and $N=\num{e6}$ measurements. We repeat the induction $\num{25}$ times with different random seeds. As a result, we obtain $\num{25}$ Q-trees $QT_{\mathrm{sim}}^1,\dots,QT_{\mathrm{sim}}^{25}$ (each parameterized by a corresponding compressed decision configuration, \cref{eqn:C}). The mean distribution of splitting feature indices in each depth is shown in \cref{fig:experiments:sim:features:q}. We find that the splitting index $15$ occurs most often for the root, whereas the most often occurring splitting indices over all depths are $1$ and $15$.\par
A final sampling with $N=\num{e6}$ measurements is performed for each Q-tree to obtain an estimate for the label probability distribution $\hat{p}(y_i \,|\,\mathcal{C}_{\nu})$, \cref{eqn:pyi:est}, for all leaves $\nu$, \cref{eqn:nu}, with the corresponding set of conditions $\mathcal{C}_{\nu}$, \cref{eqn:Cnu}. Based on this estimate, predictions of query data can be performed using \cref{eqn:pyi:pred:cl}. In case that we find zero samples for a particular leaf (\ie, $n(\mathcal{C}_{\nu}) = 0$), the probability distribution $p(y_1 \,|\, \mathcal{C}_{\nu})$, \cref{eqn:pyi:est}, is undefined. To be able to still make predictions, we assign the agnostic default value $p(y_1 \,|\, \mathcal{C}_{\nu}) = \frac{1}{2}$. Furthermore, to resolve the resulting ambiguity of $\hat{y}_1$, \cref{eqn:pyi:pred:cl:argmax}, we instead predict the majority label present in the training data set.\par
\begin{figure}
	\centering
	\begin{subfigure}[t]{.49\linewidth}
		\centering
		\begin{tikzpicture} 
[
level 1/.style          = {sibling distance=8.8cm},
level 2/.style			= {sibling distance=4.4cm},
level 3/.style			= {sibling distance=2.2cm},
level 4/.style			= {sibling distance=1.1cm},
grow                    = right,
level distance          = 1.00cm,
edge from parent/.style = {draw,thick},
every node/.style       = {},
]
\node[treenode] (root) {} % root
child { node[treenode] (n0) {} % 0
	child { node[treenode] (n00)  {} % 00
		child { node[treenode] (n000) {}  % 000
			child { node[leafnode] (n0000) {} } % 0000
			child { node[leafnode] (n0001) {} } % 0001		
		}
		child { node[treenode] (n001) {} % 001 
			child { node[leafnode] (n0010) {} } % 0010
			child { node[leafnode] (n0011) {} } % 0011		
		}
	}
	child { node[treenode] (n01) {} % 01
		child { node[leafnode] (n010) {} % 010
			child { node[leafnode] (n0100) {} } % 0100
			child { node[leafnode] (n0101) {} } % 0101	
		}
		child { node[treenode] (n011) {} % 011
			child { node[leafnode] (n0110) {} } % 0110
			child { node[leafnode] (n0111) {} } % 0111	
		}
	}
}
child { node[treenode] (n1) {} % 1
	child { node[treenode] (n10) {} % 10
		child { node[treenode] (n100) {} % 100
			child { node[leafnode] (n1000) {} } % 1000
			child { node[leafnode] (n1001) {} } % 1001	
		}
		child { node[treenode] (n101) {} % 101
			child { node[leafnode] (n1010) {} } % 1010
			child { node[leafnode] (n1011) {} } % 1011	
		}
	}
	child { node[treenode] (n11)  {} % 11
		child { node[leafnode] (n110) {} % 110
			child { node[leafnode] (n1100) {} } % 1100
			child { node[leafnode] (n1101) {} } % 1101	
		}
		child { node[treenode] (n111) {} % 111
			child { node[leafnode] (n1110) {} } % 1110
			child { node[leafnode] (n1111) {} } % 1111	
		}
	}
}
;
\node[edgenode,anchor=center] at ($.5*(root)+.5*(n0)$) {$x_{15} = 0$}; % r-0
\node[edgenode,anchor=center] at ($.5*(root)+.5*(n1)$) {$x_{15} = 1$}; % r-1
\node[edgenode,anchor=center] at ($.5*(n0)+.5*(n00)$) {$x_{10} = 0$}; % 0-00
\node[edgenode,anchor=center] at ($.5*(n0)+.5*(n01)$) {$x_{10} = 1$}; % 0-01
\node[edgenode,anchor=center] at ($.5*(n1)+.5*(n10)$) {$x_{2} = 0$}; % 1-10
\node[edgenode,anchor=center] at ($.5*(n1)+.5*(n11)$) {$x_{2} = 1$}; % 1-11		
\node[edgenode,anchor=center] at ($.5*(n00)+.5*(n000)$) {$x_{6} = 0$}; % 00-000	
\node[edgenode,anchor=center] at ($.5*(n00)+.5*(n001)$) {$x_{6} = 1$}; % 00-001
\node[edgenode,anchor=center] at ($.5*(n01)+.5*(n010)$) {$x_{14} = 0$}; % 01-010
\node[edgenode,anchor=center] at ($.5*(n01)+.5*(n011)$) {$x_{14} = 1$}; % 01-011
\node[edgenode,anchor=center] at ($.5*(n10)+.5*(n100)$) {$x_{1} = 0$}; % 10-100
\node[edgenode,anchor=center] at ($.5*(n10)+.5*(n101)$) {$x_{1} = 1$}; % 10-101
\node[edgenode,anchor=center] at ($.5*(n11)+.5*(n110)$) {$x_{8} = 0$}; % 11-110
\node[edgenode,anchor=center] at ($.5*(n11)+.5*(n111)$) {$x_{8} = 1$}; % 11-111		
\node[edgenode,anchor=center] at ($.5*(n000)+.5*(n0000)$) {$x_{12} = 0$}; % 000-0000	
\node[edgenode,anchor=center] at ($.5*(n000)+.5*(n0001)$) {$x_{12} = 1$}; % 000-0001
\node[edgenode,anchor=center] at ($.5*(n001)+.5*(n0010)$) {$x_{1} = 0$}; % 001-0010
\node[edgenode,anchor=center] at ($.5*(n001)+.5*(n0011)$) {$x_{1} = 1$}; % 001-0011
\node[edgenode,anchor=center] at ($.5*(n010)+.5*(n0100)$) {$x_{1} = 0$}; % 010-0100
\node[edgenode,anchor=center] at ($.5*(n010)+.5*(n0101)$) {$x_{1} = 1$}; % 010-0101
\node[edgenode,anchor=center] at ($.5*(n011)+.5*(n0110)$) {$x_{9} = 0$}; % 011-0110
\node[edgenode,anchor=center] at ($.5*(n011)+.5*(n0111)$) {$x_{9} = 1$}; % 011-0111	
\node[edgenode,anchor=center] at ($.5*(n100)+.5*(n1000)$) {$x_{9} = 0$}; % 100-1000	
\node[edgenode,anchor=center] at ($.5*(n100)+.5*(n1001)$) {$x_{9} = 1$}; % 100-1001
\node[edgenode,anchor=center] at ($.5*(n101)+.5*(n1010)$) {$x_{6} = 0$}; % 101-1010
\node[edgenode,anchor=center] at ($.5*(n101)+.5*(n1011)$) {$x_{6} = 1$}; % 101-1011
\node[edgenode,anchor=center] at ($.5*(n110)+.5*(n1100)$) {$x_{9} = 0$}; % 110-1100
\node[edgenode,anchor=center] at ($.5*(n110)+.5*(n1101)$) {$x_{9} = 1$}; % 110-1101
\node[edgenode,anchor=center] at ($.5*(n111)+.5*(n1110)$) {$x_{3} = 0$}; % 111-1110
\node[edgenode,anchor=center] at ($.5*(n111)+.5*(n1111)$) {$x_{3} = 1$}; % 111-1111			
\node[leafnodelabel] at (n0000) {$\treepval{0}{0.262}$ \\ $\Treepval{1}{0.738}$}; % 0000
\node[leafnodelabel] at (n0001) {$\Treepval{0}{0.518}$ \\ $\treepval{1}{0.482}$}; % 0001
\node[leafnodelabel] at (n0010) {$\Treepval{0}{0.611}$ \\ $\treepval{1}{0.389}$}; % 0010
\node[leafnodelabel] at (n0011) {$\treepval{0}{0.000}$ \\ $\Treepval{1}{1.000}$}; % 0011
\node[leafnodelabel] at (n0100) {$\Treepval{0}{0.605}$ \\ $\treepval{1}{0.395}$}; % 0100
\node[leafnodelabel] at (n0101) {$\treepval{0}{0.000}$ \\ $\Treepval{1}{1.000}$}; % 0101
\node[leafnodelabel] at (n0110) {$\treepval{0}{0.495}$ \\ $\Treepval{1}{0.505}$}; % 0110
\node[leafnodelabel] at (n0111) {$\treepval{0}{0.319}$ \\ $\Treepval{1}{0.681}$}; % 0111
\node[leafnodelabel] at (n1000) {$\treepval{0}{0.129}$ \\ $\Treepval{1}{0.871}$}; % 1000
\node[leafnodelabel] at (n1001) {$\treepval{0}{0.175}$ \\ $\Treepval{1}{0.825}$}; % 1001
\node[leafnodelabel] at (n1010) {$\treepval{0}{0.200}$ \\ $\Treepval{1}{0.800}$}; % 1010
\node[leafnodelabel] at (n1011) {$\Treepval{0}{0.543}$ \\ $\treepval{1}{0.457}$}; % 1011
\node[leafnodelabel] at (n1100) {$\treepval{0}{0.301}$ \\ $\Treepval{1}{0.699}$}; % 1100
\node[leafnodelabel] at (n1101) {$\treepval{0}{0.418}$ \\ $\Treepval{1}{0.582}$}; % 1101
\node[leafnodelabel] at (n1110) {$\Treepval{0}{0.533}$ \\ $\treepval{1}{0.467}$}; % 1110
\node[leafnodelabel] at (n1111) {$\treepval{0}{0.104}$ \\ $\Treepval{1}{0.896}$}; % 1111
\end{tikzpicture}
		\caption{Best Q-tree $QT_{\mathrm{sim}}$ with the compressed decision configuration $C_{\mathrm{sim}}$, \cref{eqn:experiments:sim:C}, induced via a genetic algorithm. Here, $\hat{p}_b \equiv \hat{p}(y_1=b \,|\, \mathcal{C}_{\nu})$ denote the estimated label probabilities for $N=\num{e6}$ measurements, \cref{eqn:pyi:est}, which are also shown in \cref{fig:experiments:sim:probas:km}.}
		\label{fig:experiments:sim:trees:q}
	\end{subfigure}%
	\hfill%
	\begin{subfigure}[t]{.49\linewidth}
		\centering
		\begin{tikzpicture} 
[
level 1/.style          = {sibling distance=8.8cm},
level 2/.style			= {sibling distance=4.4cm},
level 3/.style			= {sibling distance=2.2cm},
level 4/.style			= {sibling distance=1.1cm},
grow                    = right,
level distance          = 1.00cm,
edge from parent/.style = {draw,thick},
every node/.style       = {},
]
\node[treenode] (root) {} % root
child { node[treenode] (n0) {} % 0
	child { node[treenode] (n00)  {} % 00
		child { node[treenode] (n000) {}  % 000
			child { node[leafnode] (n0000) {} } % 0000
			child { node[leafnode] (n0001) {} } % 0001		
		}
		child { node[treenode] (n001) {} % 001 
			child { node[leafnode] (n0010) {} } % 0010
			child { node[leafnode] (n0011) {} } % 0011		
		}
	}
	child { node[leafnode] (n01) {} % 01
		%child { node[leafnode] (n010) {} % 010
		%	child { node[leafnode] (n0100) {} } % 0100
		%	child { node[leafnode] (n0101) {} } % 0101	
		%}
		%child { node[treenode] (n011) {} % 011
		%	child { node[leafnode] (n0110) {} } % 0110
		%	child { node[leafnode] (n0111) {} } % 0111	
		%}
	}
}
child { node[treenode] (n1) {} % 1
	child { node[treenode] (n10) {} % 10
		child { node[treenode] (n100) {} % 100
			child { node[leafnode] (n1000) {} } % 1000
			child { node[leafnode] (n1001) {} } % 1001	
		}
		child { node[treenode] (n101) {} % 101
			child { node[leafnode] (n1010) {} } % 1010
			child { node[leafnode] (n1011) {} } % 1011	
		}
	}
	child { node[treenode] (n11)  {} % 11
		child { node[leafnode] (n110) {} % 110
			child { node[leafnode] (n1100) {} } % 1100
			child { node[leafnode] (n1101) {} } % 1101	
		}
		child { node[treenode] (n111) {} % 111
			child { node[leafnode] (n1110) {} } % 1110
			child { node[leafnode] (n1111) {} } % 1111	
		}
	}
}
;
\node[edgenode,anchor=center] at ($.5*(root)+.5*(n0)$) {$x_{15} = 0$}; % r-0
\node[edgenode,anchor=center] at ($.5*(root)+.5*(n1)$) {$x_{15} = 1$}; % r-1
\node[edgenode,anchor=center] at ($.5*(n0)+.5*(n00)$) {$x_{1} = 0$}; % 0-00
\node[edgenode,anchor=center] at ($.5*(n0)+.5*(n01)$) {$x_{1} = 1$}; % 0-01
\node[edgenode,anchor=center] at ($.5*(n1)+.5*(n10)$) {$x_{2} = 0$}; % 1-10
\node[edgenode,anchor=center] at ($.5*(n1)+.5*(n11)$) {$x_{2} = 1$}; % 1-11		
\node[edgenode,anchor=center] at ($.5*(n00)+.5*(n000)$) {$x_{5} = 0$}; % 00-000	
\node[edgenode,anchor=center] at ($.5*(n00)+.5*(n001)$) {$x_{5} = 1$}; % 00-001
%\node[edgenode,anchor=center] at ($.5*(n01)+.5*(n010)$) {X}; % 01-010
%\node[edgenode,anchor=center] at ($.5*(n01)+.5*(n011)$) {X}; % 01-011
\node[edgenode,anchor=center] at ($.5*(n10)+.5*(n100)$) {$x_{1} = 0$}; % 10-100
\node[edgenode,anchor=center] at ($.5*(n10)+.5*(n101)$) {$x_{1} = 1$}; % 10-101
\node[edgenode,anchor=center] at ($.5*(n11)+.5*(n110)$) {$x_{4} = 0$}; % 11-110
\node[edgenode,anchor=center] at ($.5*(n11)+.5*(n111)$) {$x_{4} = 1$}; % 11-111		
\node[edgenode,anchor=center] at ($.5*(n000)+.5*(n0000)$) {$x_{4} = 0$}; % 000-0000	
\node[edgenode,anchor=center] at ($.5*(n000)+.5*(n0001)$) {$x_{4} = 1$}; % 000-0001
\node[edgenode,anchor=center] at ($.5*(n001)+.5*(n0010)$) {$x_{9} = 0$}; % 001-0010
\node[edgenode,anchor=center] at ($.5*(n001)+.5*(n0011)$) {$x_{9} = 1$}; % 001-0011
%\node[edgenode,anchor=center] at ($.5*(n010)+.5*(n0100)$) {X}; % 010-0100
%\node[edgenode,anchor=center] at ($.5*(n010)+.5*(n0101)$) {X}; % 010-0101
%\node[edgenode,anchor=center] at ($.5*(n011)+.5*(n0110)$) {X}; % 011-0110
%\node[edgenode,anchor=center] at ($.5*(n011)+.5*(n0111)$) {X}; % 011-0111	
\node[edgenode,anchor=center] at ($.5*(n100)+.5*(n1000)$) {$x_{3} = 0$}; % 100-1000	
\node[edgenode,anchor=center] at ($.5*(n100)+.5*(n1001)$) {$x_{3} = 1$}; % 100-1001
\node[edgenode,anchor=center] at ($.5*(n101)+.5*(n1010)$) {$x_{10} = 0$}; % 101-1010
\node[edgenode,anchor=center] at ($.5*(n101)+.5*(n1011)$) {$x_{10} = 1$}; % 101-1011
\node[edgenode,anchor=center] at ($.5*(n110)+.5*(n1100)$) {$x_{8} = 0$}; % 110-1100
\node[edgenode,anchor=center] at ($.5*(n110)+.5*(n1101)$) {$x_{8} = 1$}; % 110-1101
\node[edgenode,anchor=center] at ($.5*(n111)+.5*(n1110)$) {$x_{8} = 0$}; % 111-1110
\node[edgenode,anchor=center] at ($.5*(n111)+.5*(n1111)$) {$x_{8} = 1$}; % 111-1111			
\node[leafnodelabel] at (n0000) {$\treepval{0}{0.333}$ \\ $\Treepval{1}{0.667}$}; % 0000
\node[leafnodelabel] at (n0001) {$\Treepval{0}{0.566}$ \\ $\treepval{1}{0.434}$}; % 0001
\node[leafnodelabel] at (n0010) {$\Treepval{0}{0.725}$ \\ $\treepval{1}{0.275}$}; % 0010
\node[leafnodelabel] at (n0011) {$\Treepval{0}{0.500}$ \\ $\treepval{1}{0.500}$}; % 0011
\node[leafnodelabel] at (n01) {$\treepval{0}{0.000}$ \\ $\Treepval{1}{1.000}$}; % 01
%\node[leafnodelabel] at (n0100) {$\treepval{0}{0.001}$ \\ $\treepval{1}{0.001}$}; % 0100
%\node[leafnodelabel] at (n0101) {$\treepval{0}{0.001}$ \\ $\treepval{1}{0.001}$}; % 0101
%\node[leafnodelabel] at (n0110) {$\treepval{0}{0.001}$ \\ $\treepval{1}{0.001}$}; % 0110
%\node[leafnodelabel] at (n0111) {$\treepval{0}{0.001}$ \\ $\treepval{1}{0.001}$}; % 0111
\node[leafnodelabel] at (n1000) {$\treepval{0}{0.062}$ \\ $\Treepval{1}{0.938}$}; % 1000
\node[leafnodelabel] at (n1001) {$\treepval{0}{0.283}$ \\ $\Treepval{1}{0.717}$}; % 1001
\node[leafnodelabel] at (n1010) {$\Treepval{0}{0.538}$ \\ $\treepval{1}{0.462}$}; % 1010
\node[leafnodelabel] at (n1011) {$\treepval{0}{0.125}$ \\ $\Treepval{1}{0.875}$}; % 1011
\node[leafnodelabel] at (n1100) {$\treepval{0}{0.300}$ \\ $\Treepval{1}{0.700}$}; % 1100
\node[leafnodelabel] at (n1101) {$\Treepval{0}{0.535}$ \\ $\treepval{1}{0.465}$}; % 1101
\node[leafnodelabel] at (n1110) {$\treepval{0}{0.407}$ \\ $\Treepval{1}{0.593}$}; % 1110
\node[leafnodelabel] at (n1111) {$\treepval{0}{0.105}$ \\ $\Treepval{1}{0.895}$}; % 1111
\end{tikzpicture}
		\caption{Classical decision tree $T_{\mathrm{cl}}$ induced via scikit-learn using a top-down approach with a splitting criterion based on the information entropy. Here, $\hat{p}_b \equiv p(y_1=b \,|\, \mathcal{C}_{\nu})$ denote the exact label probabilities, \cref{eqn:pn:y}.}
		\label{fig:experiments:sim:trees:s}
	\end{subfigure}%
	\caption{Induced decision trees with maximum depth $d=3$ for the tic-tac-toe training data set $\mathbf{D}_{\mathrm{train}}$, where all quantum computations have been performed on a simulator. Starting from the root on the left, we show the decisions $x_i=b$ along all paths between the internal nodes (\treenode) to the leaves on the right (\treeleaf), where $i \in \{1,\dots,15\}$ and $b \in \mathbb{B}$. For each leaf, we also show the probability $\hat{p}_b$ of predicting $y_1=b$. The predicted label is indicated with a star (${}^*$) at the corresponding probability.}
	\label{fig:experiments:sim:trees}
\end{figure}
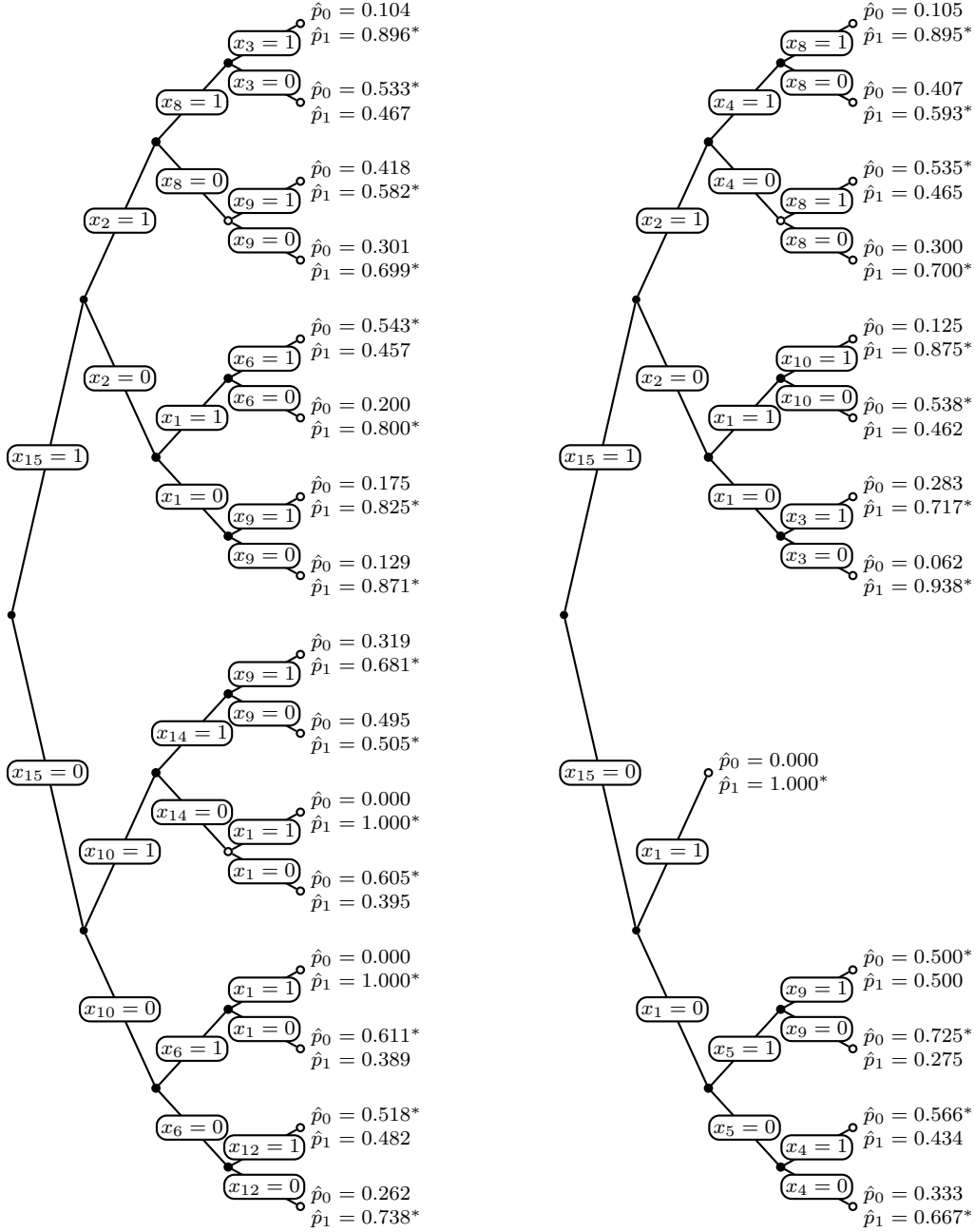

For comparison purposes, we employ \emph{scikit-learn} \cite{scikit-learn2011} to induce a classical decision tree $T_{\mathrm{cl}}$ of the same maximum depth using a top-down approach with information entropy as a splitting criterion. The resulting tree is shown in \cref{fig:experiments:sim:trees:s}. The corresponding distribution of splitting feature indices in each depth is visualized in \cref{fig:experiments:sim:features:s}. Clearly, there is a close resemblance between the distributions in \cref{fig:experiments:sim:features:q} and \cref{fig:experiments:sim:features:s}, but there are also certain differences. For example, the splitting index for the root is $15$ in \cref{fig:experiments:sim:features:s}, which corresponds to the majority of splitting indices in \cref{fig:experiments:sim:features:q}. On the other hand, the most often occurring splitting indices over all depths in \cref{fig:experiments:sim:features:s} are $1$, $4$, and $8$ as opposed to $1$ and $15$ in \cref{fig:experiments:sim:features:q}. Here and in the following, we use the definition imposed by scikit-learn that balanced label probabilities $p(y_1=0 \,|\, \mathcal{C}_{\nu}) = p(y_1=0 \,|\, \mathcal{C}_{\nu}) = \frac{1}{2}$, \cref{eqn:pn:y}, lead to a prediction $y_1=0$ for $T_{\mathrm{cl}}$.\par
\begin{figure}
	\centering
	\begin{subfigure}[t]{\linewidth}
		\centering
		\begin{tikzpicture} 
\pgfplotstableread[col sep=semicolon]{%experiment_simulator-features-qt.csv
k;d;c
0;0.000000000;0.200000000
1;0.000000000;0.080000000
2;0.000000000;0.040000000
3;0.000000000;0.000000000
4;0.000000000;0.000000000
5;0.000000000;0.040000000
6;0.000000000;0.000000000
7;0.000000000;0.000000000
8;0.000000000;0.000000000
9;0.000000000;0.000000000
10;0.000000000;0.000000000
11;0.000000000;0.000000000
12;0.000000000;0.000000000
13;0.000000000;0.000000000
14;0.000000000;0.640000000
0;1.000000000;0.200000000
1;1.000000000;0.120000000
2;1.000000000;0.160000000
3;1.000000000;0.240000000
4;1.000000000;0.120000000
5;1.000000000;0.120000000
6;1.000000000;0.120000000
7;1.000000000;0.160000000
8;1.000000000;0.200000000
9;1.000000000;0.120000000
10;1.000000000;0.040000000
11;1.000000000;0.040000000
12;1.000000000;0.000000000
13;1.000000000;0.000000000
14;1.000000000;0.360000000
0;2.000000000;0.520000000
1;2.000000000;0.440000000
2;2.000000000;0.520000000
3;2.000000000;0.160000000
4;2.000000000;0.280000000
5;2.000000000;0.360000000
6;2.000000000;0.280000000
7;2.000000000;0.200000000
8;2.000000000;0.240000000
9;2.000000000;0.200000000
10;2.000000000;0.120000000
11;2.000000000;0.200000000
12;2.000000000;0.120000000
13;2.000000000;0.200000000
14;2.000000000;0.160000000
0;3.000000000;0.520000000
1;3.000000000;0.720000000
2;3.000000000;0.640000000
3;3.000000000;0.680000000
4;3.000000000;0.680000000
5;3.000000000;0.720000000
6;3.000000000;0.400000000
7;3.000000000;0.760000000
8;3.000000000;0.600000000
9;3.000000000;0.800000000
10;3.000000000;0.160000000
11;3.000000000;0.400000000
12;3.000000000;0.520000000
13;3.000000000;0.120000000
14;3.000000000;0.280000000
0;4.000000000;1.440000000
1;4.000000000;1.360000000
2;4.000000000;1.360000000
3;4.000000000;1.080000000
4;4.000000000;1.080000000
5;4.000000000;1.240000000
6;4.000000000;0.800000000
7;4.000000000;1.120000000
8;4.000000000;1.040000000
9;4.000000000;1.120000000
10;4.000000000;0.320000000
11;4.000000000;0.640000000
12;4.000000000;0.640000000
13;4.000000000;0.320000000
14;4.000000000;1.440000000
}\data;
\pgfplotsset{colormap/blackwhite}
\begin{axis} [width=10.25cm,
	scale mode=scale uniformly,
	enlargelimits=false,
	xmin=-.5,xmax=14.5,
	ymin=-.5,ymax=4.5,	
	xtick={0,1,...,14},
	xticklabels={1,2,...,15},
	minor xtick={.5,1.5,...,13.5},
	minor ytick={.5,1.5,...,3.5},
	grid=minor,
	minor grid style={black},
	ytick={0,1,2,3,4},	
	yticklabels={0,1,2,3,all},
	xlabel={splitting index $i$},
	ylabel={depth $d$},	
	tick style={draw=none},
	point meta=explicit,
	nodes near coords,
	nodes near coords style={font=\scriptsize\bfseries,/pgf/number format/fixed,/pgf/number format/fixed zerofill,/pgf/number format/precision=2},
	coordinate style/.condition={meta<0.01}{white},
	coordinate style/.condition={meta>=0.01}{Qv},	
	nodes near coords align={anchor=center},
	colormap={reverse blackwhite}{indices of colormap={\pgfplotscolormaplastindexof{blackwhite},...,0 of blackwhite}},
	point meta min=0, point meta max=4.32, % range * 3
	]
	\addplot [matrix plot, mesh/rows=5, mesh/cols=15] table [x=k,y=d,meta=c] {\data};
\end{axis} 
\end{tikzpicture}
		\caption{Mean splitting feature index distribution for the Q-trees $QT_{\mathrm{sim}}^1,\dots,QT_{\mathrm{sim}}^{25}$.}
		\label{fig:experiments:sim:features:q}
	\end{subfigure}%
	\\[.5cm]%
	\begin{subfigure}[t]{\linewidth}
		\centering
		\begin{tikzpicture} 
\pgfplotstableread[col sep=semicolon]{%experiment_simulator-features-sl.csv
k;d;c
0;0;0
1;0;0
2;0;0
3;0;0
4;0;0
5;0;0
6;0;0
7;0;0
8;0;0
9;0;0
10;0;0
11;0;0
12;0;0
13;0;0
14;0;1
0;1;1
1;1;1
2;1;0
3;1;0
4;1;0
5;1;0
6;1;0
7;1;0
8;1;0
9;1;0
10;1;0
11;1;0
12;1;0
13;1;0
14;1;0
0;2;1
1;2;0
2;2;0
3;2;1
4;2;1
5;2;0
6;2;0
7;2;0
8;2;0
9;2;0
10;2;0
11;2;0
12;2;0
13;2;0
14;2;0
0;3;0
1;3;0
2;3;1
3;3;1
4;3;0
5;3;0
6;3;0
7;3;2
8;3;1
9;3;1
10;3;0
11;3;0
12;3;0
13;3;0
14;3;0
0;4;2
1;4;1
2;4;1
3;4;2
4;4;1
5;4;0
6;4;0
7;4;2
8;4;1
9;4;1
10;4;0
11;4;0
12;4;0
13;4;0
14;4;1
}\data;
\pgfplotsset{colormap/blackwhite}
\begin{axis} [width=10.25cm,
	scale mode=scale uniformly,
	enlargelimits=false,
	xmin=-.5,xmax=14.5,
	ymin=-.5,ymax=4.5,	
	xtick={0,1,...,14},
	xticklabels={1,2,...,15},
	minor xtick={.5,1.5,...,13.5},
	minor ytick={.5,1.5,...,3.5},
	grid=minor,
	minor grid style={black},
	ytick={0,1,2,3,4},	
	yticklabels={0,1,2,3,all},
	xlabel={splitting index $i$},
	ylabel={depth $d$},	
	tick style={draw=none},
	point meta=explicit,
	nodes near coords,
	nodes near coords style={font=\scriptsize\bfseries},
	coordinate style/.condition={meta<1}{white},
	coordinate style/.condition={meta>=1}{Qv},	
	nodes near coords align={anchor=center},
	colormap={reverse blackwhite}{indices of colormap={\pgfplotscolormaplastindexof{blackwhite},...,0 of blackwhite}},
	point meta min=0, point meta max=8, % range * 3
	]
	\addplot [matrix plot, mesh/rows=5, mesh/cols=15] table [x=k,y=d,meta=c] {\data};
\end{axis} 
\end{tikzpicture}
		\caption{Splitting feature index distribution for the classical decision tree $T_{\mathrm{cl}}$, which is also shown in \cref{fig:experiments:sim:trees:s}.}
		\label{fig:experiments:sim:features:s}
	\end{subfigure}%
	\caption{Distribution of splitting feature indices for considered decision trees. Each column represents a feature index $i \in \{1,\dots,15\}$ and each row a depth $d \in \{0,1,2,3\}$. The values in each cell denote the number of splitting decisions with feature index $i$ in depth $d$, where we omit zero values. The last row represents the number of splitting decisions cumulated over all depths. Darker backgrounds of the cells indicate higher numbers.}
	\label{fig:experiments:sim:features}
\end{figure}
As a primary performance metric for the decision trees, we consider the balanced accuracy
\begin{align} \label{eqn:experiments:metric:bacc}
	\mathrm{bac} \equiv \frac{1}{2} \left( \frac{\mathrm{tp}_0}{\mathrm{tp}_0 + \mathrm{fn}_0} + \frac{\mathrm{tn}_0}{\mathrm{tn}_0 + \mathrm{fp}_0} \right) = \frac{1}{2} \left( \frac{\mathrm{tp}_1}{\mathrm{tp}_1 + \mathrm{fn}_1} + \frac{\mathrm{tn}_1}{\mathrm{tn}_1 + \mathrm{fp}_1} \right),
\end{align}
which is based on the number of true positives
\begin{align} \label{eqn:experiments:metric:tp}
	\mathrm{tp}_b \equiv \vert\{ \mathbf{d} \,|\, (\mathbf{x}^q,(y_1)) = \mathbf{d} \in \mathbf{D} \land y_1 = b \land \hat{y}_1(\mathbf{x}^q) = b \}\vert,
\end{align}
the number of true negatives
\begin{align} \label{eqn:experiments:metric:tn}
	\mathrm{tn}_b \equiv \vert\{ \mathbf{d} \,|\, (\mathbf{x}^q,(y_1)) = \mathbf{d} \in \mathbf{D} \land y_1 = 1-b \land \hat{y}_1(\mathbf{x}^q) = 1-b \}\vert,
\end{align}
the number of false positives
\begin{align} \label{eqn:experiments:metric:fp}
	\mathrm{fp}_b \equiv \vert\{ \mathbf{d} \,|\, (\mathbf{x}^q,(y_1)) = \mathbf{d} \in \mathbf{D} \land y_1 = 1-b \land \hat{y}_1(\mathbf{x}^q) = b \}\vert,
\end{align}
and the number of false negatives
\begin{align} \label{eqn:experiments:metric:fn}
	\mathrm{fn}_b \equiv \vert\{ \mathbf{d} \,|\, (\mathbf{x}^q,(y_1)) = \mathbf{d} \in \mathbf{D} \land y_1 = b \land \hat{y}_1(\mathbf{x}^q) = 1-b \}\vert,
\end{align}
respectively, where $b \in \mathbb{B}$ denotes the labels treated as positives and $\mathbf{D}$ the data set of interest. Here, $\hat{y}_1(\mathbf{x}^q) \in \mathbb{B}$ represents a prediction of the label query data $\mathbf{x}^q \in \mathbb{B}^{15}$. For the Q-tree we use \cref{eqn:pyi:pred:cl:argmax} to perform predictions.\par
To obtain the ``best'' decision tree out of our induced sequence of Q-trees, we choose the tree with the highest balanced accuracy with respect to the training data set $\mathbf{D}_{\mathrm{train}}$. The resulting tree $QT_{\mathrm{sim}}$ is associated with the decision configuration
\begin{align} \label{eqn:experiments:sim:C}
	C_{\mathrm{sim}} = ((15), (10,2), (6,14,15,8), (12,15,15,9,9,6,9,4))
\end{align}
and shown \cref{fig:experiments:sim:trees:q}. Four exemplary predictions for $QT_{\mathrm{sim}}$ based on \cref{eqn:pyi:pred:cl} are presented in \cref{fig:experiments:board:sim}. Two correct and two incorrect predictions are used for this purpose. Since the predictions are performed on the basis of previously sampled probability distributions, no additional quantum computations are necessary. In addition, we also demonstrate two uncertain query predictions using $\hat{p}_{\mathrm{qm}}(y_i \,|\, p^q)$, \cref{eqn:pq:pyi:pred}, in \cref{fig:experiments:board:sim:q}. They require additional quantum computations for the estimation of $\hat{p}^{q}(\mathcal{C}_{\nu} \,|\, p^q)$, \cref{eqn:pq:pn:est}. The predicted probability distribution from the sampling, \cref{eqn:pq:pyi:pred}, coincides with the true probability distribution calculated from the data proportions, \cref{eqn:pq:pyi:pred:approx}. However, since the two uncertain queries represent a mixture of boards with opposite labels, we cannot assess whether or not the resulting predictions are correct in the sense of a correct classification.\par
\begin{figure}
	\centering
	\begin{subfigure}[t]{.24\linewidth}
		\centering
		\begin{tikzpicture}%experiment_simulator.ipynb
\tttboard%
{\tttX}{\tttO}{\tttO}%
{\tttX}{\tttO}{\tttX}%
{}{\tttO}{\tttX}%
{(0,0)}{}{}
\node[anchor=south,font=\small] (y) at ($(boardcenter)+(0,.75)$)  {$y_1 = 0$};
\node[anchor=north,font=\small] (phat) at ($(boardcenter)+(0,-.75)$)  {$\hat{p}(y_1=0) = \num{0.611}$};
%\node[anchor=north,font=\small,align=center] (cnu) at ($(phat.south)$)  {$\bar{x}_{0}=0$, $\bar{x}_{1}=0$ \\ $\bar{x}_{2}=1$, $\bar{x}_{3}=0$};
\node[anchor=north,font=\small,align=center] (cnu) at ($(phat.south)$)  {$\bar{x}_{0}\bar{x}_{1}\bar{x}_{2}\bar{x}_{3}= 0 0 1 0$};
%---------------- idx = 5 ----------------
%x:  [0 0 0 1 0 1 1 1 1 0 0 1 0 1 0]
%x3: [0 1 1 0 1 0 2 1 0]
%y:  False
%y^: False
%p^: [0.61107866 0.38892134]
%_____
%|XOO|
%|XOX|
%| OX|
%`````
%prediction details:
%predict_proba: depth=0, split_index=14
%predict_proba: x[14]==0 split
%predict_proba: depth=1, split_index=9
%predict_proba: x[9]==0 split
%predict_proba: depth=2, split_index=5
%predict_proba: x[5]==1 split
%predict_proba: depth=3, split_index=0
%predict_proba: x[0]==0 split
%predict_proba: depth=4, split_index=None
%predict_proba: prediction=[[0.6110786558863393], [0.3889213441136608]]
\end{tikzpicture}
		\caption{correct}
		\label{fig:experiments:board:sim:3}
	\end{subfigure}%
	\hfill%
	\begin{subfigure}[t]{.24\linewidth}
		\centering
		\begin{tikzpicture}%experiment_simulator.ipynb
\tttboard%
{}{\tttO}{\tttX}%
{}{\tttO}{\tttX}%
{\tttX}{\tttO}{}%
{(0,0)}{}{}
\node[anchor=south,font=\small] (y) at ($(boardcenter)+(0,.75)$)  {$y_1 = 0$};
\node[anchor=north,font=\small] (phat) at ($(boardcenter)+(0,-.75)$)  {$\hat{p}(y_1=0) = \num{0.301}$};
%\node[anchor=north,font=\small,align=center] (cnu) at ($(phat.south)$)  {$\bar{x}_{0}=0$, $\bar{x}_{1}=0$ \\ $\bar{x}_{2}=0$, $\bar{x}_{3}=0$};
\node[anchor=north,font=\small,align=center] (cnu) at ($(phat.south)$)  {$\bar{x}_{0}\bar{x}_{1}\bar{x}_{2}\bar{x}_{3}= 1 1 0 0$};
%---------------- idx = 1 ----------------
%x:  [0 1 1 1 1 1 0 0 0 0 0 1 0 0 1]
%x3: [2 1 0 2 1 0 0 1 2]
%y:  False
%y^: True
%p^: [0.30070203 0.69929797]
%_____
%| OX|
%| OX|
%|XO |
%`````
%prediction details:
%predict_proba: depth=0, split_index=14
%predict_proba: x[14]==1 split
%predict_proba: depth=1, split_index=1
%predict_proba: x[1]==1 split
%predict_proba: depth=2, split_index=7
%predict_proba: x[7]==0 split
%predict_proba: depth=3, split_index=8
%predict_proba: x[8]==0 split
%predict_proba: depth=4, split_index=None
%predict_proba: prediction=[[0.30070203372656873], [0.6992979662734313]]
\end{tikzpicture}
		\caption{incorrect}
		\label{fig:experiments:board:sim:2}
	\end{subfigure}%
	\hfill%
	\begin{subfigure}[t]{.24\linewidth}
		\centering
		\begin{tikzpicture}%experiment_simulator.ipynb
\tttboard%
{\tttO}{}{\tttX}%
{\tttX}{\tttX}{\tttX}%
{}{\tttO}{\tttO}%
{(0,0)}{}{}
\node[anchor=south,font=\small] (y) at ($(boardcenter)+(0,.75)$)  {$y_1 = 1$};
\node[anchor=north,font=\small] (phat) at ($(boardcenter)+(0,-.75)$)  {$\hat{p}(y_1=1) = \num{0.896}$};
%\node[anchor=north,font=\small,align=center] (cnu) at ($(phat.south)$)  {$\bar{x}_{0}=1$, $\bar{x}_{1}=0$ \\ $\bar{x}_{2}=1$, $\bar{x}_{3}=1$};
\node[anchor=north,font=\small,align=center] (cnu) at ($(phat.south)$)  {$\bar{x}_{0}\bar{x}_{1}\bar{x}_{2}\bar{x}_{3}= 1 1 1 1$};
%---------------- idx = 3 ----------------
%x:  [0 1 0 1 0 1 0 1 1 0 0 1 1 0 1]
%x3: [1 2 0 0 0 0 2 1 1]
%y:  True
%y^: True
%p^: [0.10446011 0.89553989]
%_____
%|O X|
%|XXX|
%| OO|
%`````
%prediction details:
%predict_proba: depth=0, split_index=14
%predict_proba: x[14]==1 split
%predict_proba: depth=1, split_index=1
%predict_proba: x[1]==1 split
%predict_proba: depth=2, split_index=7
%predict_proba: x[7]==1 split
%predict_proba: depth=3, split_index=3
%predict_proba: x[3]==1 split
%predict_proba: depth=4, split_index=None
%predict_proba: prediction=[[0.10446010869702281], [0.8955398913029772]]
\end{tikzpicture}
		\caption{correct}
		\label{fig:experiments:board:sim:1}
	\end{subfigure}%
	\hfill%
	\begin{subfigure}[t]{.24\linewidth}
		\centering
		\begin{tikzpicture}%experiment_simulator.ipynb
\tttboard%
{\tttX}{\tttO}{\tttO}%
{\tttX}{\tttX}{\tttX}%
{\tttX}{\tttX}{\tttO}%
{(0,0)}{}{}
\node[anchor=south,font=\small] (y) at ($(boardcenter)+(0,.75)$)  {$y_1 = 1$};
\node[anchor=north,font=\small] (phat) at ($(boardcenter)+(0,-.75)$)  {$\hat{p}(y_1=1) = \num{0.395}$};
%\node[anchor=north,font=\small,align=center] (cnu) at ($(phat.south)$)  {$\bar{x}_{0}=0$, $\bar{x}_{1}=0$ \\ $\bar{x}_{2}=1$, $\bar{x}_{3}=0$};
\node[anchor=north,font=\small,align=center] (cnu) at ($(phat.south)$)  {$\bar{x}_{0}\bar{x}_{1}\bar{x}_{2}\bar{x}_{3}= 0 1 0 0$};
%---------------- idx = 32 ----------------
%x:  [0 0 0 1 0 1 1 0 1 1 0 1 0 0 0]
%x3: [0 1 1 0 0 0 0 1 1]
%y:  True
%y^: False
%p^: [0.60547966 0.39452034]
%_____
%|XOO|
%|XXX|
%|XOO|
%`````
%prediction details:
%predict_proba: depth=0, split_index=14
%predict_proba: x[14]==0 split
%predict_proba: depth=1, split_index=9
%predict_proba: x[9]==1 split
%predict_proba: depth=2, split_index=13
%predict_proba: x[13]==0 split
%predict_proba: depth=3, split_index=0
%predict_proba: x[0]==0 split
%predict_proba: depth=4, split_index=None
%predict_proba: prediction=[[0.6054796588421768], [0.39452034115782325]]
\end{tikzpicture}
		\caption{incorrect}
		\label{fig:experiments:board:sim:4}
	\end{subfigure}%
	\caption{Four exemplary predictions using the Q-tree $QT_{\mathrm{sim}}$, \cref{eqn:experiments:sim:C,fig:experiments:sim:trees:q}. For each of the four examples, we visualize the query data $\mathbf{x}^q = \mathbf{x}$ in form of the corresponding board configuration based on \cref{eqn:experiments:sim:xconversion}, where \tttX and \tttO represent a field taken by the starting player and the other player, respectively. Also shown is the true label $y_1 = b$ with $b \in \mathbb{B}$, where $b=1$ represents a victory of the starting player and $b=0$ a victory of the other player. The predicted probability of the true label is denoted by $\hat{p}(y_1 = b) \equiv \hat{p}_{\mathrm{cl}}(y_1=b \,|\, \mathbf{x}^q)$, \cref{eqn:pyi:pred:cl}. The corresponding bit string $\bar{x}_{0} \bar{x}_{1} \bar{x}_{2} \bar{x}_{3}$ represents the traversed path in the tree, \cref{eqn:Cnu:q}. We indicate for each example whether the prediction is correct or incorrect using \cref{eqn:pyi:pred:argmax}.} \label{fig:experiments:board:sim}		
\end{figure}
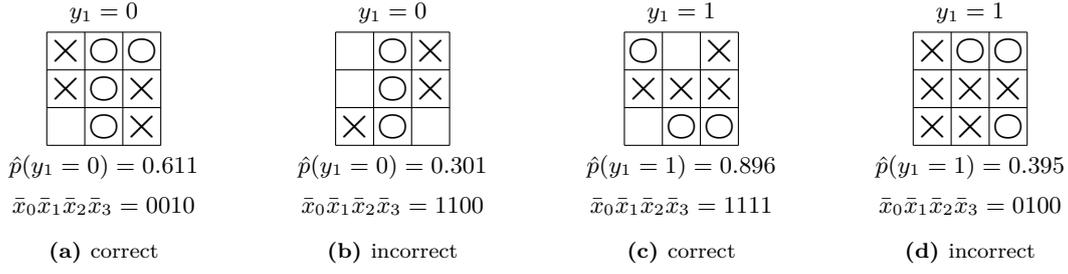
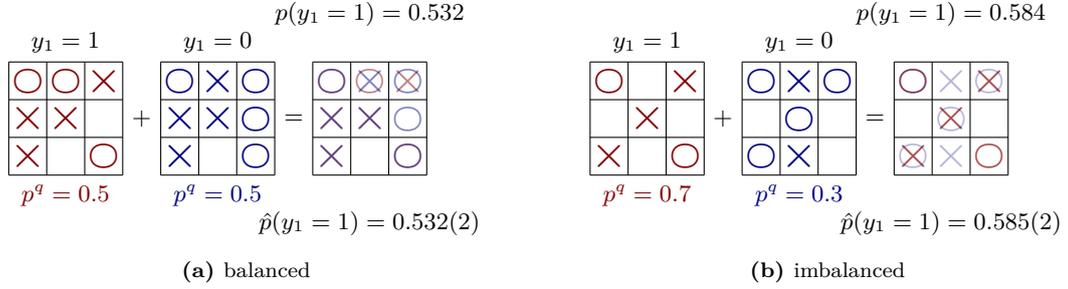
\begin{figure}
	\centering
	\begin{subfigure}[t]{.49\linewidth}
		\centering
		\begin{tikzpicture}%experiment_backend-4_ibmq_ehningen
\tttboard%
{\tttO}{\tttO}{\tttX}%
{\tttX}{\tttX}{}%
{\tttX}{}{\tttO}%
{(0,0)}{boardcol1}{}
\coordinate (c1) at (boardcenter);
\node[anchor=south,font=\small] (y1) at ($(c1)+(0,.75)$)  {$y_1 = 1$};
\node[anchor=north,font=\small,boardcol1] (p1) at ($(c1)+(0,-.75)$)  {$p^q = \num{0.5}$};
\tttboard%
{\tttO}{\tttX}{\tttO}%
{\tttX}{\tttX}{\tttO}%
{\tttX}{}{\tttO}%
{(2,0)}{boardcol2}{}
\coordinate (c2) at (boardcenter);
\node[anchor=south,font=\small] (y2) at ($(c2)+(0,.75)$)  {$y_1 = 0$};
\node[anchor=north,font=\small,boardcol2] (p2) at ($(c2)+(0,-.75)$)  {$p^q = \num{0.5}$};
\tttboard%
{\tttO}{\tttO}{\tttX}%
{\tttX}{\tttX}{}%
{\tttX}{}{\tttO}%
{(4,0)}{boardcol1,opacity=.5}{draw=none}
\tttboard%
{\tttO}{\tttX}{\tttO}%
{\tttX}{\tttX}{\tttO}%
{\tttX}{}{\tttO}%
{(4,0)}{boardcol2,opacity=.5}{draw=none}
\tttboard%
{}{}{}%
{}{}{}%
{}{}{}%
{(4,0)}{draw=none}{}
\coordinate (c3) at (boardcenter);
\node[anchor=south,font=\small] at ($(c3)+(0,.75)+(0,.35)$)  {$p(y_1=1) = \num{0.532}$};
\coordinate (p3) at ($(c3)+(0,-.75)$);
\node[anchor=north,font=\small] (phat) at ($.5*(p3)+.5*(p3)+(0,-.35)$) {$\hat{p}(y_1=1) = \num[separate-uncertainty=false]{0.532 +- 0.002}$};
\node[font=\small] at ($.5*(c1)+.5*(c2)$) {$+$};
\node[font=\small] at ($.5*(c2)+.5*(c3)$) {$=$};
%
%_____
%|OOX|
%|XX |
%|X O|
%`````
%_____
%|OXO|
%|XXO|
%|X O|
%`````
%query_probabilities: {'010001001101001': 0.5, '001110010011100': 0.5}
\end{tikzpicture}
		\caption{balanced}
		\label{fig:experiments:board:hard:1}
	\end{subfigure}%
	\hfill%
	\begin{subfigure}[t]{.49\linewidth}
		\centering
		\begin{tikzpicture}%experiment_backend-4_ibmq_ehningen
\tttboard%
{\tttO}{}{\tttX}%
{}{\tttX}{}%
{\tttX}{}{\tttO}%
{(0,0)}{boardcol1}{}
\coordinate (c1) at (boardcenter);
\node[anchor=south,font=\small] (y1) at ($(c1)+(0,.75)$)  {$y_1 = 1$};
\node[anchor=north,font=\small,boardcol1] (p1) at ($(c1)+(0,-.75)$)  {$p^q = \num{0.7}$};
\tttboard%
{\tttO}{\tttX}{\tttO}%
{}{\tttO}{}%
{\tttO}{\tttX}{}%
{(2,0)}{boardcol2}{}
\coordinate (c2) at (boardcenter);
\node[anchor=south,font=\small] (y2) at ($(c2)+(0,.75)$) {$y_1 = 0$};
\node[anchor=north,font=\small,boardcol2] (p2) at ($(c2)+(0,-.75)$) {$p^q = \num{0.3}$};
\tttboard%
{\tttO}{}{\tttX}%
{}{\tttX}{}%
{\tttX}{}{\tttO}%
{(4,0)}{boardcol1,opacity=.7}{draw=none}
\tttboard%
{\tttO}{\tttX}{\tttO}%
{}{\tttO}{}%
{\tttO}{\tttX}{}%
{(4,0)}{boardcol2,opacity=.3}{draw=none}
\tttboard%
{}{}{}%
{}{}{}%
{}{}{}%
{(4,0)}{draw=none}{}
\coordinate (c3) at (boardcenter);
\node[anchor=south,font=\small] at ($(c3)+(0,.75)+(0,.35)$)  {$p(y_1=1) = \num{0.584}$};
\coordinate (p3) at ($(c3)+(0,-.75)$);
\node[anchor=north,font=\small] (phat) at ($.5*(p3)+.5*(p3)+(0,-.35)$) {$\hat{p}(y_1=1) = \num[separate-uncertainty=false]{0.585 +- 0.002}$};
\node[font=\small] at ($.5*(c1)+.5*(c2)$) {$+$};
\node[font=\small] at ($.5*(c2)+.5*(c3)$) {$=$};
%
%_____
%|O X|
%| X |
%|X O|
%`````
%_____
%|XXO|
%| O |
%|OX |
%`````
%query_probabilities: {'010110011011010': 0.7, '000010101010001': 0.3}
\end{tikzpicture}
		\caption{imbalanced}
		\label{fig:experiments:board:hard:2}
	\end{subfigure}%, 
	\caption{Two exemplary uncertain query data predictions using the Q-tree $QT_{\mathrm{sim}}$, \cref{eqn:experiments:sim:C,fig:experiments:sim:trees:q}. Using the same notation as in \cref{fig:experiments:board:sim}, we show an uncertain overlay of two boards weighted by balanced and imbalanced probabilities $p^q$, respectively. The opacity of the resulting overlay on the right (with respect to the two colors \tttXc{boardcol1}/\tttOc{boardcol1} and \tttXc{boardcol2}/\tttOc{boardcol2}) reflects the corresponding values of $p^q$ (the more transparent, the smaller). For each example, we show the true labels of the query components $y_1 = b$ with $b \in \mathbb{B}$, the true predicted probability distribution for the query label $p(y_1=b) \equiv p(y_1=b \,|\, p^q)$, \cref{eqn:pq:pyi:pred:approx}, as well as its prediction $\hat{p}(y_1=b) \equiv \hat{p}_{\mathrm{qm}}(y_1=b \,|\, p^q)$, \cref{eqn:pq:pyi:pred}, with the standard deviation $\stdev{\hat{p}_{\mathrm{qm}}(y_1 \,|\, p^q)}$, \cref{eqn:app:pq:pyi:pred:unc}, in brackets. All quantum computations have been performed on a simulator.}
	\label{fig:experiments:board:sim:q}
\end{figure}
Next, we perform a direct comparison of the classification performance between $QT_{\mathrm{sim}}^1,\dots,QT_{\mathrm{sim}}^{25}$ and $T_{\mathrm{cl}}$. In addition to the balanced accuracy, \cref{eqn:experiments:metric:bacc}, we also consider the unbalanced accuracy
\begin{align} \label{eqn:experiments:metric:acc}
	\mathrm{acc} \equiv \frac{\mathrm{tp}_0 + \mathrm{tn}_0}{\mathrm{tp}_0 + \mathrm{tn}_0 + \mathrm{fp}_0 + \mathrm{fn}_0} = \frac{\mathrm{tp}_1 + \mathrm{tn}_1}{\mathrm{tp}_1 + \mathrm{tn}_1 + \mathrm{fp}_1 + \mathrm{fn}_1},
\end{align}
the precision (fraction of relevant instances among the retrieved instances)
\begin{align} \label{eqn:experiments:metric:pre}
	\mathrm{pre}_b \equiv \frac{\mathrm{tp}_b}{\mathrm{tp}_b + \mathrm{fp}_b},
\end{align}
the recall (fraction of relevant instances that were retrieved)
\begin{align} \label{eqn:experiments:metric:rec}
	\mathrm{rec}_b \equiv \frac{\mathrm{tp}_b}{\mathrm{tp}_b + \mathrm{fn}_b},
\end{align}
and the balanced F-score (harmonic mean of precision and recall)
\begin{align} \label{eqn:experiments:metric:f1}
	\mathrm{f1}_b \equiv \frac{2 \mathrm{pre}_b \mathrm{rec}_b}{\mathrm{pre}_b + \mathrm{rec}_b},
\end{align}
respectively, with $b \in \mathbb{B}$, where we recall \cref{eqn:experiments:metric:tp,eqn:experiments:metric:tn,eqn:experiments:metric:fp,eqn:experiments:metric:fn}. The performance metrics are determined for the training and test data set (\ie, $\mathbf{D} \in \{ \mathbf{D}_{\mathrm{train}}, \mathbf{D}_{\mathrm{test}}\}$).\par
The results are listed in \cref{tab:experiments:sim:scores}. We find that for the training data, certain best metrics of the Q-trees ($\mathrm{acc}$, $\mathrm{pre}_0$, $\mathrm{rec}_1$, and $\mathrm{f1}_1$) are superior to the corresponding metrics of the classical tree $T_{\mathrm{cl}}$, whereas others ($\mathrm{bac}$, $\mathrm{pre}_1$, $\mathrm{rec}_0$, and $\mathrm{f1}_0$) are inferior. For the test data, the best metrics of the Q-trees are always superior or at least as good as the corresponding metrics of the classical tree. In all cases, the mean Q-tree metrics and the metrics of the exemplary Q-tree $QT_{\mathrm{sim}}$, \cref{eqn:experiments:sim:C}, are worse than or equal to the corresponding metrics of the classical tree.\par
\begin{table}
	\centering
	\caption{Classification performance metrics, \cref{eqn:experiments:metric:bacc,eqn:experiments:metric:acc,eqn:experiments:metric:pre,eqn:experiments:metric:rec,eqn:experiments:metric:f1}, of the Q-trees $QT_{\mathrm{sim}}^1,\dots,QT_{\mathrm{sim}}^{25}$ and the classical decision tree $T_{\mathrm{cl}}$, respectively. We list the mean and standard deviation (in brackets) over all Q-trees, the best metric over all trees and the metrics of the exemplary Q-tree $QT_{\mathrm{sim}}$, \cref{eqn:experiments:sim:C,fig:experiments:sim:trees:q}. The best results (with respect to two digits) are highlighted in bold.} \label{tab:experiments:sim:scores}
	\begin{tabular}{lcccccccc}%experiment_simulator.ipynb
		\hline\hline
		& $\mathrm{bac}$ & $\mathrm{acc}$ & $\mathrm{pre}_0$ & $\mathrm{pre}_1$ & $\mathrm{rec}_0$ & $\mathrm{rec}_1$ & $\mathrm{f1}_0$& $\mathrm{f1}_1$ \\
		\hline
		\multicolumn{9}{l}{Training data} \\
		Mean Q-tree metric               & \tnum{0.65 +- 0.02} & \tnum{0.70 +- 0.01} & \tnum{0.58 +- 0.03} & \tnum{0.75 +- 0.02} & \tnum{0.50 +- 0.07} & \tnum{0.80 +- 0.05} & \tnum{0.53 +- 0.04} & \tnum{0.77 +- 0.01} \\		 
		Best Q-tree metric               & \tnum{0.68} & \Tnum{0.72} & \Tnum{0.66} & \tnum{0.78} & \tnum{0.61} & \Tnum{0.90} & \tnum{0.58} & \Tnum{0.81} \\		
		Q-tree $QT_{\mathrm{sim}}$       & \tnum{0.68} & \tnum{0.71} & \tnum{0.57} & \tnum{0.78} & \tnum{0.59} & \tnum{0.77} & \tnum{0.58} & \tnum{0.77} \\	
		Classical tree $T_{\mathrm{cl}}$ & \Tnum{0.70} & \tnum{0.71} & \tnum{0.58} & \Tnum{0.80} & \Tnum{0.64} & \tnum{0.75} & \Tnum{0.61} & \tnum{0.78} \\
		\hline
		\multicolumn{9}{l}{Test data} \\
		Mean Q-tree metric               & \tnum{0.63 +- 0.03} & \tnum{0.69 +- 0.02} & \tnum{0.57 +- 0.04} & \tnum{0.74 +- 0.02} & \tnum{0.44 +- 0.07} & \tnum{0.82 +- 0.04} & \tnum{0.50 +- 0.05} & \tnum{0.78 +- 0.02} \\		
		Best Q-tree metric               & \Tnum{0.68} & \Tnum{0.72} & \Tnum{0.63} & \Tnum{0.78} & \Tnum{0.58} & \Tnum{0.88} & \Tnum{0.58} & \Tnum{0.80} \\	
		Q-tree $QT_{\mathrm{sim}}$       & \tnum{0.62} & \tnum{0.68} & \tnum{0.55} & \tnum{0.73} & \tnum{0.43} & \tnum{0.81} & \tnum{0.48} & \tnum{0.77} \\	
		Classical tree $T_{\mathrm{cl}}$ & \tnum{0.67} & \Tnum{0.72} & \tnum{0.61} & \tnum{0.76} & \tnum{0.52} & \tnum{0.83} & \tnum{0.56} & \tnum{0.79} \\		
		\hline\hline
	\end{tabular}
\end{table}
To quantify the performance of the tree induction, we consider a set of randomly generated decision trees. For this purpose, we uniformly draw $\num{25}$ decision configurations, \cref{eqn:C}. Subsequently, we perform a sampling with $N=\num{e6}$ measurements for each corresponding Q-tree to obtain an estimate for the label probability distribution. We determine the balanced training and test accuracy, \cref{eqn:experiments:metric:bacc}, for each tree and obtain a mean and standard deviation (in brackets) of $\num{0.53 +- 0.04}$ for both test and training data. This result is significantly worse than the mean balanced accuracy over all induced trees listed in \cref{tab:experiments:sim:scores}. We consequently conclude that our induction method is clearly superior to a random selection of trees for this particular example.\par
Finally, we compare the effect of using different numbers of experiments $N \in \{\num{e3},\num{e4},\num{e5},\num{e6}\}$ to estimate the probability distribution of the decision set $\hat{p}(\mathcal{C}_{\nu})$, \cref{eqn:pn:est}, and the corresponding label probability distribution $\hat{p}(y_i \,|\,\mathcal{C}_{\nu})$, \cref{eqn:pyi:est}, for the Q-tree $QT_{\mathrm{sim}}$. We compare the estimates with the exact probability distributions based on data proportions $p(\mathcal{C}_{\nu})$, \cref{eqn:pn}, and $p(\mathbf{y} \,|\, \mathcal{C}_{\nu})$, \cref{eqn:pn:y}, respectively, and also consider the statistical uncertainties as discussed in \cref{sec:app:uncertainty}.\par
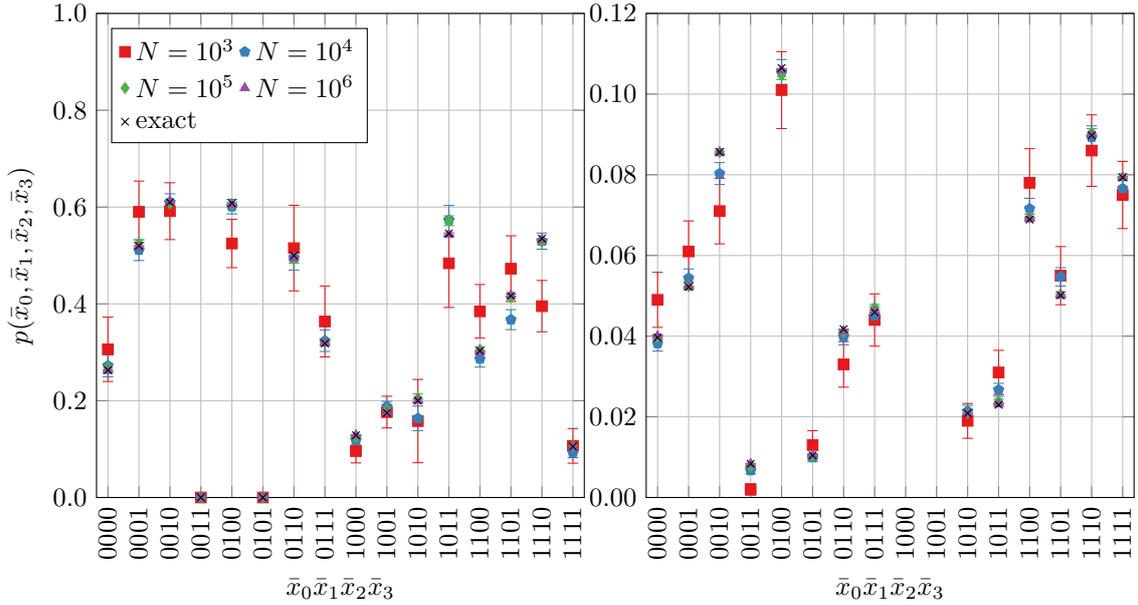
\begin{figure}
	\centering
	\begin{subfigure}[t]{.49\linewidth}
		\centering
		\begin{tikzpicture} 
\pgfplotstableread[col sep=semicolon]{%experiment_simulator-proba-km-1000-list.csv
idx;mean1;std1
0;0.3061224490;0.0665028503
1;0.5901639344;0.0634678265
2;0.5915492958;0.0587270825
3;0.0000000000;0.0000000000
4;0.5247524752;0.0499157774
5;0.0000000000;0.0000000000
6;0.5151515152;0.0883468067
7;0.3636363636;0.0733411491
8;0.0958904110;0.0244401862
9;0.1764705882;0.0327944698
10;0.1578947368;0.0860506146
11;0.4838709677;0.0912452377
12;0.3846153846;0.0554186589
13;0.4727272727;0.0679169252
14;0.3953488372;0.0530080329
15;0.1066666667;0.0358693011
}\uncA;
\pgfplotstableread[col sep=semicolon]{%experiment_simulator-proba-km-10000-list.csv
idx;mean1;std1
0;0.2722513089;0.0228030475
1;0.5110294118;0.0214507928
2;0.6102117061;0.0172205116
3;0.0000000000;0.0000000000
4;0.6009478673;0.0150831258
5;0.0000000000;0.0000000000
6;0.4949748744;0.0250918324
7;0.3237250554;0.0220558000
8;0.1183591123;0.0083794529
9;0.1877862595;0.0107938314
10;0.1635514019;0.0253419409
11;0.5730337079;0.0303268480
12;0.2863128492;0.0169044089
13;0.3674588665;0.0206315268
14;0.5296752520;0.0167109000
15;0.0926892950;0.0104843302
}\uncB;
\pgfplotstableread[col sep=semicolon]{%experiment_simulator-proba-km-100000-list.csv
idx;mean1;std1
0;0.2693178958;0.0070386612
1;0.5251908397;0.0068990891
2;0.6069384418;0.0052791446
3;0.0000000000;0.0000000000
4;0.6055475849;0.0047800015
5;0.0000000000;0.0000000000
6;0.4920751036;0.0078076684
7;0.3201949566;0.0067923226
8;0.1238366959;0.0027149456
9;0.1815168799;0.0033840829
10;0.2057279236;0.0088336626
11;0.5716673631;0.0101176479
12;0.3057196369;0.0055302811
13;0.4135876043;0.0069417617
14;0.5283789753;0.0052459355
15;0.1066750630;0.0034645841
}\uncC;
\pgfplotstableread[col sep=semicolon]{%experiment_simulator-proba-km-1000000-list.csv
idx;mean1;std1
0;0.2618280778;0.0021967192
1;0.5178434529;0.0021875694
2;0.6110786559;0.0016663849
3;0.0000000000;0.0000000000
4;0.6054796588;0.0015012388
5;0.0000000000;0.0000000000
6;0.4945007026;0.0024608744
7;0.3192305193;0.0021697971
8;0.1287845111;0.0008759736
9;0.1752911624;0.0010490198
10;0.2003755236;0.0027774542
11;0.5434265704;0.0032751992
12;0.3007020337;0.0017446582
13;0.4183080884;0.0022036412
14;0.5325160745;0.0016641273
15;0.1044601087;0.0010861140
}\uncD;
\pgfplotstableread[col sep=semicolon]{%experiment_simulator-proba-km-exact-list.csv
idx;mean1;std1
0;0.2631578947;0.0000000000
1;0.5200000000;0.0000000000
2;0.6097560976;0.0000000000
3;0.0000000000;0.0000000000
4;0.6078431373;0.0000000000
5;0.0000000000;0.0000000000
6;0.5000000000;0.0000000000
7;0.3181818182;0.0000000000
8;0.1285714286;0.0000000000
9;0.1746031746;0.0000000000
10;0.2000000000;0.0000000000
11;0.5454545455;0.0000000000
12;0.3030303030;0.0000000000
13;0.4166666667;0.0000000000
14;0.5348837209;0.0000000000
15;0.1052631579;0.0000000000
}\uncE;
\begin{axis} [width=8cm, height=8cm,
	ymin=0, ymax=1,
	xmin=0, xmax=15,
	xtick={0,...,15},
	xticklabels={0000,0001,0010,0011,0100,0101,0110,0111,1000,1001,1010,1011,1100,1101,1110,1111},
	ytick={0.0,0.2,0.4,0.6,0.8,1.0},
	yticklabels={0.0,0.2,0.4,0.6,0.8,1.0},
	x tick label style={rotate=90},
	grid=both,
	enlarge x limits={abs=.15cm},
	xlabel={$\bar{x}_{0}\bar{x}_{1}\bar{x}_{2}\bar{x}_{3}$},
	ylabel={$p(\bar{x}_{0},\bar{x}_{1},\bar{x}_{2},\bar{x}_{3})$},	
	legend cell align=left,
	legend pos=north west,
	legend columns=2,
	]
	\addplot [color51,only marks,mark=square*] plot [error bars/.cd, y dir=both, y explicit] table [y=mean1, y error plus=std1, y error minus=std1] {\uncA};
	\addplot [color52,only marks,mark=pentagon*] plot [error bars/.cd, y dir=both, y explicit] table [y=mean1, y error plus=std1, y error minus=std1] {\uncB};
	\addplot [color53,only marks,mark=diamond*] plot [error bars/.cd, y dir=both, y explicit] table [y=mean1, y error plus=std1, y error minus=std1] {\uncC};	
	\addplot [color54,only marks,mark=triangle*] plot [error bars/.cd, y dir=both, y explicit] table [y=mean1, y error plus=std1, y error minus=std1] {\uncD};		
	%\addplot [black,only marks,mark=x] plot [error bars/.cd, y dir=both, y explicit] table [y=mean1, y error plus=std1, y error minus=std1] {\uncE};	
	\addplot [black,only marks,mark=x] plot [] table [y=mean1] {\uncE};
	\legend{$N=\num{e3}$,$N=\num{e4}$,$N=\num{e5}$,$N=\num{e6}$,exact}
\end{axis} 
\end{tikzpicture}
		\caption{Label probabilities from Q-tree samples $p(\bar{x}_{0},\bar{x}_{1},\bar{x}_{2},\bar{x}_{3})\equiv\hat{p}(y_1 = 0 \,|\, \mathcal{C}_{\nu_4(\bar{x}_{0},\bar{x}_{1},\bar{x}_{2},\bar{x}_{3})})$, \cref{eqn:pyi:est}, with standard deviation (one sigma) $\stdeva{\hat{p}'(k,m,N)}$, \cref{eqn:app:unc:pp:std}, and from exact calculations $p(\bar{x}_{0},\bar{x}_{1},\bar{x}_{2},\bar{x}_{3})\equiv p(y_1 = 0 \,|\, \mathcal{C}_{\nu_4(\bar{x}_{0},\bar{x}_{1},\bar{x}_{2},\bar{x}_{3})})$, \cref{eqn:pn:y}, respectively.}
		\label{fig:experiments:sim:probas:km}
	\end{subfigure}%
	\hfill%
	\begin{subfigure}[t]{.49\linewidth}
		\centering
		\begin{tikzpicture} 
\pgfplotstableread[col sep=semicolon]{%experiment_simulator-proba-mN-1000-list.csv
idx;mean1;std1
0;0.0490000000;0.0068263460
1;0.0610000000;0.0075682891
2;0.0710000000;0.0081215146
3;0.0020000000;0.0014127986
4;0.1010000000;0.0095288509
5;0.0130000000;0.0035820385
6;0.0330000000;0.0056489822
7;0.0440000000;0.0064856765
8;0.1460000000;0.0111661990
9;0.1360000000;0.0108399262
10;0.0190000000;0.0043172908
11;0.0310000000;0.0054807846
12;0.0780000000;0.0084803302
13;0.0550000000;0.0072093689
14;0.0860000000;0.0088658897
15;0.0750000000;0.0083291656
}\uncA;
\pgfplotstableread[col sep=semicolon]{%experiment_simulator-proba-mN-10000-list.csv
idx;mean1;std1
0;0.0382000000;0.0019167879
1;0.0544000000;0.0022680529
2;0.0803000000;0.0027175708
3;0.0067000000;0.0008157886
4;0.1055000000;0.0030719660
5;0.0099000000;0.0009900500
6;0.0398000000;0.0019548903
7;0.0451000000;0.0020752347
8;0.1487000000;0.0035579251
9;0.1310000000;0.0033740036
10;0.0214000000;0.0014471365
11;0.0267000000;0.0016120518
12;0.0716000000;0.0025782444
13;0.0547000000;0.0022739373
14;0.0893000000;0.0028517628
15;0.0766000000;0.0026595571
}\uncB;
\pgfplotstableread[col sep=semicolon]{%experiment_simulator-proba-mN-100000-list.csv
idx;mean1;std1
0;0.0397300000;0.0006176692
1;0.0524000000;0.0007046576
2;0.0856100000;0.0008847651
3;0.0076800000;0.0002760619
4;0.1045500000;0.0009675707
5;0.0102600000;0.0003186649
6;0.0410100000;0.0006271218
7;0.0471900000;0.0006705453
8;0.1472100000;0.0011204428
9;0.1297400000;0.0010625796
10;0.0209500000;0.0004528918
11;0.0239300000;0.0004832945
12;0.0694100000;0.0008036931
13;0.0503400000;0.0006914180
14;0.0905600000;0.0009075180
15;0.0794000000;0.0008549599
}\uncC;
\pgfplotstableread[col sep=semicolon]{%experiment_simulator-proba-mN-1000000-list.csv
idx;mean1;std1
0;0.0400530000;0.0001960835
1;0.0521760000;0.0002223818
2;0.0855880000;0.0002797547
3;0.0083610000;0.0000910554
4;0.1059920000;0.0003078274
5;0.0105070000;0.0001019637
6;0.0412780000;0.0001989325
7;0.0461610000;0.0002098337
8;0.1462210000;0.0003533276
9;0.1313700000;0.0003378046
10;0.0207710000;0.0001426168
11;0.0231310000;0.0001503195
12;0.0690850000;0.0002535986
13;0.0501090000;0.0002181699
14;0.0898940000;0.0002860298
15;0.0793030000;0.0002702111
}\uncD;
\pgfplotstableread[col sep=semicolon]{%experiment_simulator-proba-mN-exact-list.csv
idx;mean1;std1
0;0.0396659708;0.0000000000
1;0.0521920668;0.0000000000
2;0.0855949896;0.0000000000
3;0.0083507307;0.0000000000
4;0.1064718163;0.0000000000
5;0.0104384134;0.0000000000
6;0.0417536534;0.0000000000
7;0.0459290188;0.0000000000
8;0.1461377871;0.0000000000
9;0.1315240084;0.0000000000
10;0.0208768267;0.0000000000
11;0.0229645094;0.0000000000
12;0.0688935282;0.0000000000
13;0.0501043841;0.0000000000
14;0.0897703549;0.0000000000
15;0.0793319415;0.0000000000
}\uncE;
\begin{axis} [width=8cm, height=8cm,
	ymin=0, ymax=.12,
	xmin=0, xmax=15,
	xtick={0,...,15},
	xticklabels={0000,0001,0010,0011,0100,0101,0110,0111,1000,1001,1010,1011,1100,1101,1110,1111},
	ytick={0.00,0.02,0.04,0.06,0.08,0.10,0.12},
	yticklabels={0.00,0.02,0.04,0.06,0.08,0.10,0.12},
	x tick label style={rotate=90},
	grid=both,
	enlarge x limits={abs=.15cm},
	xlabel={$\bar{x}_{0}\bar{x}_{1}\bar{x}_{2}\bar{x}_{3}$},
	legend cell align=left,
	legend pos=north west,
	]
	\addplot [color51,only marks,mark=square*] plot [error bars/.cd, y dir=both, y explicit] table [y=mean1, y error plus=std1, y error minus=std1] {\uncA};
	\addplot [color52,only marks,mark=pentagon*] plot [error bars/.cd, y dir=both, y explicit] table [y=mean1, y error plus=std1, y error minus=std1] {\uncB};
	\addplot [color53,only marks,mark=diamond*] plot [error bars/.cd, y dir=both, y explicit] table [y=mean1, y error plus=std1, y error minus=std1] {\uncC};	
	\addplot [color54,only marks,mark=triangle*] plot [error bars/.cd, y dir=both, y explicit] table [y=mean1, y error plus=std1, y error minus=std1] {\uncD};		
	%\addplot [black,only marks,mark=x] plot [error bars/.cd, y dir=both, y explicit] table [y=mean1, y error plus=std1, y error minus=std1] {\uncE};	
	\addplot [black,only marks,mark=x] plot [] table [y=mean1] {\uncE};	
	%\legend{$N=\num{1000}$,$N=\num{10000}$,$N=\num{100000}$,$N=\num{1000000}$,exact}	
\end{axis} 
\end{tikzpicture}
		\caption{Probabilities of the decision set from Q-tree samples $p(\bar{x}_{0},\bar{x}_{1},\bar{x}_{2},\bar{x}_{3})\equiv\hat{p}(\mathcal{C}_{\nu_4(\bar{x}_{0},\bar{x}_{1},\bar{x}_{2},\bar{x}_{3})})$, \cref{eqn:pn:est}, with standard deviation (one sigma) $\stdeva{\hat{p}(n,N)}$, \cref{eqn:app:unc:p:std}, and from exact calculations $p(\bar{x}_{0},\bar{x}_{1},\bar{x}_{2},\bar{x}_{3})\equiv p(\mathcal{C}_{\nu_4(\bar{x}_{0},\bar{x}_{1},\bar{x}_{2},\bar{x}_{3})})$, \cref{eqn:pn}, respectively. We use the same symbols as in \subref{fig:experiments:sim:probas:km}.}
		\label{fig:experiments:sim:probas:mn}
	\end{subfigure}%
	\caption{Sampled probability distributions of the Q-tree $QT_{\mathrm{sim}}$, \cref{eqn:experiments:sim:C,fig:experiments:sim:trees:q}, using $N \in \{\num{e3},\num{e4},\num{e5},\num{e6}\}$ measurements. All quantum computations have been performed on a simulator. For comparison, we also show the exact probability distributions based on the data proportions.}
	\label{fig:experiments:sim:probas}		
\end{figure}

The results are shown in \cref{fig:experiments:sim:probas}. As expected, the sampled probabilities converge to the exact results for a sufficiently high number of experiments, whereas their standard deviations decrease with an increasing number of experiments. Moreover, the standard deviations of vanishing probabilities also vanish.

\subsection{Quantum hardware} \label{sec:experiments:quantum hardware experiments}
A major challenge of quantum algorithms is to run them on noisy intermediate-scale quantum (NISQ) devices \cite{preskill2018}. To perform quantum hardware experiments, we remotely access a physical quantum computer via Qiskit using the cloud-based quantum computing service provided by \emph{IBM Quantum} \cite{ibmq2021}. This service allows online requests for quantum experiments using a high-level quantum circuit model of computation \cite{larose2019}. The quantum experiments are then realized and executed sequentially on physical hardware that operates on superconducting transmon qubits. Specifically, we access the quantum computer \emph{ibmq\_ehningen} (version 2.2.1), which represents a so-called \emph{Falcon r4} processor with $\num{27}$ qubits and a quantum volume \cite{cross2019} of $\num{32}$.\par
\begin{figure}
	\begin{center}
		\begin{tikzpicture}
\newcommand{\qubit}[3]{%
%\node[draw=colorA#1,fill=colorB#1,very thick,circle,minimum width=.5cm,inner sep=0pt,outer sep=0pt,font=\small] (Q#1) at ($(.6*#2,.6*#3)$) {\color{white}#1};
\node[draw,fill=white,rounded corners=.1cm,minimum width=.4cm,minimum height=.4cm,inner sep=0pt,outer sep=0pt,font=\small] (Q#1) at ($(.55*#2,.55*#3)$) {};
\node[inner sep=0pt,outer sep=0pt,font=\small] at ($(Q#1)+(0,0)$) {#1};}
\newcommand{\connection}[2]{\draw[very thick] (Q#1) -- (Q#2);}
\qubit{0}{0}{3}
\qubit{1}{1}{3}
\qubit{2}{1}{2}
\qubit{3}{1}{1}
\qubit{4}{2}{3}
\qubit{5}{2}{1}
\qubit{6}{3}{4}
\qubit{7}{3}{3}
\qubit{8}{3}{1}
\qubit{9}{3}{0}
\qubit{10}{4}{3}
\qubit{11}{4}{1}
\qubit{12}{5}{3}
\qubit{13}{5}{2}
\qubit{14}{5}{1}
\qubit{15}{6}{3}
\qubit{16}{6}{1}
\qubit{17}{7}{4}
\qubit{18}{7}{3}
\qubit{19}{7}{1}
\qubit{20}{7}{0}
\qubit{21}{8}{3}
\qubit{22}{8}{1}
\qubit{23}{9}{3}
\qubit{24}{9}{2}
\qubit{25}{9}{1}
\qubit{26}{10}{1}
\connection{0}{1}
\connection{1}{2}
\connection{1}{4}
\connection{2}{3}
\connection{3}{5}
\connection{4}{7}
\connection{5}{8}
\connection{6}{7}
\connection{7}{10}
\connection{8}{9}
\connection{8}{11}
\connection{10}{12}
\connection{11}{14}
\connection{12}{13}
\connection{12}{15}
\connection{13}{14}
\connection{14}{16}
\connection{15}{18}
\connection{16}{19}
\connection{17}{18}
\connection{18}{21}
\connection{19}{20}
\connection{19}{22}
\connection{21}{23}
\connection{22}{25}
\connection{23}{24}
\connection{24}{25}
\connection{25}{26}
\end{tikzpicture}
		\caption{Connectivity diagram of the quantum computer \emph{ibmq\_ehningen} with $\num{27}$ qubits. Qubits are shown as boxes with their respective hardware index $\{0,\dots,26\}$ as provided by IBM Quantum. Pairs of qubits that support a mutual $\CNOT$ gate are connected with lines.}
		\label{fig:experiments:hard:backend}
	\end{center}
\end{figure}
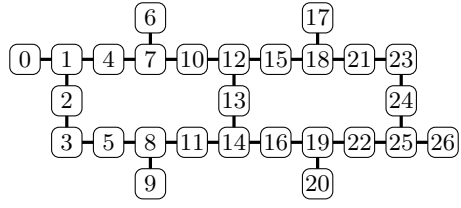

A quantum experiment is in this context defined as a circuit to be evaluated on the quantum computer, whereas the returned result consist of a histogram of counts for all measured qubits. The quantum computer we use is only capable of executing a specific set of gates ($\NOT$, $\RZ$, $\SX$, and $\CNOT$) and has a limited connectivity of qubits (\ie, $\CNOT$ gates can only be applied to specific pairs of qubits) as sketched in \cref{fig:experiments:hard:backend}. In a process called \emph{transpilation}, Qiskit can algorithmically translate a general circuit into one that can be run on a specific hardware. We use this functionality to realize our quantum experiments for the Q-tree. Furthermore, we make use of Qiskit's built-in state preparation approach \cite{shende2006} to find the data encoding operator $\hat{U}_{\mathrm{data}}(\mathbf{D}_{\mathrm{train}})$, \cref{eqn:psi:d}.\par
In total, a Q-tree of maximum depth $d$ consists of $k+m$ qubits and exhibits a complexity of $\BigO{2^{k+m} + d 2^d}$ gates. It is possible to reduce this complexity by only encoding data in such qubits that are actually measured. This effectively corresponds to performing the marginalization of $p(\mathbf{x},\mathbf{y})$, \cref{eqn:p:est}, beforehand and leads to a circuit consisting of $d+1+m$ qubits and $\BigO{2^d( d + 2^m)}$ gates. In addition, the first $\SWAP$ gate can be omitted by a suitable re-ordering of the data (\ie, by assigning the corresponding feature of the root decision to the first qubit). The data preparation consequently depends on the tree structure in these cases. To simplify our circuits, we make use of these complexity reductions in the following.\par
The tic-tac-toe endgame data set considered in \cref{sec:experiments:simulator} for the simulator turns out to be too complex for the actual quantum hardware. Therefore, we instead employ the simpler toy data set, \cref{eqn:experiments:XY}, where we use the total data set for both training and testing purposes. In analogy to our simulator experiments, we start by inducing a decision tree of maximum depth $d=\num{1}$ as described in \cref{sec:quantum representation:tree induction} using the genetic algorithm from \cref{sec:app:genetic induction algorithm}. We use the hyperparameters $\theta=(\num{31},\num{6},\num{4},\num{.5},\num{.5},\num{.5},\num{1},\num{1})$, \cref{eqn:app:ga:theta}, and $N=\num{8192}$ measurements (which is the maximum number of measurements -- so-called \emph{shots} -- that can be performed on the chosen hardware in a single run). We repeat the induction $\num{10}$ times with different random seeds and consequently obtain $\num{10}$ Q-trees $QT_{\mathrm{qc}}^1,\dots,QT_{\mathrm{qc}}^{10}$. Again, we perform a final sampling with $N=\num{8192}$ measurements to obtain an estimate for the label probability distribution $\hat{p}(y_i \,|\,\mathcal{C}_{\nu})$, \cref{eqn:pyi:est}. In addition, we use scikit-learn to induce a classical decision tree $T'_{\mathrm{cl}}$ of the same maximum depth for comparison purposes, which is shown in \cref{fig:experiments:hard:trees:s}.\par 
We visualize the mean distribution of splitting feature indices in each depth in \cref{fig:experiments:hard:features} and find a close resemblance between the distributions in \cref{fig:experiments:hard:features:q} and \cref{fig:experiments:hard:features:s}. Furthermore, the first three splitting indices, which refer to informative features, are chosen over the remaining splitting indices, which refer to uninformative features (in all except for one case in \cref{fig:experiments:hard:features:q}). The most often occurring splitting index over all depths is $2$ in \cref{fig:experiments:hard:features:q}, whereas in \cref{fig:experiments:hard:features:q} the splitting indices $2$ and $3$ are both occurring once in the tree.\par
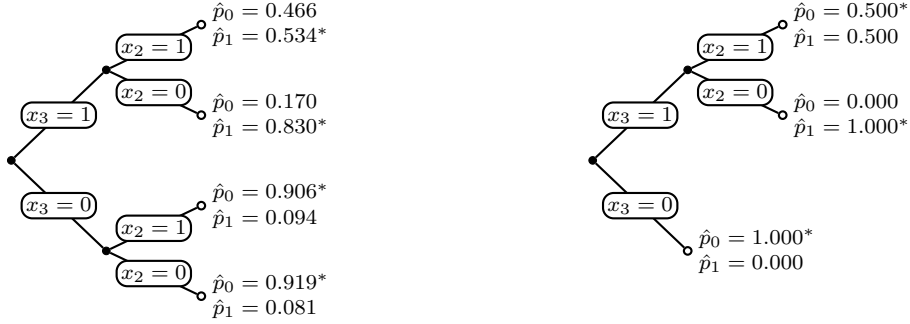
\begin{figure}
	\centering
	\begin{subfigure}[t]{.49\linewidth}
		\centering
		\begin{tikzpicture} 
%experiment_backend-2_ibmq_ehningen.ipynb
[
level 1/.style          = {sibling distance=2.4cm},
level 2/.style			= {sibling distance=1.2cm},
level 3/.style			= {sibling distance=0.6cm},
grow                    = right,
level distance          = 1.25cm,
edge from parent/.style = {draw,thick},
every node/.style       = {},
]
\node[treenode] (root) {} % root
child { node[treenode] (n0) {} % 0
	child { node[leafnode] (n00) {} } % 00	
	child { node[leafnode] (n01) {} } % 01
}
child { node[treenode] (n1) {} % 1
	child { node[leafnode] (n10) {} } % 10	
	child { node[leafnode] (n11) {} } % 11
}	
;
\node[edgenode,anchor=center] at ($.5*(root)+.5*(n0)$) {$x_{3} = 0$}; % r-0
\node[edgenode,anchor=center] at ($.5*(root)+.5*(n1)$) {$x_{3} = 1$}; % r-1
\node[edgenode,anchor=center] at ($.5*(n0)+.5*(n00)$) {$x_{2} = 0$}; % 0-00
\node[edgenode,anchor=center] at ($.5*(n0)+.5*(n01)$) {$x_{2} = 1$}; % 0-01
\node[edgenode,anchor=center] at ($.5*(n1)+.5*(n10)$) {$x_{2} = 0$}; % 1-10
\node[edgenode,anchor=center] at ($.5*(n1)+.5*(n11)$) {$x_{2} = 1$}; % 1-11				
\node[leafnodelabel] at (n00) {$\Treepval{0}{0.919}$ \\ $\treepval{1}{0.081}$}; % 00
\node[leafnodelabel] at (n01) {$\Treepval{0}{0.906}$ \\ $\treepval{1}{0.094}$}; % 01
\node[leafnodelabel] at (n10) {$\treepval{0}{0.170}$ \\ $\Treepval{1}{0.830}$}; % 10
\node[leafnodelabel] at (n11) {$\treepval{0}{0.466}$ \\ $\Treepval{1}{0.534}$}; % 11
\useasboundingbox ($(root)+(-.25,-2.25)$) rectangle ($(root)+(4.25,2.25)$);
\end{tikzpicture}
		\caption{Best Q-tree $QT_{\mathrm{qc}}$ with the compressed decision configuration $C_{\mathrm{qc}}$, \cref{eqn:experiments:hard:C}, induced via a genetic algorithm.}
		\label{fig:experiments:hard:trees:q}
	\end{subfigure}%
	\hfill%
	\begin{subfigure}[t]{.49\linewidth}
		\centering
		\begin{tikzpicture} 
%experiment_backend-1_ibmq_ehningen.ipynb
[
level 1/.style          = {sibling distance=2.4cm},
level 2/.style			= {sibling distance=1.2cm},
level 3/.style			= {sibling distance=0.6cm},
grow                    = right,
level distance          = 1.25cm,
edge from parent/.style = {draw,thick},
every node/.style       = {},
]
\node[treenode] (root) {} % root
child { node[leafnode] (n0) {} % 0
}
child { node[treenode] (n1) {} % 1
	child { node[leafnode] (n10) {} } % 10	
	child { node[leafnode] (n11) {} } % 11
}	
;
\node[edgenode,anchor=center] at ($.5*(root)+.5*(n0)$) {$x_{3} = 0$}; % r-0
\node[edgenode,anchor=center] at ($.5*(root)+.5*(n1)$) {$x_{3} = 1$}; % r-1
\node[edgenode,anchor=center] at ($.5*(n1)+.5*(n10)$) {$x_{2} = 0$}; % 1-10
\node[edgenode,anchor=center] at ($.5*(n1)+.5*(n11)$) {$x_{2} = 1$}; % 1-11					
\node[leafnodelabel] at (n0) {$\Treepval{0}{1.000}$ \\ $\treepval{1}{0.000}$}; % 0
\node[leafnodelabel] at (n10) {$\treepval{0}{0.000}$ \\ $\Treepval{1}{1.000}$}; % 10
\node[leafnodelabel] at (n11) {$\Treepval{0}{0.500}$ \\ $\treepval{1}{0.500}$}; % 11
\useasboundingbox ($(root)+(-.25,-2.25)$) rectangle ($(root)+(4.25,2.25)$);
\end{tikzpicture}
		\caption{Classical decision tree $T'_{\mathrm{cl}}$ induced via scikit-learn using a top-down approach with a splitting criterion based on the information entropy.}
		\label{fig:experiments:hard:trees:s}
	\end{subfigure}%
	\caption{Induced decision trees with maximum depth $d=\num{1}$ for the toy data set, \cref{eqn:experiments:XY}, where all quantum computations have been performed on actual quantum hardware. We use the same notation as in \cref{fig:experiments:sim:trees}.}
	\label{fig:experiments:hard:trees}		
\end{figure}
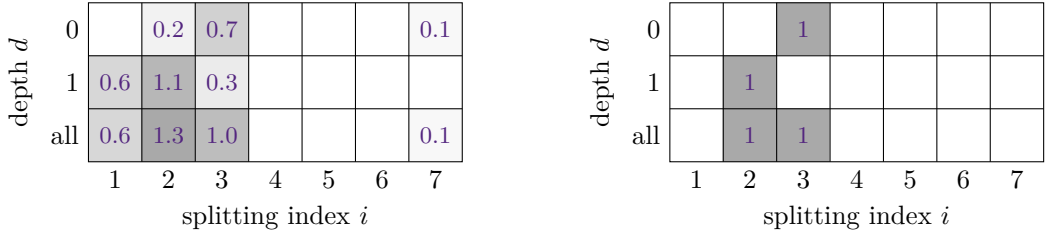
\begin{figure}
	\centering
	\begin{subfigure}[t]{.49\linewidth}
		\centering
		\begin{tikzpicture} 
\pgfplotstableread[col sep=semicolon]{%experiment_backend-1_ibmq_ehningen-features-qt.csv
k;d;c
0;0.000000000;0.000000000
1;0.000000000;0.200000000
2;0.000000000;0.700000000
3;0.000000000;0.000000000
4;0.000000000;0.000000000
5;0.000000000;0.000000000
6;0.000000000;0.100000000
0;1.000000000;0.600000000
1;1.000000000;1.100000000
2;1.000000000;0.300000000
3;1.000000000;0.000000000
4;1.000000000;0.000000000
5;1.000000000;0.000000000
6;1.000000000;0.000000000
0;2.000000000;0.600000000
1;2.000000000;1.300000000
2;2.000000000;1.000000000
3;2.000000000;0.000000000
4;2.000000000;0.000000000
5;2.000000000;0.000000000
6;2.000000000;0.100000000
}\data;
\pgfplotsset{colormap/blackwhite}
\begin{axis} [width=6.5cm,
	scale mode=scale uniformly,
	enlargelimits=false,
	xmin=-.5,xmax=6.5,
	ymin=-.5,ymax=2.5,	
	xtick={0,1,...,6},
	xticklabels={1,2,...,7},
	minor xtick={.5,1.5,...,5.5},
	minor ytick={.5,1.5},
	grid=minor,
	minor grid style={black},
	ytick={0,1,2},	
	yticklabels={0,1,all},
	xlabel={splitting index $i$},
	ylabel={depth $d$},	
	tick style={draw=none},
	point meta=explicit,
	nodes near coords,
	nodes near coords style={font=\small\bfseries,/pgf/number format/fixed,/pgf/number format/fixed zerofill,/pgf/number format/precision=1},
	coordinate style/.condition={meta<0.01}{white},
	coordinate style/.condition={meta>=0.01}{Qv},	
	nodes near coords align={anchor=center},
	colormap={reverse blackwhite}{indices of colormap={\pgfplotscolormaplastindexof{blackwhite},...,0 of blackwhite}},
	point meta min=0, point meta max=3.9, % range * 3
	]
	\addplot [matrix plot, mesh/rows=3, mesh/cols=7] table [x=k,y=d,meta=c] {\data};
\end{axis} 
\end{tikzpicture}
		\caption{Mean splitting feature index distribution for the Q-trees $QT_{\mathrm{qc}}^1,\dots,QT_{\mathrm{qc}}^{10}$.}
		\label{fig:experiments:hard:features:q}
	\end{subfigure}%
	\hfill%
	\begin{subfigure}[t]{.49\linewidth}
		\centering
		\begin{tikzpicture} 
\pgfplotstableread[col sep=semicolon]{%experiment_backend-1_ibmq_ehningen-features-sl.csv
k;d;c
0;0.000000000;0.000000000
1;0.000000000;0.000000000
2;0.000000000;1.000000000
3;0.000000000;0.000000000
4;0.000000000;0.000000000
5;0.000000000;0.000000000
6;0.000000000;0.000000000
0;1.000000000;0.000000000
1;1.000000000;1.000000000
2;1.000000000;0.000000000
3;1.000000000;0.000000000
4;1.000000000;0.000000000
5;1.000000000;0.000000000
6;1.000000000;0.000000000
0;2.000000000;0.000000000
1;2.000000000;1.000000000
2;2.000000000;1.000000000
3;2.000000000;0.000000000
4;2.000000000;0.000000000
5;2.000000000;0.000000000
6;2.000000000;0.000000000

}\data;
\pgfplotsset{colormap/blackwhite}
\begin{axis} [width=6.5cm,
	scale mode=scale uniformly,
	enlargelimits=false,
	xmin=-.5,xmax=6.5,
	ymin=-.5,ymax=2.5,	
	xtick={0,1,...,6},
	xticklabels={1,2,...,7},
	minor xtick={.5,1.5,...,5.5},
	minor ytick={.5,1.5},
	grid=minor,
	minor grid style={black},
	ytick={0,1,2},	
	yticklabels={0,1,all},
	xlabel={splitting index $i$},
	ylabel={depth $d$},	
	tick style={draw=none},
	point meta=explicit,
	nodes near coords,
	nodes near coords style={font=\small\bfseries},
	coordinate style/.condition={meta<1}{white},
	coordinate style/.condition={meta>=1}{Qv},	
	nodes near coords align={anchor=center},
	colormap={reverse blackwhite}{indices of colormap={\pgfplotscolormaplastindexof{blackwhite},...,0 of blackwhite}},
	point meta min=0, point meta max=3, % range * 3
	]
	\addplot [matrix plot, mesh/rows=3, mesh/cols=7] table [x=k,y=d,meta=c] {\data};
\end{axis} 
\end{tikzpicture}
		\caption{Splitting feature index distribution for the classical decision tree $T'_{\mathrm{cl}}$, which is also shown in \cref{fig:experiments:hard:trees:s}.}
		\label{fig:experiments:hard:features:s}
	\end{subfigure}%
	\caption{Distribution of splitting feature indices for considered decision trees in analogy to \cref{fig:experiments:sim:features}.}
	\label{fig:experiments:hard:features}
\end{figure}
We choose the Q-tree with the highest balanced accuracy, \cref{eqn:experiments:metric:bacc}, for demonstration purposes. This ``best'' Q-tree $QT_{\mathrm{qc}}$ is associated with the decision configuration
\begin{align} \label{eqn:experiments:hard:C}
	C_{\mathrm{qc}} = ((3), (2,2))
\end{align}
and shown in \cref{fig:experiments:hard:trees:q}. The structure of this Q-tree is particularly equivalent to the structure of the corresponding classical decision tree $T'_{\mathrm{cl}}$ except for the additional split at depth $d=1$. We compare the previously introduced classification performance metrics of the Q-trees $QT_{\mathrm{qc}}^1,\dots,QT_{\mathrm{qc}}^{10}$ and the classical decision tree $T_{\mathrm{cl}}$. \par
The results are shown in \cref{tab:experiments:hard:scores}. We find that all best metrics of the Q-trees are superior to the corresponding metrics of the classical tree $T'_{\mathrm{cl}}$. Furthermore, the majority of the metrics of the exemplary Q-tree $QT_{\mathrm{qc}}$ ($\mathrm{bac}$, $\mathrm{acc}$, $\mathrm{pre}_0$, $\mathrm{rec}_1$, and $\mathrm{f1}_1$) are better than or equal to the corresponding metrics of the classical tree, whereas the others ($\mathrm{pre}_0$, $\mathrm{rec}_1$, and $\mathrm{f1}_1$) are worse. Due to their similar structure, the only difference between the predictions of $QT_{\mathrm{qc}}$ and $T'_{\mathrm{cl}}$ comes from the path (or bit string) $\bar{x}_{0}\bar{x}_{1}=11$. For the corresponding leaf, the Q-tree predicts $y_1=1$, whereas the classical tree estimates balanced label probabilities that lead to a prediction $y_1=0$ (according to the scikit-learn default behavior). All advantages of $QT_{\mathrm{qc}}$ are therefore a consequence of the noise from the quantum sampling process. Given these facts, the apparent superiority that can be inferred from \cref{tab:experiments:hard:scores} is questionable, and there is also no fundamental reason why our quantum representation should lead to better predictions than the classical representation. Given the simplicity of the problem, this is no surprise. However, our results at least indicate that the quantum representation running on a NISQ device is (for this particular example) not significantly inferior to the classical representation.\par
\begin{table}
	\centering
	\caption{Classification performance metrics, \cref{eqn:experiments:metric:bacc,eqn:experiments:metric:acc,eqn:experiments:metric:pre,eqn:experiments:metric:rec,eqn:experiments:metric:f1}, of the Q-trees $QT_{\mathrm{qc}}^1,\dots,QT_{\mathrm{qc}}^{10}$, the exemplary Q-tree $QT_{\mathrm{qc}}$, \cref{eqn:experiments:hard:C}, and the classical decision tree $T'_{\mathrm{cl}}$, respectively, in analogy to \cref{tab:experiments:sim:scores}. When calculating the mean and standard deviation, we ignore all metrics that are not well-defined due to a division by zero. The best results (with respect to two digits) are highlighted in bold.} \label{tab:experiments:hard:scores}
	\begin{tabular}{lccccccccc}%experiment_backend-1_ibmq_ehningen,experiment_backend-2_ibmq_ehningen.ipynb
		\hline\hline
		& $\mathrm{bac}$ & $\mathrm{acc}$ & $\mathrm{pre}_0$ & $\mathrm{pre}_1$ & $\mathrm{rec}_0$ & $\mathrm{rec}_1$ & $\mathrm{f1}_0$& $\mathrm{f1}_1$ \\
		\hline
		Mean Q-tree metric                & \tnum{0.71 +- 0.11} & \tnum{0.76 +- 0.08} & \tnum{0.74 +- 0.10} & \tnum{0.96 +- 0.11} & \tnum{0.97 +- 0.10} & \tnum{0.45 +- 0.27} & \tnum{0.83 +- 0.04} & \tnum{0.68 +- 0.04} \\		 
		Best Q-tree metric                & \Tnum{0.83} & \Tnum{0.80} & \Tnum{1.00} & \Tnum{1.00} & \Tnum{1.00} & \Tnum{1.00} & \Tnum{0.86} & \Tnum{0.80} \\	
		Q-tree $QT_{\mathrm{qc}}$         & \Tnum{0.83} & \Tnum{0.80} & \Tnum{1.00} & \tnum{0.67} & \tnum{0.67} & \Tnum{1.00} & \tnum{0.80} & \Tnum{0.80} \\		
		Classical tree $T'_{\mathrm{cl}}$ & \tnum{0.75} & \Tnum{0.80} & \tnum{0.75} & \Tnum{1.00} & \Tnum{1.00} & \tnum{0.50} & \Tnum{0.86} & \tnum{0.78} \\	
		\hline\hline
	\end{tabular}
\end{table}
To verify that our tree induction approach is superior to a random guess, we create a set of $\num{10}$ randomly generated decision trees by uniformly drawing decision configurations, \cref{eqn:C}, and subsequently sample with $N=\num{8192}$ measurements to obtain an estimate for the label probability distribution. The resulting a mean and standard deviation (in brackets) for the balanced accuracy, \cref{eqn:experiments:metric:bacc}, over all trees reads $\num{0.51 +- 0.02}$ and is therefore significantly worse than the mean balanced accuracy over all induced trees listed in \cref{tab:experiments:hard:scores}.\par
\begin{figure}
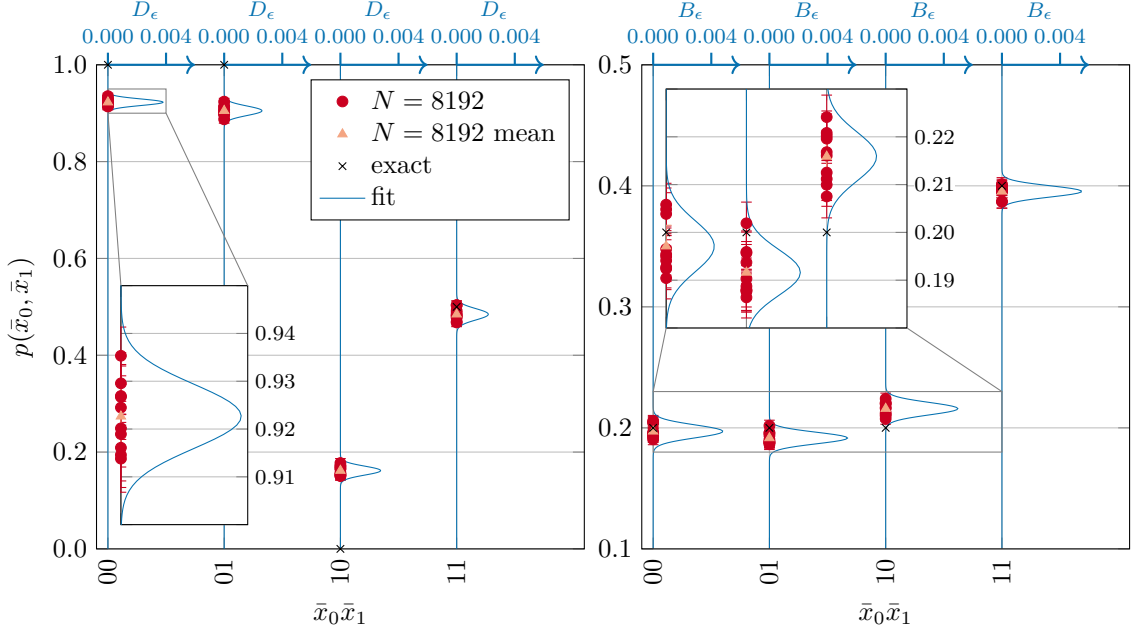

	\centering
	\begin{subfigure}[t]{.49\linewidth}
		\centering
		\includestandalone{fig_experiments-hard-probas-km}
		\caption{Label probabilities from Q-tree samples $p(\bar{x}_{0},\bar{x}_{1})\equiv\hat{p}(y_1 = 0 \,|\, \mathcal{C}_{\nu_2(\bar{x}_{0},\bar{x}_{1})})$, \cref{eqn:pyi:est}, with standard deviation (one sigma) $\stdeva{\hat{p}'(k,m,N)}$, \cref{eqn:app:unc:pp:std}, and from exact calculations $p(\bar{x}_{0},\bar{x}_{1})\equiv p(y_1 = 0 \,|\, \mathcal{C}_{\nu_2(\bar{x}_{0},\bar{x}_{1})})$, \cref{eqn:pn:y}, respectively. The effective probability distributions (``fit'') corresponds to $D_{\epsilon}(k, m; N, p', q)$, \cref{eqn:De}, with the parameters $\mu'_{\epsilon}$ and $\sigma'_{\epsilon}$ shown in \cref{tab:experiments:hard:proba} and is plotted with respect to one of the four separate axes shown on top. An inset plot shows the results for $\bar{x}_{0}\bar{x}_{1}=00$ in more detail.}
		\label{fig:experiments:hard:probas:km}
	\end{subfigure}%
	\hfill%
	\begin{subfigure}[t]{.49\linewidth}
		\centering
		\includestandalone{fig_experiments-hard-probas-mn}
		\caption{Probabilities of the decision set from Q-tree samples $p(\bar{x}_{0},\bar{x}_{1})\equiv\hat{p}(\mathcal{C}_{\nu_2(\bar{x}_{0},\bar{x}_{1})})$, \cref{eqn:pn:est}, with standard deviation (one sigma) $\stdeva{\hat{p}(n,N)}$, \cref{eqn:app:unc:p:std}, and from exact calculations $p(\bar{x}_{0},\bar{x}_{1})\equiv p(\mathcal{C}_{\nu_2(\bar{x}_{0},\bar{x}_{1})})$, \cref{eqn:pn}, respectively. The effective probability distributions (``fit'') corresponds to $B_{\epsilon}( n; N, p )$, \cref{eqn:Be}, with the parameters $\mu_{\epsilon}$ and $\sigma_{\epsilon}$ shown in \cref{tab:experiments:hard:proba} and is plotted with respect to one of the four separate axes shown on top. An inset plot shows the results for $\bar{x}_{0}\bar{x}_{1}=00$, $\bar{x}_{0}\bar{x}_{1}=01$, and $\bar{x}_{0}\bar{x}_{1}=10$ in more detail. We use the same symbols as in \subref{fig:experiments:hard:probas:km}.}
		\label{fig:experiments:hard:probas:mn}
	\end{subfigure}%
	\caption{Sampled probability distributions of the Q-tree $QT_{\mathrm{qc}}$, which is also shown in \cref{fig:experiments:hard:trees:q}, using $N=\num{8192}$ measurements. We use a similar notation as in \cref{fig:experiments:sim:probas}. All quantum computations have been performed on actual quantum hardware. For comparison, we also show the exact probability distributions based on the data proportions as well as estimations for the hardware-induced noise (``fit'') based on the effective probability distributions from \cref{eqn:Be,eqn:De} with the fitted parameters from \cref{tab:experiments:hard:proba}. For the effective probability distributions, a separate axis is provided on top of the plot for each bit string $\bar{x}_{0}\bar{x}_{1}$.}
	\label{fig:experiments:hard:probas}		
\end{figure}

Next, we consider the estimations for the probability distributions for the best Q-tree $QT_{\mathrm{qc}}$ in analogy to our experiments on the simulator. The results are shown in \cref{fig:experiments:hard:probas}. We find that some of the sampled probabilities deviate significantly from the corresponding exact probabilities. We suppose that these deviations result from the hardware-induced noise.\par
To further quantify our observation, we take a closer look at the deviations. For this purpose, we assume that statistical errors of the estimated probability distributions are negligible due to a sufficiently large sampling size in comparison with the hardware-induced noise as discussed in \cref{sec:app:uncertainty}. Consequently, we can make use of the approximations in \cref{eqn:app:unc:Be:eps,eqn:app:unc:De:eps}, respectively. We furthermore presume that the hardware-induced noise, \cref{eqn:app:unc:eps:p,eqn:app:unc:eps:pp}, can be described by a truncated normal distribution $\mathcal{N}_0^1(x; \mu, \sigma)$ within the interval $(0,1)$ with mean $\mu$ and standard deviation $\sigma$, where $x \in (0,1)$. Consequently, we arrive at the effective probability distributions
\begin{align} \label{eqn:Be}
	B_{\epsilon}( n; N, p ) \approx \mathcal{N}_0^1\left(\frac{n}{N}; p+\mu_{\epsilon}, \sigma_{\epsilon}\right)
\end{align}
and
\begin{align} \label{eqn:De}
	D_{\epsilon}(k, m; N, p', q) \approx \mathcal{N}_0^1\left(\frac{k}{m}; p'+\mu'_{\epsilon}, \sigma'_{\epsilon}\right),
\end{align}
respectively. Here we introduce the expectation value shifts $\mu_{\epsilon}$ and $\mu'_{\epsilon}$, which represent hardware-induced systematic errors, and the standard deviations $\sigma_{\epsilon}$ and $\sigma'_{\epsilon}$, which represent hardware-induced statistical uncertainties. These parameters can be obtained by fitting \cref{eqn:Be,eqn:De} to the estimated probability distributions from the measurements.\par
The fitted parameters are listed in \cref{tab:experiments:hard:proba}. The resulting effective probability distributions are also shown in \cref{fig:experiments:hard:probas}. As expected, we find both non-vanishing statistical and systematic errors for all bit strings. The systematic errors are neither strictly positive or strictly negative. The largest overall statistical error $\sigma'_{\epsilon}=\num{0.0117}$ occurs for the bit string $\bar{x}_{0}\bar{x}_{1}=11$, whereas the largest absolute overall systematic error $|\mu'_{\epsilon}|=\num{0.1617}$ occurs for the bit string $\bar{x}_{0}\bar{x}_{1}=10$. However, a discussion of possible physical origins of these hardware-induced errors goes beyond the scope of this manuscript.\par
\begin{table}
	\centering
	\caption{Fitted parameters for the effective probability distributions from \cref{eqn:Be,eqn:De}, respectively. The expectation value shifts $\mu_{\epsilon}$ and $\mu'_{\epsilon}$ represent hardware-induced systematic errors, whereas the standard deviations $\sigma_{\epsilon}$ and $\sigma'_{\epsilon}$ represent hardware-induced statistical uncertainties. The resulting effective probability distributions are also shown in \cref{fig:experiments:hard:probas}.} \label{tab:experiments:hard:proba}
	\begin{tabular}{ccccc}%experiment_backend-3_ibmq_ehningen-appendix.ipynb
		\hline\hline
		$\bar{x}_{0}\bar{x}_{1}$ & $\mu_{\epsilon}$ & $\sigma_{\epsilon}$ & $\mu'_{\epsilon}$ & $\sigma'_{\epsilon}$ \\
		\hline
		00 & \num[retain-explicit-plus]{-0.0029} & \num{0.0053} & \num[retain-explicit-plus]{-0.0773} & \num{0.0067} \\
		01 & \num[retain-explicit-plus]{-0.0084} & \num{0.0048} & \num[retain-explicit-plus]{-0.0950} & \num{0.0098} \\
		10 & \num[retain-explicit-plus]{+0.0159} & \num{0.0051} & \num[retain-explicit-plus]{+0.1617} & \num{0.0092} \\
		11 & \num[retain-explicit-plus]{-0.0046} & \num{0.0047} & \num[retain-explicit-plus]{-0.0154} & \num{0.0117} \\			
		\hline\hline
	\end{tabular}
\end{table}
\section{Conclusions} \label{eqn:conclusions}
We have proposed a quantum representation of classification trees in form of a quantum circuit, called Q-tree, which allows us to perform probabilistic tree traversals in the sense that it can sample from the corresponding probability distribution of the tree leaves. The proposed circuit consists of $d+k+m$ qubits and $\BigO{2^{k+m} + d 2^d}$ gates (where $d$ represents the tree depth, $k$ the number of binary features and $m$ the number of binary labels). Furthermore, we have also discussed how the sampled data can be used to feed a fitness function of a genetic algorithm to induce Q-trees in form of parameterized circuits. In addition, we have provided three approaches how Q-trees can be used for the prediction of labels for query data. First, by gathering samples and estimating the corresponding leaf probability distributions. And second, by modifying the circuit such that it directly draws samples from the distribution of leaves that a query distribution reaches, which allows to predict the labels for uncertain queries. As a third method, we have proposed an on-demand sampling approach to perform Q-tree predictions with a constant number of classical memory slots, independent of the tree depth. Using simple data sets, we have studied the induction and prediction capabilities of Q-trees in comparison with a classical benchmark using both a simulator and actual IBM quantum hardware. To our knowledge, this is the first realization of a decision tree classifier on a quantum device. Based on our experiments, we have found that the performance of Q-trees is comparable to the considered classical benchmark.\par
However, we have also found that the field of application of Q-trees is very limited on NISQ devices and hardware-induced noise can have a major impact on the quality of the results. Investigation of the hardware and data requirements necessary for the application of Q-trees could be the subject of future work. In addition, explicitly modifying the Q-trees to better fit the demands of the hardware could be one way to further improve the results and broaden the field of application. For example, we have mentioned how the circuit complexity can be reduced by introducing ancilla qubits (for both the data encoding and the structure encoding). An application of these approaches could lead to shallower Q-trees, which might be more suitable for NISQ devices. Furthermore, our proposed on-demand sampling approach can also be modified to mitigate hardware-induced noise by employing the approach presented in \cref{sec:quantum representation:tree predictions}. Further research is necessary to evaluate this topic in more detail.\par
There are various possibilities to extend or alter our proposed approach for different applications. For example, it can be generalized by considering decision trees with missing splits by an appropriate modification of the compressed decision configuration. Another obvious extension is the consideration of non-binary features and labels, respectively, which requires a suitable encoding with possibly more than one qubit per feature and label. Our approach could also be modified to allow decision trees for regression problems by employing a suitable probabilistic representation of the problem. All of these conceptional ideas can be considered as possible research directions for further work. Moreover, instead of single decision trees, another extension of our approach are ensembles of Q-trees in the sense of random forests \cite{breiman2001}. In \cref{sec:app:random forests}, we briefly discuss a possible strategy for such an extension.\par 
In this work, we have only discussed heuristic tree induction methods. We have specifically proposed a genetic algorithm in \cref{sec:app:genetic induction algorithm}, for which we have also suggested possible variations. However, quantum computers can in principle also be used to induce global optimal decision trees \cite{bertsimas2017}. For this purpose, the corresponding mixed-integer optimization problem can for example be converted \cite{chang2020} into an unconstrained binary quadratic programming problem \cite{kochenberger2014}, which can then be solved with an appropriate quantum algorithm \cite{zahedinejad2017}. We consider such or a similar approach as another promising research direction for further studies.

\section*{Acknowledgements}
We thank Michael Helmling, Michael Bortz, and Peter Klein for informative discussion and acknowledge the use of IBM Quantum services for this work. Part of this work has been funded by the Competence Center Quantum Computing Rhineland-Palatinate.

\section*{Authors' contributions}
Raoul~Heese developed the theoretical formalism, designed and conducted the experiments, and wrote the manuscript. Patricia~Bickert contributed to the evaluation of the experiments and the writing. Astrid~Elisa~Niederle contributed to the overall research process.

\appendix
\section{Compressed decision configuration} \label{sec:app:compressed decision configuration}
In this appendix section, we discuss the bijective transformation, \cref{eqn:EC}, between the decision configuration $E$, \cref{eqn:E}, and the compressed decision configuration $C$, \cref{eqn:C}, where we implicitly assume that $E$ fulfills \cref{eqn:E:condition}. The intuition behind the compressed decision configuration is that each element $c^i_j$ represents a swap of the elements of a $k$-dimensional index vector
\begin{align} \label{eqn:app:J}
	\mathbf{J} \equiv (1,\dots,k) \in \{1,\dots,k\}^k
\end{align}
such that the actual feature index $e^i_j$ corresponds to the $(i+1)$th element of this vector. We therefore call $c^i_j$ a \emph{swap index} and $\mathbf{c}^i$ the vector of \emph{layer swaps}. For the root, there exists a direct correspondence
\begin{align} \label{eqn:app:EC:0}
	c^0_1 = e^0_1
\end{align}
between the two parameterizations.\par
Furthermore, for any depth $l>0$, one has the transformation rule
\begin{align} \label{eqn:app:EC:ce}
	c^l_{\nu_l(\bar{x}_{0}\cdots\bar{x}_{l-1})} \equiv w(\mathbf{T}^0(l-1;\bar{x}_{0}\cdots\bar{x}_{l-1}), e^l_{\nu_l(\bar{x}_{0}\cdots\bar{x}_{l-1})})
\end{align}
for $E \mapsto C$,  where we denote the index of a value $n \in \{1,\dots,k\}$ in a vector $\mathbf{I} \in \{1,\dots,k\}^k$ with
\begin{align}
	w(\mathbf{I}, n) \equiv \argmax_{i \in \{1,\dots,k\}} \vert I_i - n \vert
\end{align}
and require that $n$ exists exactly once in $\mathbf{I}$. Moreover, we make use of the vector transformation
\begin{align} \label{eqn:app:C:T}
	\mathbf{T}^{\omega}(u;\bar{x}_{0}\cdots\bar{x}_{l-1}) \equiv \left( \composition{v=0}{u} \mathbf{V}^{\omega}(\cdot, v; \bar{x}_{0}\cdots\bar{x}_{l-1}) \right)(\mathbf{J})
\end{align}
with $\omega \in \{0,1\}$ and $u \in \{l-1,l\}$. It contains the composition
\begin{align}
	\composition{v=0}{u} \mathbf{V}^{\omega} (\cdot, v) \equiv \mathbf{V}^{\omega} (\cdot, u) \circ \cdots \circ \mathbf{V}^{\omega} (\cdot, 0)
\end{align}
of
\begin{align} \label{eqn:app:C:V}
	\mathbf{V}^{\omega}(\mathbf{I}, v) \equiv \mathbf{V}^{\omega}(\mathbf{I}, v; \bar{x}_{0}\cdots\bar{x}_{l-1}) \equiv \begin{cases} \mathbf{S}(\mathbf{I}, v+1, w(\mathbf{I}, e^v_{\nu'_v(\bar{x}_{0}\cdots\bar{x}_{l-1})})) & \text{for}\,\, \omega = 0 \\ \mathbf{S}(\mathbf{I}, v+1, c^v_{\nu'_v(\bar{x}_{0}\cdots\bar{x}_{l-1})}) & \text{for}\,\, \omega = 1 \end{cases}
\end{align}
with $0 \leq v \leq l$. The latter is based on the permutation (element swap)
\begin{align}
	\mathbf{S}(\mathbf{I},l,n) = \begin{cases} (I_1,\dots,I_{l-1},I_n,I_{l+1},\dots,I_{n-1},I_l,I_{n+1},\dots,I_k) & \text{for}\,\, l<n \\ \mathbf{I} & \text{for}\,\, n=l \\ (I_1,\dots,I_{n-1},I_l,I_{n+1},\dots,I_{l-1},I_n,I_{l+1},\dots,I_k) & \text{for}\,\, l>n \end{cases}
\end{align}
of $\mathbf{I}$ with $l,n \in \{1,\dots,k\}$. We also recall \cref{eqn:nu} and introduce
\begin{align}
	\nu'_v(\bar{x}_{0}\cdots\bar{x}_{l-1}) \equiv \begin{cases} \nu_v(\bar{x}_{0}\cdots\bar{x}_{v-1}) & \text{for}\,\, v \geq 1 \\ 1 & \text{for}\,\, v = 0 \end{cases}
\end{align}
with $0 \leq v \leq l$ to identify the nodes along the subpath (with depth $v < l$) of a node of depth $l$, which is identified by a bit string $\bar{x}_{0}\cdots\bar{x}_{l-1}$. In particular, one has $\nu'_l(\bar{x}_{0}\cdots\bar{x}_{l-1}) = \nu_l(\bar{x}_{0}\cdots\bar{x}_{l-1})$.\par
Conversely, the inverse transformation rule $C \mapsto E$ for any depth $l>0$ reads
\begin{align} \label{eqn:app:EC:ec}
	e^l_{\nu_l(\bar{x}_{0}\cdots\bar{x}_{l-1})} = T^1_{l+1}(l;\bar{x}_{0}\cdots\bar{x}_{l-1})
\end{align}
such that the feature indices $e^i_j$ are defined by the corresponding elements of \cref{eqn:app:C:T}. These elements represent a permutation of the elements of the index vector $\mathbf{J}$, \cref{eqn:app:J}. We briefly illustrate the transformations of $\mathbf{J}$ in \cref{fig:app:sample-tree-encs}.\par
\begin{figure}
	\begin{center}
		\begin{tikzpicture}
\coordinate (n00) at (0,0);
\coordinate (n10) at ($(n00)+(-3,-.5)$);
\coordinate (n11) at ($(n00)+(3,-.5)$);
\coordinate (n20) at ($(n10)+(-1.25,-.5)$);
\coordinate (n21) at ($(n10)+(1.25,-.5)$);
\coordinate (n22) at ($(n11)+(-1.25,-.5)$);
\coordinate (n23) at ($(n11)+(1.25,-.5)$);
\coordinate (n30) at ($(n20)+(-.5,-.5)$);
\coordinate (n31) at ($(n20)+(.5,-.5)$);
\coordinate (n32) at ($(n21)+(-.5,-.5)$);
\coordinate (n33) at ($(n21)+(.5,-.5)$);
\coordinate (n34) at ($(n22)+(-.5,-.5)$);
\coordinate (n35) at ($(n22)+(.5,-.5)$);
\coordinate (n36) at ($(n23)+(-.5,-.5)$);
\coordinate (n37) at ($(n23)+(.5,-.5)$);
%
%\draw[thick,blue,dotted] (-4,-.25) -- (5,-.25);
%\draw[thick,blue,dotted] (-4,-.75) -- (5,-.75);
%\draw[thick,blue,dotted] (-4,-1.25) -- (5,-1.25);
%
\draw[thick,<-,>=latex] (n00) -- (n10) node [draw,fill=white,circle,midway,font=\tiny,inner sep=1pt] {$0$};
\draw[thick,<-,>=latex] (n00) -- (n11) node [draw,fill=white,circle,midway,font=\tiny,inner sep=1pt] {$1$};
\draw[thick,<-,>=latex] (n10) -- (n20) node [draw,fill=white,circle,midway,font=\tiny,inner sep=1pt] {$0$};
\draw[thick,<-,>=latex] (n10) -- (n21) node [draw,fill=white,circle,midway,font=\tiny,inner sep=1pt] {$1$};
\draw[thick,<-,>=latex] (n11) -- (n22) node [draw,fill=white,circle,midway,font=\tiny,inner sep=1pt] {$0$};
\draw[thick,<-,>=latex] (n11) -- (n23) node [draw,fill=white,circle,midway,font=\tiny,inner sep=1pt] {$1$};
\draw[thick,dotted,<-,>=latex] (n20) -- (n30);% node [draw,fill=white,circle,midway,font=\tiny,inner sep=1pt] {$0$};
\draw[thick,dotted,<-,>=latex] (n20) -- (n31);% node [draw,fill=white,circle,midway,font=\tiny,inner sep=1pt] {$1$};
\draw[thick,dotted,<-,>=latex] (n21) -- (n32);% node [draw,fill=white,circle,midway,font=\tiny,inner sep=1pt] {$0$};
\draw[thick,dotted,<-,>=latex] (n21) -- (n33);% node [draw,fill=white,circle,midway,font=\tiny,inner sep=1pt] {$1$};
\draw[thick,dotted,<-,>=latex] (n22) -- (n34);% node [draw,fill=white,circle,midway,font=\tiny,inner sep=1pt] {$0$};
\draw[thick,dotted,<-,>=latex] (n22) -- (n35);% node [draw,fill=white,circle,midway,font=\tiny,inner sep=1pt] {$1$};
\draw[thick,dotted,<-,>=latex] (n23) -- (n36);% node [draw,fill=white,circle,midway,font=\tiny,inner sep=1pt] {$0$};
\draw[thick,dotted,<-,>=latex] (n23) -- (n37);% node [draw,fill=white,circle,midway,font=\tiny,inner sep=1pt] {$1$};
\node[draw,circle,fill=black,inner sep=0pt,minimum width=.1cm] at (n00) {};
\node[draw,circle,fill=black,inner sep=0pt,minimum width=.1cm] at (n10) {};
\node[draw,circle,fill=black,inner sep=0pt,minimum width=.1cm] at (n11) {};
\node[draw,circle,fill=black,inner sep=0pt,minimum width=.1cm] at (n20) {};
\node[draw,circle,fill=black,inner sep=0pt,minimum width=.1cm] at (n21) {};
\node[draw,circle,fill=black,inner sep=0pt,minimum width=.1cm] at (n22) {};
\node[draw,circle,fill=black,inner sep=0pt,minimum width=.1cm] at (n23) {};
\node[anchor=south,fill=white,inner sep=0pt,outer sep=0pt,yshift=.15cm,align=center] at (n00) {$\mathbf{J}$};
\node[anchor=south east,fill=white,inner sep=0pt,outer sep=0pt,yshift=.1cm,align=center] at (n10) {$\mathbf{T}^{\alpha}(u;0)$};
\node[anchor=south west,fill=white,inner sep=0pt,outer sep=0pt,yshift=.1cm,align=center] at (n11) {$\mathbf{T}^{\alpha}(u;1)$};
\node[anchor=south east,fill=white,inner sep=0pt,outer sep=0pt,yshift=.1cm,align=center] at (n20) {$\mathbf{T}^{\alpha}(u;00)$};
\node[anchor=south west,fill=white,inner sep=0pt,outer sep=0pt,yshift=.1cm,align=center] at (n21) {$\mathbf{T}^{\alpha}(u;01)$};
\node[anchor=south east,fill=white,inner sep=0pt,outer sep=0pt,yshift=.1cm,align=center] at (n22) {$\mathbf{T}^{\alpha}(u;10)$};
\node[anchor=south west,fill=white,inner sep=0pt,outer sep=0pt,yshift=.1cm,align=center] at (n23) {$\mathbf{T}^{\alpha}(u;11)$};
\end{tikzpicture}
		\caption{Illustration of the transformations $\mathbf{T}^{\alpha}(u;\bar{x}_{0}\cdots\bar{x}_{l-1})$, \cref{eqn:app:C:T}, of the index vector $\mathbf{J}$, \cref{eqn:app:J}, by mapping them to the corresponding nodes of a decision tree $T$, \cf \cref{fig:sample-tree}. In each layer, a permutation of the vector elements according to \cref{eqn:app:C:V} is performed (with the arrows indicating dependencies on previous permutations). The compressed decision configuration $C$, \cref{eqn:C}, describes the corresponding index swaps in form of a sequence of mutually independent swap indices $c^i_j$, \cref{eqn:c}, with $j \in \{1,\dots,2^i\}$. In contrast, the decision configuration $E$, \cref{eqn:E}, directly describes the feature indices of splits in form of mutually dependent layer decisions $e^i_j$, \cref{eqn:e}, which have to fulfill \cref{eqn:E:condition}. The two representations can be converted into each other, \cref{eqn:EC}, according to \cref{eqn:app:EC:0,eqn:app:EC:ce,eqn:app:EC:ec}.} \label{fig:app:sample-tree-encs}
	\end{center}
\end{figure}
Summarized, a decision tree $T$ can be parameterized by either $E$, \cref{eqn:E,eqn:E:condition}, or $C$, \cref{eqn:C}. The two representations can be converted into each other, \cref{eqn:EC}, according to \cref{eqn:app:EC:0,eqn:app:EC:ce,eqn:app:EC:ec}.

\section{Genetic induction algorithm} \label{sec:app:genetic induction algorithm}
In the present appendix section, we describe a concrete realization of a genetic algorithm for the induction of a decision tree $T$, \cref{eqn:induction:C}, as introduced in \cref{sec:classical representation:tree induction}. For this purpose, we assume that a constant maximum depth $d$ is given for the tree to be induced. Furthermore, one can either choose to use the classical or the quantum representation of the tree $T$ as described in \cref{sec:classical representation,sec:quantum representation}, respectively.\par
The proposed genetic algorithm follows the typically used ``traditional'' structure and makes use of well-known genetic operators \cite{katoch2021}. It is outlined in \cref{alg:app:ga} and particularly depends on the hyperparameters
\begin{align} \label{eqn:app:ga:theta}
	\theta \equiv ( P, G, t, p_{\mathrm{c}}, p_{\mathrm{m}}, p_{\mathrm{a}}, f_{\mathrm{ent}}, f_{\mathrm{bac}})
\end{align}
consisting of the total population size $P \in \mathbb{N}$, the number of iterations (or generations) $G \in \mathbb{N}$, the selection subset size $t \in \mathbb{N}$, the probability of mating two individuals $p_{\mathrm{c}} \in [0,1]$, the probability of mutating an individual $p_{\mathrm{m}} \in [0,1]$, the probability for each attribute to be mutated independently $p_{\mathrm{a}} \in [0,1]$, and the fitness function weights $f_{\mathrm{ent}}, f_{\mathrm{bac}} \in \mathbb{R}_{0+}$. We denote a population of $P' \in \mathbb{N}$ chromosomes (\ie, candidate solutions) of the form defined in \cref{eqn:ga:sol} by the $P'$-tuple $\chi \in \mathcal{S}^{P'}$, where we recall the solution domain $\mathcal{S}$, \cref{eqn:ga:S}. The chosen maximum depth $d$ consequently defines the dimensionality of the chromosomes. Furthermore, we denote the $i$th chromosome by $\chi^{P'}_i \in \mathcal{S}$ with $i \in \{1,\dots,P'\}$.\par
\begin{algorithm}
	\caption{Genetic tree induction algorithm} \label{alg:app:ga}
	\begin{algorithmic}[1]
		\Function{GeneticGrow}{$\theta$}
		% intitalize
		\State $\chi^{\mathrm{best}} \gets \Call{Initalize}{1}$
		\If{$P > 1$}
		% intitalize
		\State $\chi \gets \Call{Initalize}{P}$
		\State $g \gets 0$
		% iterate
		\While{$g < G$}
		% best
		\State $\chi^{\mathrm{best}}_1 \gets \Call{Best}{\chi, f_{\mathrm{ent}}, f_{\mathrm{bac}}}$
		% select
		\State $\chi^{\mathrm{gen}} \gets \Call{Select}{\chi, t, P-1, f_{\mathrm{ent}}, f_{\mathrm{bac}}}$
		% cross
		\ForAll{$u \in \{ 2j \,|\, j \in \mathbb{N} \land 2 \leq 2j \leq P-1 \}$}
		\ForAll{$v \in \{ 2j-1 \,|\, j \in \mathbb{N} \land 1 \leq 2j-1 \leq P-1 \}$}
		\State $\chi^{\mathrm{gen}}_u, \chi^{\mathrm{gen}}_v \gets \Call{Crossover}{\chi^{\mathrm{gen}}_u, \chi^{\mathrm{gen}}_v, p_{\mathrm{c}}}$
		\EndFor
		\EndFor
		% mutate
		\ForAll{$u \in \{ 1 , \dots, P-1 \}$}
		\State $\chi^{\mathrm{gen}}_u \gets \Call{Mutate}{\chi^{\mathrm{gen}}_u, p_{\mathrm{m}}, p_{\mathrm{a}}}$
		\EndFor
		% rejoin
		\State $\chi \gets \chi^{\mathrm{gen}} \cup \chi^{\mathrm{best}}$
		\State $g \gets g + 1$ \label{alg:app:ga:gloop}
		\EndWhile
		\Else
		\State $\chi \gets \chi^{\mathrm{best}}$
		\EndIf
		% end
		\State \Return $\Call{Best}{\chi, f_{\mathrm{ent}}, f_{\mathrm{bac}}}$
		\EndFunction
	\end{algorithmic}
\end{algorithm}
A central component of the algorithm is the fitness function, which we define as
\begin{align} \label{eqn:app:ga:F}
	F_{\mathrm{cl/qm}} \equiv F_{\mathrm{cl/qm}}(f_{\mathrm{ent}}, f_{\mathrm{bac}}) \equiv f_{\mathrm{ent}} F^{\mathrm{ent}}_{\mathrm{cl/qm}} + f_{\mathrm{bac}} F^{\mathrm{bac}}_{\mathrm{cl/qm}}.
\end{align}
It consists of two parts weighted by $f_{\mathrm{ent}}$ and $f_{\mathrm{bac}}$, respectively. The first part
\begin{align} \label{eqn:app:ga:F:ent}
	F^{\mathrm{ent}}_{\mathrm{cl/qm}} \equiv F^{\mathrm{ent}}_{\mathrm{cl/qm}}[ p_{\mathrm{cl/qm}}(\mathcal{C}_{\nu}), p_{\mathrm{cl/qm}}(y_i \,|\, \mathcal{C}_{\nu})] \equiv \frac{-1}{m} \sum_{\nu=1}^{2^d} p_{\mathrm{cl/qm}}(\mathcal{C}_{\nu}) \sum_{i=1}^{m} S[p_{\mathrm{cl/qm}}(y_i \,|\, \mathcal{C}_{\nu})]
\end{align}
is a measure for the average entropy of the probability distributions of the leaves of the tree, where the first sum iterates over the indices $\nu$ of all leaves of the tree $T$, \cref{eqn:nu}, with the corresponding set of conditions $\mathcal{C}_{\nu}$, \cref{eqn:Cnu}. Here we introduce the information entropy \cite{shannon1948}
\begin{align} \label{eqn:app:ga:Spy}
	S[p(y)] \equiv - p(y=0) \log_2 p(y=0) - p(y=1) \log_2 p(y=1)
\end{align}
of a probability distribution $p(y)$ with support $\mathbb{B}$. The second part
\begin{align} \label{eqn:app:ga:F:acc}
	F^{\mathrm{bac}}_{\mathrm{cl/qm}} & \equiv F^{\mathrm{bac}}_{\mathrm{cl/qm}}(\mathbf{D}_{\mathrm{train}}) \equiv \frac{1}{2 m} \sum_{i=1}^m \sum_{b=0}^1 \frac{\mathrm{t}^b_{\mathrm{cl/qm}}(i,\mathbf{D}_{\mathrm{train}})}{\mathrm{t}^b_{\mathrm{cl/qm}}(i,\mathbf{D}_{\mathrm{train}}) + \mathrm{f}^b_{\mathrm{cl/qm}}(i,\mathbf{D}_{\mathrm{train}})}
\end{align}
represents the mean balanced accuracy in accordance with \cref{eqn:experiments:metric:bacc}. Here we make use of the abbreviations
\begin{align}
	\mathrm{t}^b_{\mathrm{cl/qm}}(i,\mathbf{D}_{\mathrm{train}}) \equiv \vert\{ \mathbf{d} \,|\, (\mathbf{x}^q,\mathbf{y}) = \mathbf{d} \in \mathrm{t}_{\mathrm{cl/qm}}(\mathbf{D}_{\mathrm{train}}) \land y_i = b \}\vert
\end{align}
and
\begin{align}
	\mathrm{f}^b_{\mathrm{cl/qm}}(i,\mathbf{D}_{\mathrm{train}}) \equiv \vert\{ \mathbf{d} \,|\, (\mathbf{x}^q,\mathbf{y}) = \mathbf{d} \in \mathrm{f}_{\mathrm{cl/qm}}(\mathbf{D}_{\mathrm{train}}) \land y_i = b \}\vert,
\end{align}
which are based on
\begin{align}
	\mathrm{t}_{\mathrm{cl/qm}}(i,\mathbf{D}_{\mathrm{train}}) \equiv \vert\{ \mathbf{d} \,|\, (\mathbf{x}^q,\mathbf{y}) = \mathbf{d} \in \mathbf{D}_{\mathrm{train}} \land y_i = \hat{y}_{\mathrm{cl/qm}}(i,\mathbf{x}^q) \}\vert
\end{align}
and
\begin{align}
	\mathrm{f}_{\mathrm{cl/qm}}(i,\mathbf{D}_{\mathrm{train}}) \equiv \vert\{ \mathbf{d} \,|\, (\mathbf{x}^q,\mathbf{y}) = \mathbf{d} \in \mathbf{D}_{\mathrm{train}} \land y_i \neq \hat{y}_{\mathrm{cl/qm}}(i,\mathbf{x}^q) \}\vert,
\end{align}
respectively, where $b \in \mathbb{B}$ and $i \in \{1,\dots,m\}$.\par
Depending on whether the classical or the quantum representation of the tree $T$ is considered, one has
\begin{align} \label{eqn:app:ga:F:clqm}
	F_{\mathrm{cl/qm}} \equiv \begin{cases} F(\mathbf{C}, \mathbf{D}_{\mathrm{train}}) & \text{for the classical representation} \\ \hat{F}(b_1,\dots,b_N,\mathbf{D}_{\mathrm{train}}) & \text{for the quantum representation} \end{cases}
\end{align}
with the exact classical fitness function $F(\mathbf{C}, \mathbf{D}_{\mathrm{train}})$, \cref{eqn:ga:F}, and the measured fitness function from the quantum representation $\hat{F}(b_1,\dots,b_N,\mathbf{D}_{\mathrm{train}})$, \cref{eqn:ga:F:q}, respectively. Furthermore, one has
\begin{align}
	p_{\mathrm{cl/qm}}(\mathcal{C}_{\nu}) \equiv \begin{cases} p(\mathcal{C}_{\nu}) & \text{for the classical representation} \\ \hat{p}(\mathcal{C}_{\nu}) & \text{for the quantum representation} \end{cases}
\end{align}
for the probability distribution of reaching a certain node, where we recall \cref{eqn:pn,eqn:pn:est}, and
\begin{align}
	p_{\mathrm{cl/qm}}(y_i \,|\, \mathcal{C}_{\nu}) \equiv \begin{cases} p(y_i \,|\, \mathcal{C}_{\nu}) & \text{for the classical representation} \\ \hat{p}(y_i \,|\,\mathcal{C}_{\nu}) & \text{for the quantum representation} \end{cases}
\end{align}
for the probability distribution of the $i$th label with $i \in \{1,\dots,m\}$, where we recall \cref{eqn:pn:yi,eqn:pyi:est}, respectively. Finally, one has
\begin{align}
	\hat{y}_{\mathrm{cl/qm}}(i, \mathbf{x}^q) \equiv \argmax_{y_i} \begin{cases} p(y_i \,|\, \mathcal{C}_{q_d(\mathbf{x}^q)}) & \text{for the classical representation} \\ \hat{p}_{\mathrm{cl}}(y_i \,|\, \mathbf{x}^q) & \text{for the quantum representation} \end{cases}
\end{align}
for the prediction of the $i$th label according to \cref{eqn:pyi:pred:argmax,eqn:pyi:pred:cl:argmax}, respectively.\par
The information entropy, \cref{eqn:app:ga:Spy}, attains its smallest value $0$ for $p(y=0) \in \{0,1\}$ and its largest value $1$ for $p(y=0) = \frac{1}{2}$. The choice $p(y) = p_{\mathrm{cl/qm}}(y_i \,|\, \mathcal{C}_{\nu})$ therefore leads to the following implication: Since a smaller information entropy corresponds to a less equal partitioning of data with different labels (in the sense of a smaller variance), it indicates a more favorable splitting decision with respect to its data partitioning capabilities.\par
In particular, the information entropy of two simultaneous events is no more than the sum of the information entropies of each individual event \cite{cover2012}. A splitting decision can therefore not increase the information entropy such that
\begin{equation}
	S[p(y_i)] \geq \sum_{\nu=1}^{2^d} p(\mathcal{C}_{\nu}) S[p(y_i|\mathcal{C}_{\nu})]
\end{equation}
holds true.\par
This measure is in particular closely related to the well-known information gain criterion \cite{maimon2014} for the induction of decision trees. Maximization of $F^{\mathrm{ent}}_{\mathrm{cl/qm}}$ is equivalent to finding the minimum weighted average of the information entropy in all leaves over all label dimensions and can therefore be considered as a measure for the partitioning capabilities of all splitting decisions in the tree. An increased data partitioning is favorable since it represents a more conclusive tree.\par
On the other hand, the balanced accuracy, \cref{eqn:app:ga:F:acc}, is a measure of how well the splitting decisions are able to explain the training data $\mathbf{D}_{\mathrm{train}}$. In the fitness function, we combine both the conclusiveness \cref{eqn:app:ga:F:ent} and the accuracy \cref{eqn:app:ga:F:acc} to a combined metric for the quality of the splitting decisions in the tree (defined by $\mathbf{C}$ and $\mathbf{D}_{\mathrm{train}}$ or $b_1,\dots,b_N$). By tuning the (non-negative) weights $f_{\mathrm{ent}}$ and $f_{\mathrm{bac}}$, the influence of the two parts can be controlled for a specific use case. For the sake of convenience, one can also set $f_{\mathrm{bac}} = 1-f_{\mathrm{ent}}$ to eliminate one hyperparameter.\par
\Cref{alg:app:ga} contains the following functions:
\begin{itemize}
	\item $\Call{Initalize}{P'}$: $P'$ chromosomes are created by drawing them uniformly from an appropriate set of possible values, \cref{eqn:ga:S}. The chromosomes are returned as a $P'$-tuple, where $P' \in \{1, P\}$.
	\item $\Call{Best}{\chi, f_{\mathrm{ent}}, f_{\mathrm{bac}}}$: Return the chromosome in $\chi$ with the best fitness, \cref{eqn:app:ga:F}.
	\item $\Call{Select}{\chi, t, P-1, f_{\mathrm{ent}}, f_{\mathrm{bac}}}$: Selection of $P-1$ chromosomes with the \emph{tournament selection} method. Specifically, $t$ chromosomes are drawn uniformly from $\chi$ and the one with the best fitness, \cref{eqn:app:ga:F}, is selected. This process is repeated $P-1$ times and the $(P-1)$-tuple of selected chromosomes $\chi^{\mathrm{gen}}$ is returned.
	\item $\Call{Crossover}{\chi^{\mathrm{gen}}_u, \chi^{\mathrm{gen}}_v, p_{\mathrm{c}}}$: Uniformly draw a value $r$ from $[0,1]$ and perform a \emph{single-point crossover} if $r \leq p_{\mathrm{c}}$, otherwise return the unmodified originals $\chi^{\mathrm{gen}}_u$ and $\chi^{\mathrm{gen}}_v$. The single-point crossover begins with uniformly drawing an index $i$ of the chromosome attributes from $\{2,\dots,2^d-1\}$ at which $\chi^{\mathrm{gen}}_u$ and $\chi^{\mathrm{gen}}_v$ are each splitted into two segments. Then, two new chromosomes are created. The first corresponds to a concatenation of the first segment of $\chi^{\mathrm{gen}}_u$ and the second segment of $\chi^{\mathrm{gen}}_v$, whereas the second corresponds to a concatenation of the first segment of $\chi^{\mathrm{gen}}_v$ and the second segment of $\chi^{\mathrm{gen}}_u$, respectively. These two copies are returned.
	\item $\Call{Mutate}{\chi^{\mathrm{gen}}_u, p_{\mathrm{m}}, p_{\mathrm{a}}}$: Uniformly draw a value $r$ from $[0,1]$ and perform a \emph{uniform mutation} if $r \leq p_{\mathrm{m}}$, otherwise return the unmodified original $\chi^{\mathrm{gen}}_u$. The uniform mutation begins with uniformly drawing a value $r_i$ from $[0,1]$ for each index $i \in \{1,\dots,2^d\}$ of the chromosome $\chi^{\mathrm{gen}}_u$. Subsequently, it is checked whether $r_i \leq p_{\mathrm{a}}$ is fulfilled for all indices and if yes, the corresponding attribute of the chromosome is replaced by a uniformly drawn value from the set of possible values, \cref{eqn:ga:S}. The mutated chromosome is returned.
\end{itemize}
By keeping track of the fittest chromosome in $\chi^{\mathrm{best}} \in \mathcal{S}^1$, we ensure that best fitness in each generation cannot be worsened. This approach is also known as \emph{elitism} \cite{saini2017}. The returned result of $\Call{GeneticGrow}{\theta}$ is a chromosome that maximizes $F_{\mathrm{cl/qm}}$ in the sense of \cref{eqn:ga:opt}.\par
The evaluation of the fitness function $F_{\mathrm{cl/qm}}$ in $\Call{Select}{\chi, t, P-1, f_{\mathrm{ent}}, f_{\mathrm{bac}}}$ and $\Call{Best}{\chi, f_{\mathrm{ent}}, f_{\mathrm{bac}}}$, respectively, implicitly depends on $\mathbf{D}_{\mathrm{train}}$ and possibly $b_1,\dots,b_N$, depending on whether the classical or the quantum representation of the tree $T$ is considered, \cref{eqn:app:ga:F:clqm}. In particular, $\hat{F}(b_1,\dots,b_N,\mathbf{D}_{\mathrm{train}})$ represents a noisy fitness function. To handle this case, we perform an evaluation of $\hat{F}(b_1,\dots,b_N,\mathbf{D}_{\mathrm{train}})$ and assign the fitness to the corresponding chromosome on its initialization or whenever it is affected by a crossover or a mutation. In addition, we average the fitness values over identical chromosomes in $\chi$ at the end of each iteration (\ie, after \cref{alg:app:ga:gloop}) and re-assign the mean fitness to the associated chromosomes. By doing so we can reduce the influence of statistical errors on the final outcome. The on-demand sampling proposed in \cref{sec:quantum representation:tree predictions} can also be used to reduce the classical memory required for the evaluation of the quantum fitness function.\par
The algorithm presented here can be varied in many different ways, which may lead to better or worse results depending on the specific application at hand. For example, instead of having a constant maximum depth $d$, a variable maximum depth can be achieved by considering chromosomes of different length, which can each be assigned up to a specified maximum depth. Furthermore, instead of optimizing the tree as a whole, an iterative layer-by-layer optimization could be performed by optimizing only the decision configuration of the current layer from depth $0$ to depth $d-1$ (or up to a variable depth). Finally, the fitness function $F_{\mathrm{cl/qm}}$ can be modified in various ways, for example by incorporating another common splitting criterion like the Gini index \cite{maimon2014}. One could also consider a multi-objective genetic algorithm \cite{katoch2021} with one fitness value for each of the two parts or one fitness value for each label (instead of performing the average over all labels $i \in \{1,\dots,m\}$ in \cref{eqn:app:ga:F:ent,eqn:app:ga:F:acc}).

\section{Q-tree measurements} \label{sec:app:q-tree measurements}
In \cref{sec:quantum representation:data encoding,sec:quantum representation:structure encoding}, we introduce the Q-tree data encoding and structure encoding, respectively. In the present section, we briefly show how the form of these encodings can be used to derive the probability distribution of the subsequent projective measurement, \cref{eqn:pT:p}, which is important for the discussion in \cref{sec:quantum representation:structure encoding}.\par
We begin by recognizing that $\bar{x}_{0}\cdots\bar{x}_{i-1}=\kappa(j, i, 1)\cdots\kappa(j, i, i)$ in fact represents a path within $T$. If the corresponding set of conditions $x_q=\kappa(j, i, u) \,\forall\, u \in \{1,\dots,i\}$ is fulfilled, then $a_{\Gamma}(x_1,\dots,x_k) = a(x_{V_1},\dots,x_{V_k})$, where we recall $\mathbf{V} \equiv \mathbf{V}^{1}(\mathbf{I}, i; \kappa(j, i, 1)\cdots\kappa(j, i, i))$, \cref{eqn:app:C:V}.\par
Thus,
\begin{align} \label{eqn:app:psi:t:p}
	\ket{\Psi''(T)} = \sum_{\mathbf{x},\mathbf{y}} \sqrt{p_{T}(\mathbf{x},\mathbf{y})} \ket{\mathbf{x},\mathbf{y}}
\end{align}
according to \cref{eqn:psi:t}. Here, we introduce
\begin{align} \label{eqn:app:pT}
	p_{T}(\mathbf{x},\mathbf{y}) \equiv \begin{cases} p(x_{T_1(1)},\dots,x_{T_k(1)},\mathbf{y}) & \text{for}\,\, x_q=\kappa(1, d, u) \,\forall\, u \in \{1,\dots,d\} \\ \hfil\vdots & \hfil\vdots \\ p(x_{T_1(2^d)},\dots,x_{T_k(2^d)},\mathbf{y}) & \text{for}\,\, x_q=\kappa(2^d, d, u) \,\forall\, u \in \{1,\dots,d\} \\ p(\mathbf{x},\mathbf{y}) & \text{otherwise} \end{cases}
\end{align}
and recall the vector transformation $\mathbf{T}(j) \equiv \mathbf{T}^{1}(d;\kappa(j, d, 1)\cdots\kappa(j, d, d))$, \cref{eqn:app:C:T} for all paths $j \in \{1,\dots,2^d\}$. From \cref{eqn:app:psi:t:p,eqn:app:pT}, we can straightforwardly infer \cref{eqn:pT:p}.

\section{Uncertainty} \label{sec:app:uncertainty}
In the present appendix section, we summarize how we evaluate the statistical and hardware-induced uncertainties of probabilities estimated from fractions of measurement counts \cite{part2020} using frequentist confidence intervals \cite{neyman1967}. First, we consider the ratio of a binomial random variable to a constant number of trials, which allows us to determine confidence intervals for \cref{eqn:pn:est,eqn:pq:pn:est}, respectively. Second, to determine confidence intervals of \cref{eqn:pyi:est,eqn:pyi:pred:cl}, we consider the case where the number of trials is also a binomial random variable. Finally, we use error propagation to estimate the uncertainty of \cref{eqn:pq:pyi:pred}. We conclude with a brief discussion of hardware-induced noise. A mathematical discussion of quantum sampling and its relation to classical sampling can be found, \eg, in \cite{bouman2012}.\par
First, presume a random variable $n \sim B( n; N, p )$ following the binomial distribution
\begin{align} \label{eqn:app:unc:B}
	B( n; N, p ) \equiv \binom{N}{n} p^n(1-p)^{N-n}
\end{align}
with $N \in \mathbb{N}$ independent Bernoulli trials of success probability $p \in [0,1]$ such that $n \in \{0,\dots,N\}$ represents the drawn number of successes. Since the expectation value of the fraction of successes obeys
\begin{align} \label{eqn:app:unc:exnN}
	\expectation{\frac{n}{N}} = p,
\end{align}
the success probability for a finite set of trials can be estimated by the fraction of successes according to
\begin{align} \label{eqn:app:unc:p}
	p \approx \hat{p}(n,N) \equiv \frac{n}{N}
\end{align}
with standard deviation
\begin{align} \label{eqn:app:unc:p:std}
	\stdev{\hat{p}(p,N)} = \sqrt{\frac{p(1-p)}{N}} \approx \stdeva{\hat{p}(n,N)} \equiv \sqrt{\frac{\hat{p}(n,N)(1-\hat{p}(n,N))}{N}},
\end{align}
which we use as a confidence interval \cite{brown2001}
\begin{align} \label{eqn:app:unc:pconf}
	\hat{p}(n,N) \pm z_{\alpha/2} \stdeva{\hat{p}(n,N)}
\end{align}
for $p$ with target error rate $\alpha \in [0,1]$. Here, $z_{\alpha/2}$ denotes the $1- \alpha/2$ quantile of the standard normal distribution. In particular, \cref{eqn:app:unc:pconf} can be used to estimate the confidence intervals of \cref{eqn:pn:est,eqn:pq:pn:est}, respectively. For this purpose, we set $n \equiv n(\mathcal{C}_{\nu})$ and identify $N$ with the total number of measurements.\par
Second, we consider the estimation of a fraction $k/m$ of two random variables $k, m \sim D(k, m; N, p', q)$, where the the probability distribution of jointly drawing $k$ and $m$ is given by the product
\begin{align} \label{eqn:app:unc:D}
	D(k, m; N, p', q) \equiv B( k; m, p' ) B_{1}( m; N, q )
\end{align}
with $p' \in [0,1]$, $q \in (0,1]$ and $N \in \mathbb{N}$ such that $k \in \{0,\dots,m\}$ and $m \in \{1,\dots,N\}$, respectively. Here, we recall the binomial distribution, \cref{eqn:app:unc:B}, and introduce the truncated binomial distribution
\begin{align}
	B_1( m; N, q ) \equiv \frac{B( m; N, q )}{b_0( N, q )},
\end{align}
which considers only one or more successes (\ie, $m \geq 1$) and contains the abbreviation
\begin{align}
	b_0( N, q ) \equiv 1 - B( 0; N, q ) = 1 - (1-q)^N \in (0,1]
\end{align}
with $b_0( N, q ) \approx 1$ for $N \gg 1$. One has
\begin{align} \label{eqn:app:unc:exkm}
	\expectation{\frac{k}{m}}  = p'
\end{align}
and
\begin{align} \label{eqn:app:unc:exmN}
	\expectation{\frac{m}{N}} = \frac{q}{b_0( N, q )},
\end{align}
respectively, in analogy to \cref{eqn:app:unc:exnN} and consequently
\begin{align}
	p' \approx \hat{p}'(k,m) \equiv \frac{k}{m}
\end{align}
in analogy to \cref{eqn:app:unc:p}. We use the corresponding standard deviation
\begin{align} \label{eqn:app:unc:pp:std}
	\stdev{\hat{p}'(N, p', q)} = \sqrt{p'(1-p') \expectation{\frac{1}{m}}} \approx \stdeva{\hat{p}'(k,m,N)} \equiv \sqrt{\hat{p}'(k,m)(1-\hat{p}'(k,m)) \zeta(N, m / N)}
\end{align}
with the expectation value
\begin{align}
	\expectation{\frac{1}{m}} = \frac{N q (1-q)^{N-1} {}_4F_2(\{1,1,1-N\}, \{2,2\}, \frac{q}{q-1}) }{b_0( N, q )} \equiv \zeta(N, q)
\end{align}
as a confidence interval
\begin{align} \label{eqn:app:unc:ppconf}
	\hat{p}'(k,m) \pm z_{\alpha/2} \stdeva{\hat{p}'(k,m,N)}
\end{align}
in analogy to \cref{eqn:app:unc:pconf}. Here,
\begin{align}
	{}_rF_s(\{a_1,\dots,a_r\}, \{b_1,\dots,b_s\}, z) \equiv \sum_{l=0}^{\infty} \frac{(a_1)_l \cdots (a_r)_l}{(b_1)_l \cdots (b_s)_l} \frac{z^l}{l!}
\end{align}
denotes the generalized hypergeometric function with the Pochhammer symbol $(\cdot)_l$. Furthermore, we use $q \approx m / N$ according to \cref{eqn:app:unc:exmN} with $N \gg 1$. \Cref{eqn:app:unc:ppconf} can be used to estimate the confidence interval of \cref{eqn:pyi:est,eqn:pyi:pred:cl}. For this purpose, we set $k \equiv n_i^{y_i}(\mathcal{C}_{\nu})$, $m \equiv n(\mathcal{C}_{\nu})$, and identify $N$ with the total number of measurements.\par
Finally, we determine the uncertainty of \cref{eqn:pq:pyi:pred} using standard error propagation (\ie, we neglect correlations and assume independent variables), which yields the standard deviation
\begin{align} \label{eqn:app:pq:pyi:pred:unc}
	\stdev{\hat{p}_{\mathrm{qm}}(y_i \,|\, p^q)} \equiv \sqrt{\sum_{\nu = 1}^{2^d} \Big[ \hat{p}(y_i \,|\, \mathcal{C}_{\nu})^2 \stdev{\hat{p}^{q}(\mathcal{C}_{\nu})}^2 + \hat{p}^{q}(\mathcal{C}_{\nu})^2 \stdev{\hat{p}(y_i \,|\, \mathcal{C}_{\nu})}^2 \Big]}.
\end{align}
Here, $\stdev{\hat{p}^{q}(\mathcal{C}_{\nu})}$ and $\stdev{\hat{p}(y_i \,|\, \mathcal{C}_{\nu})}$ represent the standard deviations of the corresponding probability densities, \cref{eqn:pyi:est,eqn:pq:pn:est}, which are given by \cref{eqn:app:unc:p:std,eqn:app:unc:pp:std}, respectively.\par
So far, we have only discussed statistical uncertainties. However, measurements on real quantum computers may also exhibit additional uncertainties from hardware imperfections. Such noise-induced uncertainties may, \eg, occur for \cref{eqn:pyi:pred:cl,eqn:pq:pn:est,eqn:pq:pyi:pred}. In general, noise-free measurement results from quantum computers are obtained by sampling from a probability distribution
\begin{align} \label{eqn:app:unc:P}
	P(\rho, \hat{M}) = \trace{ \hat{M}^{\dagger} \hat{M} \rho }
\end{align}
based on a density matrix $\rho$, which results from the application of gate operations on the initial state, and a measurement operator $\hat{M}$ (with Hermitian conjugate $\hat{M}^{\dagger}$). According to \cref{eqn:pT:p}, the probabilities $p$, $p'$, and $q$ from \cref{eqn:app:unc:B,eqn:app:unc:D} can be identified with an expression of the form of \cref{eqn:app:unc:P} when applied to \cref{eqn:pyi:pred:cl,eqn:pq:pn:est,eqn:pq:pyi:pred}, respectively.\par
When including noise from hardware imperfections \cite{saki2019}, one can use a master equation approach \cite{lindblad1976} to obtain 
\begin{align}
	P(\rho_{\epsilon}, \hat{M}) = \trace{ \hat{M}^{\dagger} \hat{M} \rho_{\epsilon} }
\end{align}
with the noise-perturbed density matrix $\rho_{\epsilon}$, which can be expressed by \cite{martinez2016}
\begin{align}
	\rho_{\epsilon} \approx \rho + \delta \rho
\end{align}
based on the approximated noise perturbation $\delta \rho$ such that
\begin{align}
	P(\rho_{\epsilon}, \hat{M}) \approx P(\rho, \hat{M}) + \trace{ \hat{M}^{\dagger} \hat{M} \delta \rho }.
\end{align}
This perturbation of the probability can also be included in \cref{eqn:app:unc:B,eqn:app:unc:D} if we treat the probabilities $p$, $p'$, and $q$ as random variables with a probability distribution determined by the hardware noise. Based on this presumption, we consider the generalizations
\begin{align} \label{eqn:app:unc:Be}
	B_{\epsilon}( n; N, p ) \equiv \int_{0}^{1} \epsilon(\tilde{p}; p) B( n; N, \tilde{p} ) \,\mathrm{d}\tilde{p}
\end{align}
and
\begin{align} \label{eqn:app:unc:De}
	D_{\epsilon}(k, m; N, p', q) \equiv \int_{0}^{1} \int_{0}^{1} \epsilon(\tilde{p}, \tilde{q}; p', q) D(k, m; N, \tilde{p}, \tilde{q}) \,\mathrm{d}\tilde{p} \,\mathrm{d}\tilde{q},
\end{align}
respectively. Here we introduce the probability density $\epsilon(\tilde{p}; p)$ with support $[0,1]$ and the probability density $\epsilon(\tilde{p}, \tilde{q}; p', q)$ with support $[0,1]^2$ to quantify the hardware-induced uncertainty.\par
Based on these presumptions, suitably chosen probability densities for $p$, $p'$, and $q$ can be used to model the noise (\eg, a truncated normal distributions for a heuristic approach or a distribution derived from the explicit form of $P(\rho_{\epsilon}, \hat{M})$ for a rigorous approach). Such a noise model may lead to a shift of expectation values, \cref{eqn:app:unc:exnN,eqn:app:unc:exkm}, which corresponds to hardware-induced systematic errors, and an increase of the standard deviations, \cref{eqn:app:unc:p:std,eqn:app:unc:pp:std}, which corresponds to a hardware-induced uncertainty.\par
In the noise-free case, these probability densities can be expressed in terms of Dirac delta distributions
\begin{align} \label{eqn:app:unc:eps:p}
	\epsilon(\tilde{p}; p) = \delta(\tilde{p} - p)
\end{align}
and
\begin{align} \label{eqn:app:unc:eps:pp}
	\epsilon(\tilde{p}, \tilde{q}; p', q) = \delta(\tilde{p} - p') \delta(\tilde{q} - q),
\end{align}
respectively, such that \cref{eqn:app:unc:Be,eqn:app:unc:De} reduce to \cref{eqn:app:unc:B,eqn:app:unc:D} and the statistical uncertainty becomes decisive. On the other hand, since the standard deviations from \cref{eqn:app:unc:p:std,eqn:app:unc:pp:std} decrease for an increasing number of trials $N$, we can write
\begin{align}
	B( n; N \gg 1, p ) \approx \delta\left(\frac{n}{N} - p\right)
\end{align}
and
\begin{align}
	D(k, m; N \gg 1, p', q) \approx \delta\left(\frac{k}{m} - p'\right) \delta\left(\frac{m}{N} - q'\right)
\end{align}
for a sufficiently large $N$. Hence, \cref{eqn:app:unc:Be,eqn:app:unc:De} reduce to
\begin{align} \label{eqn:app:unc:Be:eps}
	B_{\epsilon}( n; N, p ) \approx \epsilon\left(\frac{n}{N}; p\right)
\end{align}
and
\begin{align} \label{eqn:app:unc:De:eps}
	D_{\epsilon}(k, m; N, p', q) \approx \epsilon\left(\frac{k}{m}, \frac{m}{N}; p', q\right),
\end{align}
respectively. This means that in this case the statistical uncertainty is negligible and the hardware-induced uncertainty (if present) is dominating.

\section{Random forests} \label{sec:app:random forests}
Random forests represent a well-known predictive model composed of an ensemble of decision trees for which a prediction can be inferred from the (weighted) averaged predictions of the individual trees \cite{breiman2001}. Our proposed quantum representation of binary classification trees can be generalized to such ensembles using additional qubits and gates. The connection between decision tree ensembles and quantum computation has already been explored in previous works. For example, a conceptional quantum forest model based on quantum amplitude amplification is presented in \cite{khadiev2020}. A general discussion of quantum ensembles of quantum classifiers can be found in \cite{schuld2018b}. Furthermore, a quantum-inspired (\ie, involving no actual quantum computation) feature subset generation method for ensemble methods is presented in \cite{xie2017}. Furthermore, a discussion of probabilistic random forests can be found in \cite{correia2020}.\par
An ensemble of $f \in \mathbb{N}$ trees $T(1),\dots,T(f)$ of the same maximum depth $d$ can by characterized by the training data $\mathbf{D}_{\mathrm{train}}$, \cref{eqn:p:true}, and the $f$-tuple
\begin{align} \label{eqn:app:F}
	F \equiv (C(1), \dots, C(f)),
\end{align}
which consists of the compressed decision configurations of all trees, \cref{eqn:C}. For the $t$th tree $T(t)$, the compressed decision configuration is denoted by $C(t)$ with $t \in \{1,\dots,f\}$. An additional probability distribution $p_{\mathrm{trees}}$ with support $\{1,\dots,f\}$ represents the relative importance of every tree in the forest by which its prediction is weighted.\par
\begin{figure}
	\begin{center}
		\begin{tikzpicture}
\coordinate (qa) at (0,1);
\coordinate (q1) at (0,0);
\coordinate (d1) at (0,-.5);
\coordinate (qd) at (0,-1);
\coordinate (qd1) at (0,-1.5);
\coordinate (qd2) at (0,-2);
\coordinate (d2) at (0,-2.5);
\coordinate (qk) at (0,-3);
\coordinate (qy) at (0,-3.5);
\node[anchor=east] at ($(qa)+(-.25,0)$) {$\ket{\mathbf{a}}$};
\node[anchor=east] at ($(q1)+(-.25,0)$) {$\ket{x_1}$};
\node[anchor=east,xshift=-.25cm,yshift=.1cm] at ($(d1)+(-.25,0)$) {$\vdots$};
\node[anchor=east] at ($(qd)+(-.25,0)$) {$\ket{x_{d}}$};
\node[anchor=east] at ($(qd1)+(-.25,0)$) {$\ket{x_{d+1}}$};
\node[anchor=east] at ($(qd2)+(-.25,0)$) {$\ket{x_{d+2}}$};
\node[anchor=east] at ($(qk)+(-.25,0)$) {$\ket{x_k}$};
\node[anchor=east,xshift=-.25cm,yshift=.1cm] at ($(d2)+(-.25,0)$) {$\vdots$};
\node[anchor=east] (psiy) at ($(qy)+(-.25,0)$) {$\ket{\mathbf{y}}$};
\draw[thick] ($(qa)+(-.25,0)$) -- ($(qa)+(2.5,0)+(.6,0)$);
\draw[thick] ($(q1)+(-.25,0)$) -- ($(q1)+(2.5,0)+(.6,0)$);
\draw[thick] ($(qd)+(-.25,0)$) -- ($(qd)+(2.5,0)+(.6,0)$);
\draw[thick] ($(qd1)+(-.25,0)$) -- ($(qd1)+(2.5,0)+(.6,0)$);
\draw[thick] ($(qd2)+(-.25,0)$) -- ($(qd2)+(2.5,0)+(.6,0)$);
\draw[thick] ($(qk)+(-.25,0)$) -- ($(qk)+(2.5,0)+(.6,0)$);
\draw[thick] ($(qy)+(-.25,0)$) -- ($(qy)+(2.5,0)+(.6,0)$);
\draw[thick,draw=black,fill=white] ($(qa)+(.5,0)+(-.4,1.5)+(.2,0)$) rectangle ($(qa)+(.5,0)+(+.4,-.7)+(.2,0)$);
\node[rotate=90] (ua) at ($.5*(qa)+.5*(qa)+(.5,0)+(.2,0)+(0,.4)$) {$\hat{U}_{\mathrm{trees}}(p_{\mathrm{trees}})$};
\draw[thick,draw=black,fill=white] ($(q1)+(.5,0)+(-.4,.2)+(.2,0)$) rectangle ($(qy)+(.5,0)+(+.4,-.2)+(.2,0)$);
\node[rotate=90] (ui) at ($.5*(q1)+.5*(qy)+(.5,0)+(.2,0)$) {$\hat{U}_{\mathrm{data}}(\mathbf{D})$};
\draw[thick,draw=black,fill=white] ($(qa)+(1.7,0)+(-.4,.2)+(.4,0)$) rectangle ($(qk)+(1.7,0)+(+.4,-.2)+(.4,0)$);
\node[rotate=90] (ut) at ($.5*(qa)+.5*(qy)+(1.7,0)+(.4,0)$) {$\hat{U}_{\mathrm{forest}}(F)$};
\coordinate (ma) at ($(qa)+(2.5,0)+(.6,0)$);
\measurement{ma}
\node[anchor=west] at (m.east) {$\bar{\mathbf{a}}$};
\coordinate (m1) at ($(q1)+(2.5,0)+(.6,0)$);
\coordinate (md) at ($(qd)+(2.5,0)+(.6,0)$);
\coordinate (md1) at ($(qd1)+(2.5,0)+(.6,0)$);
\coordinate (my) at ($(qy)+(2.5,0)+(.6,0)$);
\measurement{m1}
\node[anchor=west] at (m.east) {$\bar{x}_1$};
\measurement{md}
\node[anchor=west] at (m.east) {$\bar{x}_{d}$};
\measurement{md1}
\node[anchor=west] at (m.east) {$\bar{x}_{d+1}$};
\draw[thick,dashed] ($(qd2)+(2.5,0)+(.6,0)$)-- ($(qd2)+(2.5,0)+(.75,0)+(.6,0)$);
\draw[thick,dashed] ($(qk)+(2.5,0)+(.6,0)$) -- ($(qk)+(2.5,0)+(.75,0)+(.6,0)$);
\measurement{my}
\node[anchor=west] at (m.east) {$\bar{\mathbf{y}}$};
\draw[decorate,decoration={brace,amplitude=10pt,raise=4pt},xshift=0pt,yshift=0pt] ($(qa)+(4,.5)+(.6,0)$) -- ($(qy)+(4,-.5)+(.6,0)$) node [anchor=west,midway,xshift=0.55cm] {$\bar{x}_1 \cdots \bar{x}_{d+1}\bar{\mathbf{y}}\bar{\mathbf{a}}$};
\node[Qv] (psi0) at ($(psiy)+(0,-.75)$) {$\ket{0}$};
\draw let \p1=(psi0), \p2=(ui) in node[Qv] (psi1) at ($(\x2,\y1)+(.7,0)$) {$\ket{\Psi_{\mathrm{forest}}'(p_{\mathrm{trees}},\mathbf{D})}$};
\draw let \p1=(psi1), \p2=(ut) in node[Qv] (psi2) at ($(\x2,\y1)+(.7,0)$) {};
\node[Qv,anchor=south] at ($(psi2)+(0,5.6)$) {$\ket{\Psi_{\mathrm{forest}}''(F,p_{\mathrm{trees}},\mathbf{D})}$};
\draw[thick,Qv,dotted] ($(psi1)+(0,.4)$) -- ($(psi1)+(0,5.4)$);
\draw[thick,Qv,dotted] ($(psi2)+(0,.6)$) -- ($(psi2)+(0,5.6)$);
\end{tikzpicture}
		\caption{Q-Forest (circuit). Layout for our proposed quantum representation of a random forest consisting of $f$ trees $T(1),\dots,T(f)$ of maximum depth $d$, which is uniquely defined by the training data $\mathbf{D}_{\mathrm{train}}$, the tuple of compressed decision configurations $F$, \cref{eqn:app:F}, and the probability distribution of the trees $p_{\mathrm{trees}}$. The Q-Forest is a generalization of the Q-tree, \cref{fig:tree-circuit}. Initially, all qubits are prepared in the ground state $\ket{0}$. The first set of unitary operators $\hat{U}_{\mathrm{data}}(\mathbf{D}_{\mathrm{train}})$ and $\hat{U}_{\mathrm{trees}}(p_{\mathrm{trees}})$ perform an encoding of the training data and decision tree distribution, respectively, that leads to the state $\ket{\Psi_{\mathrm{forest}}'(p_{\mathrm{trees}},\mathbf{D}_{\mathrm{train}})}$, \cref{eqn:app:psi:f1}. Second, $\hat{U}_{\mathrm{forest}}(F) \ket{\Psi_{\mathrm{forest}}'(p_{\mathrm{trees}},\mathbf{D}_{\mathrm{train}})}$ performs an encoding of the decision configuration that leads to the state $\ket{\Psi_{\mathrm{forest}}''(F,p_{\mathrm{trees}},\mathbf{D}_{\mathrm{train}})}$, \cref{eqn:app:psi:f2}. Third, a projective measurement on is performed to obtain a bit string, \cref{eqn:app:bitstring}, consisting of features, labels, and ancillas, which are drawn from \cref{eqn:app:pF} such that the circuit can be considered a quantum representation of the random forest.} \label{fig:forest-circuit}
	\end{center}
\end{figure}
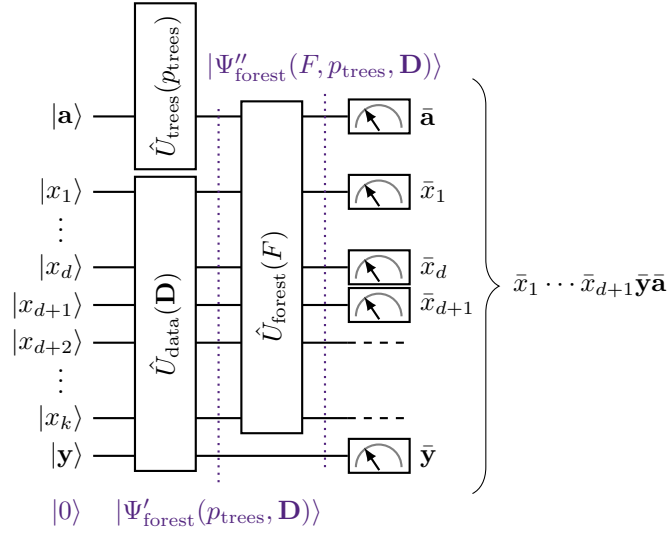

A quantum representation of such a forest can be realized in analogy to our Q-tree, \cref{sec:tree circuit}. In addition to the $k+m$ qubits that represent the features and labels, the \emph{Q-forest circuit}, which we also refer to as \emph{Q-forest} for short, requires
\begin{align} \label{eqn:app:f}
	f' \equiv \left\lceil \log_2 f \right\rceil
\end{align}
additional ancilla qubits (denoted by $\ket{a_1},\dots,\ket{a_{f'}}$ or their respective indices $k+m+1,\dots,k+m+f'$), which identify each tree. For the sake of simplicity, we assume that $f$ is chosen such that $f \geq 2$ and $\log_2 f \in \mathbb{N}$. We assume that all qubits are initially prepared in the ground state $\ket{0}$ in analogy to \cref{eqn:psi:0}. The circuit consists of three parts. First, two unitary operators responsible for the data encoding. Second, a unitary operator responsible for the structural encoding and, third, a projective measurement. The proposed Q-forest circuit layout is sketched in \cref{fig:forest-circuit}.
In the first step, a qsample encoding of the training data is performed by means of $\hat{U}_{\mathrm{data}}(\mathbf{D}_{\mathrm{train}})$ to obtain $\ket{\Psi'(F,\mathbf{D}_{\mathrm{train}})}$, \cref{eqn:psi:d}. In analogy, the distribution of trees $p_{\mathrm{trees}}$ is stored with a qsample encoding by means of $\hat{U}_{\mathrm{trees}}(p_{\mathrm{trees}})$, which only acts on the ancilla qubits such that 
\begin{align}
	\hat{U}_{\mathrm{trees}}(p_{\mathrm{trees}}) \ket{0} = \sum_{\mathbf{a}} \sqrt{p_{\mathrm{trees}}(\mathbf{a})} \ket{\mathbf{a}}.
\end{align}
Consequently, we arrive at
\begin{align} \label{eqn:app:psi:f1}
	\ket{\Psi_{\mathrm{forest}}'(p_{\mathrm{trees}},\mathbf{D}_{\mathrm{train}})} \equiv \ket{\Psi'(\mathbf{D}_{\mathrm{train}})} \otimes \hat{U}_{\mathrm{trees}}(p_{\mathrm{trees}}) \ket{\mathbf{a}}
\end{align}
for the total data encoding, where $\otimes$ denotes the tensor product.\par
\begin{figure}
	\begin{center}
		\begin{tikzpicture}
\coordinate (q0) at (0,0);
\coordinate (qx) at (0,-.5);
\coordinate (qd) at (0,-1);
\coordinate (a0) at (0,1.5);
\coordinate (ad) at (0,1);	
\node[anchor=east] at ($(q0)+(-.25,0)$) {$\ket{x_1}$};
\node[anchor=east,xshift=-.25cm,yshift=.1cm] at ($(qx)+(-.25,0)$) {$\vdots$};
\node[anchor=east] at ($(qd)+(-.25,0)$) {$\ket{x_k}$};
\node[anchor=east] at ($(a0)+(-.25,0)$) {$\ket{a_1}$};
\node[anchor=east] at ($(ad)+(-.25,0)$) {$\ket{a_2}$};	
\draw[thick] ($(q0)+(-.25,0)$) -- ($(q0)+(7.5,0)$);
\draw[thick] ($(qd)+(-.25,0)$) -- ($(qd)+(7.5,0)$);
\draw[thick] ($(a0)+(-.25,0)$) -- ($(a0)+(7.5,0)$);
\draw[thick] ($(ad)+(-.25,0)$) -- ($(ad)+(7.5,0)$);	
\draw[thick] ($(q0)+(0,2)$) rectangle ($(q0)+(7.25,-1.8)$);
% lvl 1
\draw[thick]  ($(q0)+(1,0)$) -- ($(a0)+(1,0)$);
\draw[thick]  ($(q0)+(3,0)$) -- ($(a0)+(3,0)$);
\draw[thick]  ($(q0)+(5,0)$) -- ($(a0)+(5,0)$);
\draw[thick]  ($(q0)+(6.5,0)$) -- ($(a0)+(6.5,0)$);
\node[draw,thick,fill=white,thick,minimum width=1.5cm,minimum height=.4cm,inner sep=0pt,rotate=90,font=\scriptsize] (cs1) at ($.5*(q0)+.5*(qd)+(1,0)$) {$\hat{\gamma}^0(c^0_1(1))$};
\node[draw,thick,fill=white,thick,minimum width=1.5cm,minimum height=.4cm,inner sep=0pt,rotate=90,font=\scriptsize] (cs2) at ($.5*(q0)+.5*(qd)+(3,0)$) {$\hat{\gamma}^0(c^0_1(2)$};
\node[draw,thick,fill=white,thick,minimum width=1.5cm,minimum height=.4cm,inner sep=0pt,rotate=90,font=\scriptsize] (cs3) at ($.5*(q0)+.5*(qd)+(5,0)$) {$\hat{\gamma}^0(c^0_1(3)$};
\node[draw,thick,fill=white,thick,minimum width=1.5cm,minimum height=.4cm,inner sep=0pt,rotate=90,font=\scriptsize] (cs4) at ($.5*(q0)+.5*(qd)+(6.5,0)$) {$\hat{\gamma}^0(c^0_1(4)$};
\node[draw,thick,fill=white,thick,minimum width=.4cm,minimum height=.4cm,inner sep=0pt] (t1l) at ($(a0)+(.5,0)$) {$X$};
\node[draw,thick,fill=white,thick,minimum width=.4cm,minimum height=.4cm,inner sep=0pt] at ($(ad)+(.5,0)$) {$X$};
\node[draw,thick,fill=white,thick,minimum width=.4cm,minimum height=.4cm,inner sep=0pt] at ($(a0)+(1,0)$) {$\alpha$};
\node[draw,thick,fill=white,thick,minimum width=.4cm,minimum height=.4cm,inner sep=0pt] at ($(ad)+(1,0)$) {$\alpha'$};
\node[draw,thick,fill=white,thick,minimum width=.4cm,minimum height=.4cm,inner sep=0pt] (t1r) at ($(a0)+(1.5,0)$) {$X$};
\node[draw,thick,fill=white,thick,minimum width=.4cm,minimum height=.4cm,inner sep=0pt] at ($(ad)+(1.5,0)$) {$X$};
\node[draw,thick,fill=white,thick,minimum width=.4cm,minimum height=.4cm,inner sep=0pt] (t2l) at ($(a0)+(2.5,0)$) {$X$};
\node[draw,thick,fill=white,thick,minimum width=.4cm,minimum height=.4cm,inner sep=0pt] at ($(a0)+(3,0)$) {$\alpha$};
\node[draw,thick,fill=white,thick,minimum width=.4cm,minimum height=.4cm,inner sep=0pt] at ($(ad)+(3,0)$) {$\alpha'$};
\node[draw,thick,fill=white,thick,minimum width=.4cm,minimum height=.4cm,inner sep=0pt] (t2r) at ($(a0)+(3.5,0)$) {$X$};
\node[draw,thick,fill=white,thick,minimum width=.4cm,minimum height=.4cm,inner sep=0pt] (t3l) at ($(ad)+(4.5,0)$) {$X$};
\node[draw,thick,fill=white,thick,minimum width=.4cm,minimum height=.4cm,inner sep=0pt] at ($(a0)+(5,0)$) {$\alpha$};
\node[draw,thick,fill=white,thick,minimum width=.4cm,minimum height=.4cm,inner sep=0pt] at ($(ad)+(5,0)$) {$\alpha'$};;
\node[draw,thick,fill=white,thick,minimum width=.4cm,minimum height=.4cm,inner sep=0pt] (t3r) at ($(ad)+(5.5,0)$) {$X$};
\node[draw,thick,fill=white,thick,minimum width=.4cm,minimum height=.4cm,inner sep=0pt] (t4) at ($(a0)+(6.5,0)$) {$\alpha$};
\node[draw,thick,fill=white,thick,minimum width=.4cm,minimum height=.4cm,inner sep=0pt] at ($(ad)+(6.5,0)$) {$\alpha'$};
% swap / cswap
\coordinate (b11) at ($(t1l)$);
\coordinate (b12) at ($(t1r)+(0,0)$);
\coordinate (b13) at ($(cs1.west)$);
\coordinate (b21) at ($(t2l)$);
\coordinate (b22) at ($(t2r)+(0,0)$);
\coordinate (b23) at ($(cs2.west)$);
\coordinate (b31) at ($(t3l)+(0,.5)$);
\coordinate (b32) at ($(t3r)+(0,.5)+(0,0)$);
\coordinate (b33) at ($(cs3.west)$);
\coordinate (b41) at ($(t4)$);
\coordinate (b42) at ($(t4)+(0,0)$);
\coordinate (b43) at ($(cs4.west)$);
\draw[thick] let \p1=(b11), \p2=(b12), \p3=(b13) in ($(\p1)+(-.25,.25)$) rectangle ($(\x2,\y3)+(.25,-.05)$);
\draw[thick] let \p1=(b21), \p2=(b22), \p3=(b23) in ($(\p1)+(-.25,.25)$) rectangle ($(\x2,\y3)+(.25,-.05)$);
\draw[thick] let \p1=(b31), \p2=(b32), \p3=(b33) in ($(\p1)+(-.25,.25)$) rectangle ($(\x2,\y3)+(.25,-.05)$);
\draw[thick] let \p1=(b41), \p2=(b42), \p3=(b43) in ($(\p1)+(-.25,.25)$) rectangle ($(\x2,\y3)+(.25,-.05)$);
\node[thick,font=\scriptsize] (xi1) at ($(cs1.west)+(0,-.3)$) {$\hat{\Xi}^i_j(c^0_1(1), 1)$};
\node[thick,font=\scriptsize] (xi2) at ($(cs2.west)+(0,-.3)$) {$\hat{\Xi}^i_j(c^0_1(2), 2)$};
\node[thick,font=\scriptsize] (xi3) at ($(cs3.west)+(0,-.3)$) {$\hat{\Xi}^i_j(c^0_1(3), 3)$};
\node[thick,font=\scriptsize] (xi4) at ($(cs4.west)+(0,-.3)$) {$\hat{\Xi}^i_j(c^0_1(4), 4)$};
% u
\draw [Qv,decorate,decoration={brace,amplitude=5pt}] ($(xi1)+(.75,0)+(0,-.35)$) -- ($(xi1)+(-.75,0)+(0,-.35)$) node [Qv,midway,yshift=-.45cm] (u0) {$\hat{U}_{\mathrm{forest}}^0(\mathbf{c}^0(1), 1)$};
\draw [Qv,decorate,decoration={brace,amplitude=5pt}] ($(xi2)-(.75,0)+(0,3.675)$) -- ($(xi2)-(-.75,0)+(0,3.675)$) node [Qv,midway,yshift=.45cm] (u1) {$\hat{U}_{\mathrm{forest}}^0(\mathbf{c}^0(2), 2)$};
\draw [Qv,decorate,decoration={brace,amplitude=5pt}] ($(xi3)+(.75,0)+(0,-.35)$) -- ($(xi3)+(-.75,0)+(0,-.35)$) node [Qv,midway,yshift=-.45cm] (u2) {$\hat{U}_{\mathrm{forest}}^0(\mathbf{c}^0(3), 3)$};
\draw [Qv,decorate,decoration={brace,amplitude=5pt}] ($(xi4)-(.25,0)+(0,3.675)$) -- ($(xi4)-(-.25,0)+(0,3.675)$) node [Qv,midway,yshift=.45cm] (u3) {$\hat{U}_{\mathrm{forest}}^0(\mathbf{c}^0(4), 4)$};
\draw [Qv,decorate,decoration={brace,amplitude=5pt}] ($(q0)+(7.25,0)+(0,-2.75)$) -- ($(q0)+(0,0)+(0,-2.75)$) node [Qv,midway,yshift=-.45cm] (u0) {$\hat{U}_{\mathrm{forest}}(F)$};
\end{tikzpicture}
		\caption{Example of the structural encoding of a decision forest consisting of $f=4$ trees of maximum depth $d=0$, which requires $f'=2$ ancilla qubits, \cref{eqn:app:f}. The structure is encoded by the unitary operator $\hat{U}_{\mathrm{forest}}(F)$, \cref{eqn:app:UF}, as defined in \cref{eqn:app:psi:f2}. Specifically, $\hat{U}_{\mathrm{forest}}^i(\mathbf{c}^i(t), t)$ from \cref{eqn:app:UF:ct} encodes the structure of the tree $t \in \{1,\dots,f\}$ in depth $i \in \{0,\dots,d\}$ defined by the corresponding compressed decision configuration $\mathbf{c}^i(t)$, \cref{eqn:c}. It consists of a composition of unitary operators $\hat{\Xi}^i_j(c, t)$, \cref{eqn:app:Xi}, which in turn consist of $\NOT$ gates and $\MCSWAP{2}$ gates, respectively. These $\MCSWAP{2}$ gates are effectively generalizations of $\hat{\gamma}^i(c)$, \cref{eqn:gamma}, which are additionally conditioned on the ancilla qubits. We denote $\NOT$ gates with $X$ and mark the additional control qubits of $\MCSWAP{2}$ gates with $\alpha$ and $\alpha'$, respectively, in analogy to the notation from \cref{fig:tree-compose} such that $\hat{\xi}^0(c,t) = \MCSWAP{2}( 1, c, \{ \alpha, \alpha' \})$, \cref{eqn:app:xi}. We also explicitly show consecutive $\NOT$ gates for the sake of clarity.} \label{fig:forest-compose}
	\end{center}
\end{figure}
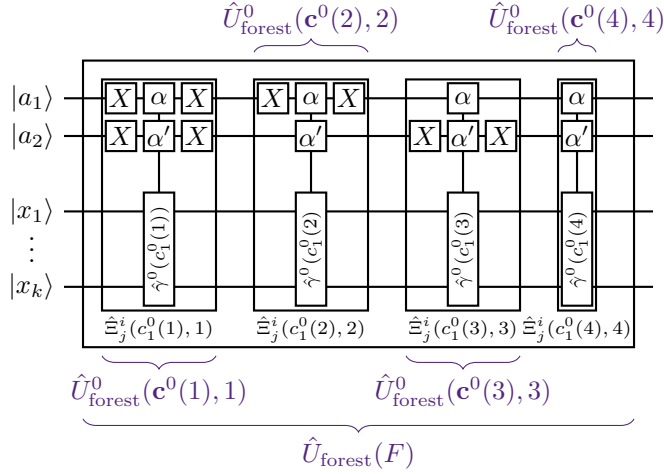

In the second step, the structure of the forest is encoded using
\begin{align} \label{eqn:app:psi:f2}
	\ket{\Psi_{\mathrm{forest}}''(F,p_{\mathrm{trees}},\mathbf{D}_{\mathrm{train}})} & = \hat{U}_{\mathrm{forest}}(F) \ket{\Psi_{\mathrm{forest}}'(p_{\mathrm{trees}},\mathbf{D}_{\mathrm{train}})} \nonumber\\
	& = \sum_{\mathbf{x},\mathbf{y},\mathbf{a}} \sqrt{p_{F}(\mathbf{x},\mathbf{y},\mathbf{a})} \ket{\mathbf{x},\mathbf{y},\mathbf{a}},
\end{align}
which only acts on the feature and ancilla qubits, respectively, where
\begin{align} \label{eqn:app:UF}
	\hat{U}_{\mathrm{forest}}(F) = \orderprod{t=1}{f} \orderprod{i=0}{d} \hat{U}_{\mathrm{forest}}^i(\mathbf{c}^i(t), t)
\end{align}
iterates over all trees $1,\dots,f$ and all layers $0,\dots,d$. Here we also introduce
\begin{align} \label{eqn:app:UF:ct}
	\hat{U}_{\mathrm{forest}}^i(\mathbf{c}^i(t), t) \equiv \orderprod{j=1}{2^i} \hat{\Xi}^i_j(c^i_j(t), t)
\end{align}
in analogy to \cref{eqn:UC:i}, where
\begin{align}\label{eqn:app:Xi}
	\hat{\Xi}^i_j(c, t) \equiv \hat{\rho}(t) \, \hat{\mu}^i_j \, \hat{\xi}^i(c,t) \, \hat{\mu}^i_j \, \hat{\rho}(t)
\end{align}
with $c \in \{i+1,\dots,k\}$ and we recall $\hat{\mu}^i_j$, \cref{eqn:mu}. The products in \cref{eqn:app:UF,eqn:app:UF:ct} are to be understood in the sense of \cref{eqn:orderprod}. Furthermore, we use the abbreviations
\begin{align} \label{eqn:app:xi}
	\hat{\xi}^i(c,t) \equiv \begin{cases} \MCSWAP{f'+i}( i+1, c, \{ k+m+1, \dots, k+m+f', 1, \dots, i \} ) & \text{for}\,\, i \geq 1 \\ \MCSWAP{f'}( i+1, c, \{ k+m+1, \dots, k+m+f' \} ) & \text{for}\,\, i = 0 \end{cases}
\end{align}
and
\begin{align}
	\hat{\rho}(t) \equiv \prod_{u=1}^{f'} \begin{cases} \NOT(u) & \text{for}\,\, \kappa(t, f', u) = 0 \\ \unitop & \text{otherwise} \end{cases}
\end{align}
for which we recall \cref{eqn:kappa} and the standard gates $\NOT$, $\MCSWAP{v}$ (with $f' \leq v \leq f'+d$), and $\unitop$, respectively, from \cref{sec:quantum representation:structure encoding}. Due to \cref{eqn:app:xi}, the structural information of each tree is additionally conditioned by the ancilla qubits. An example of this structural encoding is shown in \cref{fig:forest-compose}.\par
In the third and final measurement step, we simultaneously measure the first $d+1$ qubits (to obtain feature information), the qubits $k+1$ to $k+m$ (to obtain label information), and the qubits $k+m+1$ to $k+m+f'$ (to obtain ancilla information). Thus, we measure a bit string
\begin{align} \label{eqn:app:bitstring}
	\bar{x}_1\cdots\bar{x}_{d+1}\bar{\mathbf{y}}\bar{\mathbf{a}} \sim p_{F}(\bar{x}_{1},\dots,\bar{x}_{d+1},\bar{\mathbf{y}},\bar{\mathbf{a}})
\end{align}
similar to \cref{eqn:pT:bitstring} with
\begin{align} \label{eqn:app:pF}
	& \hphantom{{}={}} p_{F}(\bar{x}_{1},\dots,\bar{x}_{d+1},\bar{\mathbf{y}},\bar{\mathbf{a}}) \nonumber\\
	& \equiv p_{F}(x_1=\bar{x}_{1},\dots,x_{d+1}=\bar{x}_{d+1},\mathbf{y}=\bar{\mathbf{y}},\mathbf{a}=\bar{\mathbf{a}}) \nonumber\\
	& = \operatorname{Tr} \{ \braket{\bar{x}_1,\dots,\bar{x}_{d+1}, \bar{\mathbf{y}},\bar{\mathbf{a}} | \Psi_{\mathrm{forest}}''(F)} \braket{\Psi_{\mathrm{forest}}''(F) | \bar{x}_1,\dots,\bar{x}_{d+1},\bar{\mathbf{y}},\bar{\mathbf{a}}} \}_{\bar{x}_{d+2},\dots,\bar{x}_{k}} \nonumber\\
	& = p_{\mathrm{trees}}(\mathbf{a}=\bar{\mathbf{a}}) p_{T(\tau(\bar{\mathbf{a}}))}(\bar{x}_{1},\dots,\bar{x}_{d+1},\bar{\mathbf{y}}),
\end{align}
where we recall \cref{eqn:pT:p}. Since the tree structure is additionally conditioned on the ancilla, a specific tree from the forest 
\begin{align}
	\tau \equiv \tau(\bar{\mathbf{a}}) \equiv \sum_{j=1}^{f'} 2^{j-1} \bar{a}_{f'-j+1} + 1 \in \{1,\dots,f\}
\end{align}
is evaluated depending on the measured ancilla values $\bar{\mathbf{a}}$.\par
To induce a forest, the individual trees are typically induced independently on a subsample of the data that is randomly and with replacement drawn from $\mathbf{D}_{\mathrm{train}}$, a method also known as \emph{bootstrap aggregating} or \emph{bagging}. In addition, random subsets of features can optionally be selected for each tree, a method known as feature bagging. The method described in \cref{sec:quantum representation:tree induction} can therefore be applied for the induction of the Q-forest in the sense of a Q-tree ensemble, where each Q-tree is trained using its own data set. To determine the distribution of trees $p_{\mathrm{trees}}$, the relative performance of the previously induced trees can be determined (\eg, using the prediction accuracy of test data). Appropriately normalized, it can be directly used as a probability distribution such that trees with a high performance correlate with a high probability \cite{shahhosseini2021}.\par
A forest prediction can be performed by taking the weighted average of all tree predictions following the scheme in \cref{sec:quantum representation:tree predictions}. For a query request $\mathbf{x}^q \in \mathbb{B}^k$ and $N$ measurements, the measurement counts of a bit string of ancilla qubits $n(\mathbf{a})$ allow to estimate the tree distribution according to
\begin{align}
	p_{\mathrm{trees}}(\mathbf{a}) \approx \hat{p}_{\mathrm{trees}}(\mathbf{a}) \equiv \frac{n(\mathbf{a})}{N}
\end{align}
and can also be used together with the other parts of the measured bit strings to estimate the corresponding label probability distribution $\hat{p}_{T(\tau(\mathbf{a}))}(y_i \,|\, \mathcal{C}_{\nu})$, \cref{eqn:pn:yi}, for all $\nu \in \{1,\dots,2^d\}$ and all trees $\tau \in \{1,\dots,f\}$ according to \cref{eqn:pyi:est}. Here we recall the set of conditions $\mathcal{C}_{\nu}$, \cref{eqn:Cnu}. Thus, we can obtain the Q-forest prediction
\begin{align}
	\hat{p}_{\mathrm{cl}}(y_i \,|\, \mathbf{x}^q) \equiv \sum_{\mathbf{a}} \hat{p}_{\mathrm{forest}}(\bar{\mathbf{a}}) \hat{p}_{T(\tau(\mathbf{a}))}(y_i \,|\, \mathcal{C}_{q_d(\mathbf{x}^q)})
\end{align}
as a generalization of \cref{eqn:pyi:pred:cl}, where we recall $\mathcal{C}_{q_d(\mathbf{x}^q)}$, \cref{eqn:Cnu:q}.

\printbibliography

\end{document}